%% file: main.tex
\documentclass[12pt]{article}
\usepackage{wrapfig}
\usepackage{verbatim}
\usepackage{titletoc}

\input{math_commands}

\input{commands}

\definecolor{todored}{rgb}{0.75,0.10,0.10}

\IfFileExists{review/comments.tex}{\input{review/comments}}{\input{review_fallback}}
\ReviewCommentsOff

\title{From Experimental Artifacts to Solution Lineages}

\author{
    \fontsize{9.7pt}{12pt}\selectfont
    Jin Li$^{1{,}*}$, %
    Ahmed Murtadha$^{1{,}*}$, %
    Zhiyu Wang$^{1{,}*}$, %
    Qiwen Chen$^{1{,}5{,}*}$, %
    William Chen$^{1{,}*}$, %
    Yifei Wu$^{1{,}*}$, %
    Guan Wang$^{1{,}*}$, %
    Cheng Gong$^{1{,}*}$, %
    Andy L. Siy$^{1}$, %
    Jiayi Yang$^{1}$, %
    Mengsha Huang$^{3}$, %
    Wenhao Li$^{1{,}2}$, %
    Yixuan Liu$^{3}$, %
    Shuailin Pan$^{1{,}4}$, %
    Mingli Yuan$^{1}$, %
    Sen Song$^{3}$, %
    Yuhao~Sun$^{1{,}\dagger}$%
  \\[1.5ex]
  $^{1}$Sapient Intelligence
  \qquad
  $^{2}$Nanyang Technological University
  \qquad
  $^{3}$Tsinghua University
  \\
  $^{4}$Carnegie Mellon University
  \qquad
  $^{5}$University of Pennsylvania
  \\[1.2ex]
  \href{https://github.com/sapientinc/praxist}{\faGithub\;\texttt{github.com/sapientinc/praxist}}
  \qquad
  \href{https://praxist.sapient.inc/}{\faGlobe\;\texttt{praxist.sapient.inc}}
}

\begin{document}
\pagestyle{fancy}
% Front matter (title + abstract) numbered separately so hyperref anchors stay
% unique; the number itself is suppressed by the "first" page style.
\pagenumbering{roman}

% Cover page only: tighter top margin so the title block, abstract, and the
% graphical abstract all fit on page 1 (restored right after the abstract).
\newgeometry{left=1.75cm,right=1.75cm,top=1.3cm,bottom=2cm,
             headheight=20pt,headsep=10pt,footskip=25pt}

\maketitle
\thispagestyle{first}
\vspace{-1.5em}
\begingroup
\let\thefootnote\relax\footnotetext{$^{\dagger}$ Corresponding author. $^{*}$ Equal Contribution. Contact: \texttt{praxist@sapient.inc}. 
}
\endgroup

\begin{abstractpanel}
{\small\noindent\textbf{Abstract.}\hspace{0.5em}\input{sections/abstract}\par}

\vspace{0.55em}
\begingroup
\centering
\includegraphics[width=\linewidth]{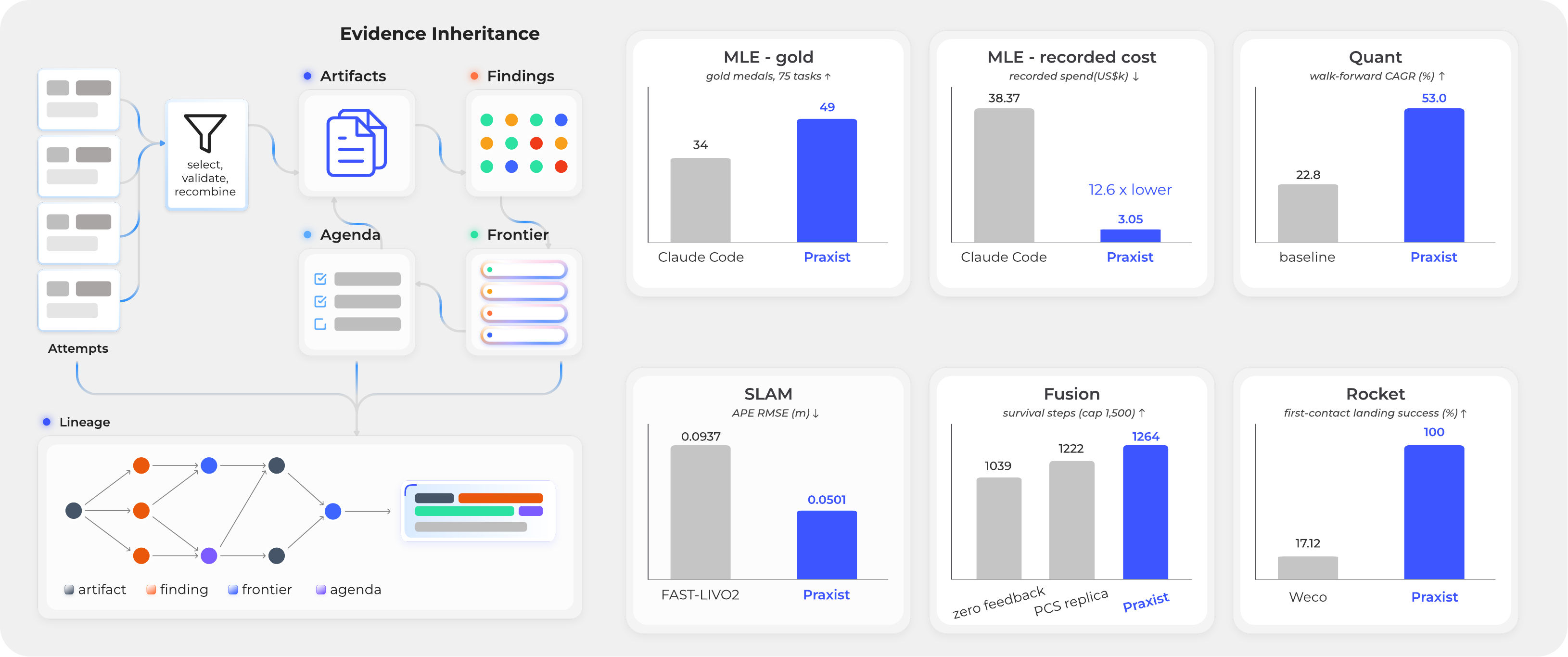}\par
\vspace{0.4em}
{\footnotesize\itshape Overview of \textsc{Praxist}: selective evidence inheritance turns
distributed, evaluated artifacts into lineage-grounded research state, improving
results on MLE-bench and in open-ended R\&D case studies.\par}
\endgroup
\end{abstractpanel}
\restoregeometry

% Page numbering starts after the abstract: the Introduction is page 1.
\clearpage
\pagenumbering{arabic}

\input{sections/1_intro}

\input{sections/2_method}

\input{sections/3_experiments}

\input{sections/7_conclusion}

\input{sections/availability}

\input{sections/acknowledgements}

\clearpage
\begin{hyphenrules}{nohyphenation}
\setlength{\bibsep}{.5ex plus .8ex}
\bibliographystyle{unsrtnat}
\bibliography{main}
\end{hyphenrules}

\clearpage
\appendix
\section*{Appendix}

\input{sections/8_supplementary}

\end{document}

%% file: math_commands.tex
\usepackage{amsmath,amsfonts,bm}

\def\Secref#1{Section~\ref{#1}}
\def\eqref#1{equation~\ref{#1}}
\def\1{\bm{1}}

\DeclareMathAlphabet{\mathsfit}{\encodingdefault}{\sfdefault}{m}{sl}
\SetMathAlphabet{\mathsfit}{bold}{\encodingdefault}{\sfdefault}{bx}{n}

%% file: commands.tex
\usepackage[left=2.5cm,
right=2.5cm,
top=2.3cm,
bottom=2.3cm,
headheight=20pt,
headsep=10pt,
footskip=25pt,
letterpaper]{geometry}
\usepackage[utf8]{inputenc}
\usepackage[T1]{fontenc}
\usepackage[english]{babel}
\usepackage{amsmath,amsfonts,amssymb,amsthm,thmtools}
\usepackage{graphicx}
\usepackage{float}
\usepackage[ruled]{algorithm}
\usepackage[hyperfootnotes=false]{hyperref}
\usepackage{fontawesome5}
\usepackage{fancyhdr}
\usepackage[normalem]{ulem}
\usepackage{graphicx}
\usepackage{stfloats}
\usepackage{wrapfig}
\usepackage{epstopdf}
\usepackage{cleveref}
\usepackage{subfloat}
\usepackage{subcaption}
\usepackage{xspace}
\usepackage{enumitem}
\usepackage{listings}
\usepackage{titlesec}
\usepackage{etoolbox}
\usepackage{setspace}
\usepackage{changepage}
\usepackage{etoolbox}
\usepackage{multirow}
\usepackage{booktabs}
\usepackage{tabularx}
\usepackage{colortbl}
\usepackage{wrapfig}
\usepackage{svg}
\usepackage[percent]{overpic}
\usepackage{tikz}
\usetikzlibrary{arrows.meta,calc,positioning}
\usepackage[round]{natbib}
\usepackage[colorinlistoftodos, shadow,color=blue!30!white
]{todonotes}
\usepackage{placeins}
\usepackage{xpatch}
\usepackage{siunitx}
\usepackage[htt]{hyphenat}
\usepackage[most]{tcolorbox}

\renewcommand{\sfdefault}{phv}

\fancypagestyle{first}{\fancyfoot[R]{}}

\setlist[itemize]{leftmargin=1em,itemsep=0ex,topsep=0ex}
\titlespacing*{\paragraph}{0pt}{0ex plus .1ex}{1ex}
\titlespacing*{\section}{0ex}{2.3ex plus .3ex minus .0ex}{.6ex plus .3ex minus .2ex}
\titlespacing*{\subsection}{0ex}{1.5ex plus .3ex minus .5ex}{.4ex plus .2ex minus .1ex}
\titlespacing*{\subsubsection}{0ex}{1.2ex plus .3ex minus .3ex}{.3ex plus .2ex minus .2ex}

\xapptocmd\normalsize{%
\abovedisplayskip=.8em plus .2em minus .2em
\belowdisplayskip=.6em plus .1em minus .1em
\abovedisplayshortskip=.8em plus .2em minus .2em
\belowdisplayshortskip=.6em plus .1em minus .1em
}{}{}

\setcitestyle{numbers}
\renewcommand{\cite}[1]{\citep{#1}}

\definecolor{medalgold}{HTML}{FDEBB2}
\definecolor{medalsilver}{HTML}{E4E7EB}
\definecolor{medalbronze}{HTML}{F0D8C0}

\definecolor{mydarkblue}{rgb}{0.0,0.15,0.7}
\definecolor{figureink}{HTML}{212121}
\definecolor{figuremuted}{HTML}{6E6E6E}
\definecolor{figureblue}{HTML}{3A86FF}
\definecolor{figurenavy}{HTML}{264653}
\definecolor{figureorange}{HTML}{FB5607}
\definecolor{figureviolet}{HTML}{5E56E8}
\definecolor{figuregreen}{HTML}{2A9D8F}
\definecolor{figurered}{HTML}{E11D48}
\definecolor{figureamber}{HTML}{F0A202}
\definecolor{figurebluefill}{HTML}{EFF5FF}
\definecolor{figurenavyfill}{HTML}{F2F4F5}
\definecolor{figureorangefill}{HTML}{FFF7F3}
\definecolor{figurevioletfill}{HTML}{F6F6FD}
\definecolor{figuregreenfill}{HTML}{F2F9F8}
\definecolor{figuregrayfill}{HTML}{FAFAFA}
\tikzset{
  case card/.style={rounded corners=3pt, line width=0.7pt, fill=white},
  case band/.style={rounded corners=3pt, line width=0.7pt, draw=figuremuted, fill=figuregrayfill},
  case flow/.style={-{Latex[length=3.2pt,width=4.2pt]}, line width=1.05pt, draw=figureink},
  case feedback/.style={-{Latex[length=2.8pt,width=3.8pt]}, line width=0.75pt, draw=figuremuted},
  case secondary/.style={-{Latex[length=2.6pt,width=3.4pt]}, densely dashed, line width=0.65pt, draw=figureink},
  case heading/.style={font=\sffamily\bfseries\fontsize{8.5}{9.5}\selectfont, align=center},
  case body/.style={font=\sffamily\fontsize{6.5}{7.5}\selectfont, text=figureink},
  case small/.style={font=\sffamily\fontsize{5.8}{6.7}\selectfont, text=figuremuted},
}
\definecolor{abstractfill}{HTML}{F2F3F8}
\definecolor{abstractframe}{HTML}{DADEE9}
\newtcolorbox{abstractpanel}{%
  enhanced, breakable,
  colback=abstractfill, colframe=abstractframe,
  boxrule=0.7pt, arc=11pt,
  left=18pt, right=18pt, top=11pt, bottom=11pt, boxsep=0pt,
}

\hypersetup{%
colorlinks=true,
linkcolor=mydarkblue,
citecolor=mydarkblue,
filecolor=mydarkblue,
urlcolor=mydarkblue}

\makeatletter
  \renewcommand{\maketitle}{%
    \begingroup
      \setlength{\parindent}{0pt}%
      \raggedright
      \includegraphics[height=1.05cm]{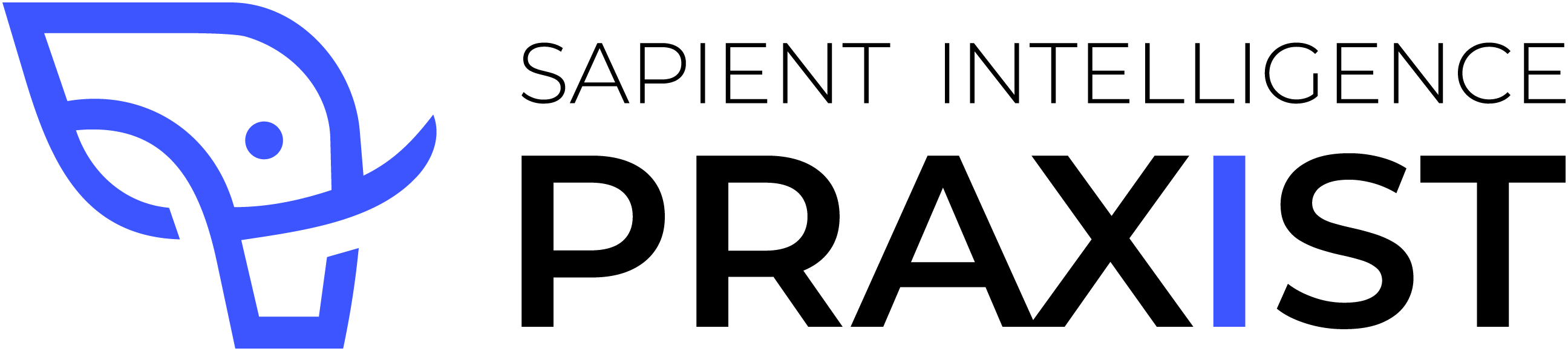}\par
      \vskip 1.0em
      {\LARGE\bfseries\@title\par}%
      \vskip 0.75em
      {\@author\par}%
      \vskip 0.28in minus 0.1in
    \endgroup
  }
\makeatother

%% file: review/comments.tex
% ======================================================================

\makeatletter

% ---- visibility switch ------------------------------------------------
\newif\ifrc@show
\rc@showtrue
\newcommand{\ReviewCommentsOff}{\rc@showfalse}
\newcommand{\ReviewCommentsOn}{\rc@showtrue}

% ---- palette (dark enough to read at \footnotesize, distinct in print)
\definecolor{rcErrorCol}{HTML}{C62828}   % red     — error
\definecolor{rcWarnCol}{HTML}{B45309}    % orange  — warning
\definecolor{rcSuggCol}{HTML}{1D4ED8}    % blue    — suggestion / proposed text
\definecolor{rcQuestCol}{HTML}{7B1FA2}   % purple  — question
\definecolor{rcExpCol}{HTML}{0F766E}     % teal    — experiment proposal
\definecolor{rcNoteCol}{HTML}{57606A}    % gray    — note
\definecolor{rcAnsCol}{HTML}{15803D}     % green   — author answer / applied
\definecolor{rcDelCol}{HTML}{C62828}     % red     — struck-out old text
\definecolor{rcMarkCol}{HTML}{D97706}    % amber   — wavy anchor underline

% ---- counters (per type, so ids read E1, E2, W1, Q1, ...) -------------
\newcounter{rc@error}
\newcounter{rc@warn}
\newcounter{rc@sugg}
\newcounter{rc@quest}
\newcounter{rc@exp}
\newcounter{rc@note}
\newcounter{rc@sub}
\newcounter{rc@app}

% ---- shared renderer --------------------------------------------------
% \rc@comment{color}{counter}{letter}{LABEL}{slug}{body}
\DeclareRobustCommand{\rc@comment}[6]{%
  \ifrc@show
    \stepcounter{#2}%
    \begingroup
      \sffamily\footnotesize\color{#1}%
      \textbf{[#3\arabic{#2}\,\textperiodcentered\,#4%
        \ifblank{#5}{}{\,\textperiodcentered\,#5}]} #6%
    \endgroup
  \fi
}

% ---- typed comments ---------------------------------------------------
\DeclareRobustCommand{\cerror}[2][]{\rc@comment{rcErrorCol}{rc@error}{E}{ERROR}{#1}{#2}}
\DeclareRobustCommand{\cwarn}[2][]{\rc@comment{rcWarnCol}{rc@warn}{W}{WARN}{#1}{#2}}
\DeclareRobustCommand{\csugg}[2][]{\rc@comment{rcSuggCol}{rc@sugg}{S}{SUGGEST}{#1}{#2}}
\DeclareRobustCommand{\cquest}[2][]{\rc@comment{rcQuestCol}{rc@quest}{Q}{QUESTION}{#1}{#2}}
\DeclareRobustCommand{\cexp}[2][]{\rc@comment{rcExpCol}{rc@exp}{X}{EXPERIMENT}{#1}{#2}}
\DeclareRobustCommand{\cnote}[2][]{\rc@comment{rcNoteCol}{rc@note}{N}{NOTE}{#1}{#2}}

% ---- author answer (used inside any comment's braces) -----------------
% Rendered in green so replies stand out from the surrounding comment.
% It only ever appears inside a comment body, so it vanishes together
% with the comment when \ReviewCommentsOff is active.
\DeclareRobustCommand{\cans}[1]{%
  \begingroup
    \sffamily\footnotesize\color{rcAnsCol}%
    \textbf{\ $\hookrightarrow$~AUTHOR:} #1%
  \endgroup
}

% ---- span anchor ------------------------------------------------------
% \chl{exact span of paper text}{comment about that span}
% Visible: span keeps normal ink but gets an amber wavy underline (the
% reviewer's squiggle), immediately followed by the comment.
% Hidden: the span prints untouched; the comment is discarded.
% Colored wave without ulem's color-leakage quirk: color only the repeated
% squiggle glyph inside \markoverwith (same construction ulem uses for
% \uwave), so the spanned text keeps its own ink.
\DeclareRobustCommand{\rc@wave}[1]{%
  \bgroup
  \markoverwith{\lower3.5\p@\hbox{\textcolor{rcMarkCol}{\sixly\char58}}}%
  \ULon{#1}% \ULon closes the \bgroup itself
}
\DeclareRobustCommand{\chl}[2]{%
  \ifrc@show
    \rc@wave{#1}\,#2%
  \else
    #1%
  \fi
}

% ---- proposed rewrite -------------------------------------------------
% \csub[slug]{old}{new}: a PROPOSAL. Old text struck out in red, new text
% in blue. Hidden mode shows the ORIGINAL text, because the proposal has
% not been accepted — hiding comments must recover the clean paper.
\DeclareRobustCommand{\csub}[3][]{%
  \ifrc@show
    \stepcounter{rc@sub}%
    {\sffamily\footnotesize\color{rcSuggCol}%
      \textbf{[R\arabic{rc@sub}\,\textperiodcentered\,REWRITE%
        \ifblank{#1}{}{\,\textperiodcentered\,#1}]}\,}%
    {\color{rcDelCol}\sout{#2}}%
    {\sffamily\footnotesize\color{rcSuggCol}\,$\to$\,}%
    {\color{rcSuggCol}#3}%
  \else
    #2%
  \fi
}

% ---- applied rewrite (visible audit trail) ----------------------------
% \capplied[slug]{old}{new}: the change has been ACCEPTED and applied;
% the new text is real paper text (normal ink), the old text stays
% visible struck-out for auditing. Hidden mode shows only the NEW text.
\DeclareRobustCommand{\capplied}[3][]{%
  \ifrc@show
    \stepcounter{rc@app}%
    {\sffamily\footnotesize\color{rcAnsCol}%
      \textbf{[A\arabic{rc@app}\,\textperiodcentered\,APPLIED%
        \ifblank{#1}{}{\,\textperiodcentered\,#1}]}\,}%
    {\color{rcDelCol}\sout{#2}}%
    {\sffamily\footnotesize\color{rcAnsCol}\,$\to$\,}%
    #3%
  \else
    #3%
  \fi
}

\makeatother

%% file: review_fallback.tex
% Fallback definitions used when the review/ submodule is NOT checked out
% (e.g. on Overleaf, whose git bridge does not fetch submodule contents, or
% after a plain "git clone" without --recurse-submodules).
%
% This is a copy of the canonical review/review_fallback.tex from the
% latex-review-skill repo. It must live in the repo root (not only in the
% submodule) because it is needed precisely when review/ is absent; keep the
% two files in sync when comments.tex gains new commands.
%
% main.tex loads review/comments.tex when present and this file otherwise.
% These no-ops reproduce the \ReviewCommentsOff semantics exactly: typed
% comments vanish, \chl prints its span untouched, \csub falls back to the
% ORIGINAL text, \capplied keeps the NEW text. The paper therefore builds
% identically to a hidden-comments build even without the submodule.
\newcommand{\ReviewCommentsOff}{}
\newcommand{\ReviewCommentsOn}{}
\newcommand{\cerror}[2][]{}
\newcommand{\cwarn}[2][]{}
\newcommand{\csugg}[2][]{}
\newcommand{\cquest}[2][]{}
\newcommand{\cexp}[2][]{}
\newcommand{\cnote}[2][]{}
\newcommand{\cans}[1]{}
\newcommand{\chl}[2]{#1}
\newcommand{\csub}[3][]{#2}
\newcommand{\capplied}[3][]{#3}

%% file: sections/abstract.tex
Autonomous R\&D agents now write, run, and improve executable artifacts under automated evaluation---but largely as laboratory instruments: shown on curated benchmarks, with gains that are hard to trace to a cause and costs well above what sustained engineering practice absorbs. The limitation is structural. Most systems treat each attempt as nearly self-contained, so logs, memories, and search trees record what happened without establishing which design element produced an improvement, whether its evidence survived validation, or how it recombines with others. Long campaigns therefore keep re-learning the same lessons.
We introduce \textsc{Praxist}, a lineage-centered generational system that converts reproducible artifacts and evaluator outcomes into a typed evidence graph of findings, lane-structured frontiers, and agendas. Separating local artifact construction from cohort-level evidence synthesis lets later attempts inherit validated mechanisms, unresolved claims, and useful constraints, and leaves results attached to an inspectable lineage.
On the standardized 75-task MLE-bench suite, the finalized official-grader results give \textsc{Praxist} 60 medals (80.0\%), 49 of them gold, against 55 medals (73.3\%) and 34 gold for a Claude Code baseline on Claude Opus 4.8---at a recorded model spend of US\$3{,}054 versus US\$38{,}370, roughly a twelfth of the cost. Five case studies---quantitative trading, LiDAR-inertial-visual SLAM, tokamak magnetic control, rocket landing, and job-shop rescheduling---carry the same process into open-ended engineering problems, improving on each task-native baseline in headline accuracy, survival, or resource cost, with the discovery path on record.
Stronger artifacts at an order of magnitude less spend, each backed by an auditable lineage, are, to our knowledge, first brought together here: the operating profile production research requires, not the one a benchmark demonstration establishes.

%% file: sections/1_intro.tex
\section{Introduction}
\label{sec:introduction}

\begin{figure}[b]
    \centering
    \includegraphics[width=0.87\textwidth]{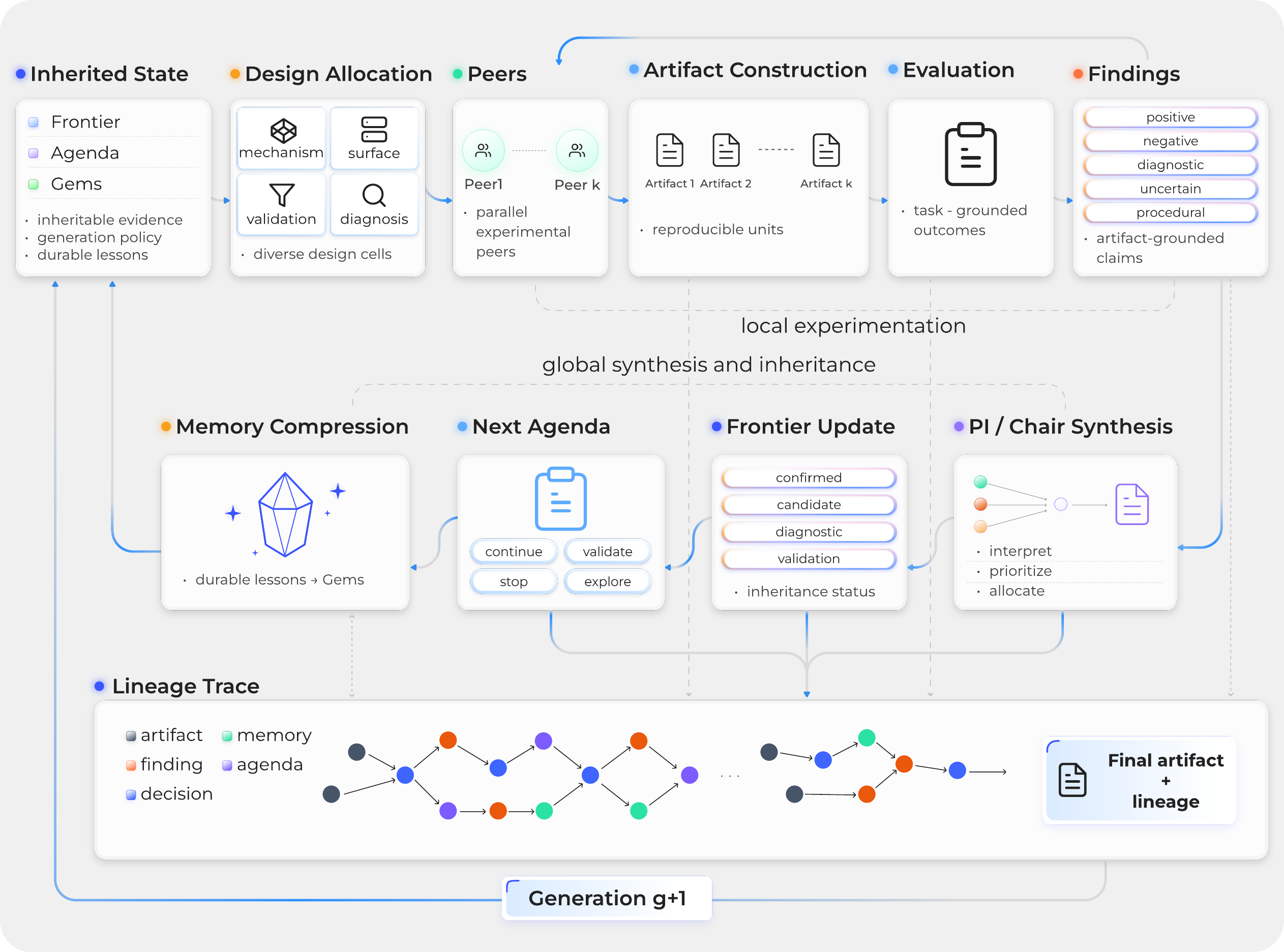}
    \caption{One generation of the \textsc{Praxist} loop. Agents inherit the accumulated evidence and run deliberately different experiments, each producing a reproducible artifact scored by an external evaluator (top). Outcomes become typed claims, pooled into updated evidence and the next generation's plan: continue, stop, validate, or explore (middle). A lineage records every step (bottom).}
    \label{fig:praxis_method_overview}
\end{figure}

Complex engineering and scientific breakthroughs are rarely found in a single attempt; they are constructed cumulatively from validated intermediate discoveries.
Recognizing this, recent autonomous R\&D systems solve problems by constructing executable artifacts, evaluating them with task-grounded feedback, and using the results to guide later attempts \citep{lu2024ai_scientist,yamada2025ai_scientist_v2,schmidgall2025agent_laboratory,romera_paredes2024funsearch,novikov2025alphaevolve,jiang2025aide}. The same pattern underpins machine-learning engineering benchmarks \citep{huang2023mlagentbench,chan2024mlebench,nathani2025mlgym,wijk2024rebench,chen2025mlrbench,jing2024dsbench}, agents for software engineering \citep{jimenez2024swebench,yang2024sweagent,wang2025openhands}, and scientific reproduction benchmarks \citep{siegel2024corebench,starace2025paperbench}: language models sustain long-horizon discovery when generation is coupled to automated evaluation.

As these systems move from isolated attempts to long-running campaigns, they must decide how previous attempts shape future construction. Most preserve campaign state as a search tree over candidate solutions \citep{jiang2025aide,lu2024ai_scientist,yamada2025ai_scientist_v2, romera_paredes2024funsearch,novikov2025alphaevolve}. As the tree grows, each node artifact becomes increasingly compositional, combining data choices, algorithms, hyper-parameters, implementation techniques, and diagnostics from many attempts. Since a mechanism's value may emerge only through combination, artifact-level search can prune components before their utility becomes visible. Campaigns therefore generate extensive evaluated experience, yet only a limited portion of what they learn is retained in a form that later attempts can build on.

Assembly theory suggests a different primitive: it characterizes complex objects through the formation histories that build them from reusable substructures, with selection determining which persist \citep{sharma2023assembly,marshall2021assemblybiosignatures,kempes2025assemblycomplexity}. Because every step operates on parts already retained, cost scales with the number of assembly operations rather than the size of the space, and each element is amortized across later constructions. Search then need not foresee which variant will matter, only retain the right elements in the right roles and keep them recombinable. We call this conversion of evaluated outcomes into actionable, artifact-grounded state evidence inheritance.

Several traditions supply relevant capabilities: reflection and memory reuse observations across attempts \citep{shinn2023reflexion,madaan2023self_refine,park2023generativeagents,wang2023voyager}; multi-agent systems distribute planning and revision across roles \citep{li2023camel,wu2023autogen,hong2023metagpt}; graph structures organize agent computation, reasoning, and memory \citep{zhuge2024gptswarm,besta2024graphofthoughts,liu2025graphaugmented,yang2026graphmemory,edge2024graphrag}; quality-diversity search maintains coverage across high-performing regions \citep{mouret2015map_elites,pugh2016quality,cully2017quality}; and provenance systems link outputs to the processes producing them \citep{groth2013prov,soilandreyes2022rocrate,zaharia2018mlflow,greff2017sacred}. Cumulative construction also requires a common interface between evaluation and inheritance: each result must state what was learned, how strongly it is supported, what grounds it, and its role in later work.

We introduce \textsc{Praxist}, a lineage-centered system that implements this interface for evaluator-grounded autonomous R\&D. \textsc{Praxist} organizes each campaign as an artifact-to-lineage process that repeats over generations, separating parallel artifact construction from cohort-level evidence synthesis; Figure~\ref{fig:praxis_method_overview} traces one such generation. Each generation begins from inherited evidence and produces \emph{design contracts}: pre-artifact plans fixing a mechanism, intervention surface, research intent, parent lineage, evidence signature, and validation hook. A cohort-level allocation rule, \emph{Quantified Diversity} (QD), spreads these contracts across distinct design cells of mechanism family, intervention surface, and intent, keeping coverage broad.

A cohort of \emph{peers} constructs reproducible artifacts against the assigned contracts, an external evaluator scores each one, and the outcomes are interpreted into typed \emph{findings} that PI roles read from complementary perspectives. Synthesis promotes eligible findings onto a \emph{frontier} with confirmed, candidate, diagnostic, and validation lanes, and a Chair turns the panel memos and updated frontier into the next generation's \emph{agenda} of continue, stop, validate, and explore decisions. Durable cross-generation lessons are compressed into \emph{Gems}. Together they form a lineage graph that stays active throughout the campaign and accompanies the final artifact as a record of its formation.

We evaluate \textsc{Praxist} on all 75 MLE-bench tasks against a local Claude Code baseline, earning 60 medals (80.0\%) and 49 golds against 55 (73.3\%) and 34 for Claude Code on Opus 4.8. In five open-ended case studies---quantitative trading, LiDAR-inertial-visual SLAM, tokamak magnetic control, rocket landing, and job-shop rescheduling---it improves the task-native baseline's headline metric in three domains, leads on survival but not on full-horizon precision in tokamak control, and, in SLAM, cuts the baseline's visual-compute cost at unchanged accuracy, while the accompanying lineages document mechanism discovery, validation, recombination, and boundary testing, so that each reported gain arrives with an auditable account of how it was reached.

Our contributions are:
\begin{itemize}
    \item We formulate \emph{evidence inheritance} as a systems requirement for long-horizon evaluator-grounded R\&D: prior evidence must be selectively retained and recombined rather than merely stored, as assembly theory describes objects built from reusable parts under selection.
    \item We develop \textsc{Praxist}, a generational system that converts evaluated artifacts into typed, inheritable evidence---findings, frontier lanes, agendas, and Gems---and keeps the resulting lineage graph active during research and inspectable afterwards.
    \item We evaluate \textsc{Praxist} on all 75 MLE-bench tasks and five open-ended case studies, scoring outcomes, lineage process measures, and cost against local baselines.
\end{itemize}

%% file: sections/2_method.tex
\section{Method}
\label{sec:method}

\subsection{Overview and Formalization}
\label{sec:method_overview}

\textsc{Praxist} runs autonomous R\&D as an \emph{artifact-to-lineage} process. Every attempt is materialized as a reproducible \emph{artifact}; an evaluated artifact is interpreted into one or more \emph{findings}; findings are promoted onto a \emph{frontier} of inheritable evidence; the frontier is synthesized into an \emph{agenda} that directs the next round of attempts; and a campaign concludes by reporting a final artifact together with the \emph{lineage} records of how it was produced:
\[
\text{Artifact} \;\rightarrow\; \text{Finding} \;\rightarrow\; \text{Frontier} \;\rightarrow\; \text{Agenda} \;\rightarrow\; \text{Lineage}.
\]
The central design choice is that evidence is never inherited as a raw transcript or a scalar score. Before any evidence is allowed to influence future work, \textsc{Praxist} assigns it an explicit \emph{operational role}: a mature parent to build on, a fragile candidate to validate, a failure to avoid, or a lesson to remember. Systems that carry state as score-ranked artifact collections---evolving program databases, solution trees---inherit an untyped object \citep{romera_paredes2024funsearch,novikov2025alphaevolve,jiang2025aide,lu2024ai_scientist,yamada2025ai_scientist_v2}; \textsc{Praxist} inherits typed findings whose frontier lanes make validation status explicit, agenda decisions that assign each element an operational role, and, where enabled, Gems that compress durable lessons across generations.

\paragraph{Problem formalization.}
We formalize \textsc{Praxist} as a generational state-transition process. Let $g=0,1,\dots,G{-}1$ index generations. A generation is the synchronization boundary at which peer-level evidence becomes shared research state. The state inherited into generation $g$ is
\begin{equation}
\label{eq:state}
\mathcal{S}_g \;=\; \bigl(\mathcal{F}_g,\; \mathcal{A}_g,\; \mathcal{G}_g,\; \mathcal{L}_g\bigr),
\end{equation}
where $\mathcal{F}_g$ is the frontier, $\mathcal{A}_g$ is the agenda for generation $g$, $\mathcal{G}_g$ is the set of Gems (durable cross-generation lessons), and $\mathcal{L}_g$ is the lineage trace accumulated so far. Table~\ref{tab:method_notation} collects the notation.
Each generation reads $\mathcal{S}_g$, executes a cohort of $C$ parallel \emph{peers}, and emits the successor state $\mathcal{S}_{g+1}$; Algorithm~\ref{alg:praxis} states the full cycle depicted in Figure~\ref{fig:praxis_method_overview}, with the peer-owned span---peers through findings---bracketed as \emph{local experimentation} in the figure (design allocation is itself a cohort-level decision). Full field-level schemas for every object are collected in Appendix~\ref{app:additional-methods} and referenced by compact pointers below; the main text remains self-contained. One convention applies throughout: the vocabularies named below---finding types, inheritance actions, frontier lanes, evidence stages, and agenda dispositions---are the \emph{method-level} roles that inheritance is defined over, not the literal field values of any one deployment. The implementation exposes a task-facing enum for finding records and lets each task realization declare its own frontier lanes---or none---and its own evidence stages, which the campaign then maps onto these roles; Appendix~\ref{app:additional-methods} records that correspondence and where a task's own labels are the operative ones.

\paragraph{One generation at a glance.}
Figure~\ref{fig:praxis_method_overview} shows a single generation end to end, and the rest of this section follows it box by box. The \textbf{top row} is \emph{local experimentation}: it turns inherited evidence into new evaluated evidence. A generation starts from the \emph{inherited state} on the left---the \emph{frontier} of evidence eligible to influence this generation, the \emph{agenda} that says what this generation should do, and the \emph{Gems} that carry durable lessons across many generations. \emph{Design allocation} (\S\ref{sec:method_alloc}) turns that state into one design contract per peer and deliberately spreads the cohort across distinct design cells---different mechanism families, intervention surfaces, and intents---so that a generation tests several hypotheses rather than crowding the one direction that currently looks best. The \emph{peers} are $C$ autonomous experimental workers that run concurrently, each owning a single contract. Each peer builds a \emph{reproducible artifact}, the self-contained unit of work that makes an attempt re-runnable and inspectable, and an \emph{external evaluator}---the task's own metric, never the agent's self-assessment---scores it. The evaluated artifact is then interpreted \emph{against the contract that produced it} into typed \emph{findings} (\S\ref{sec:method_local}), so that a result is recorded as an artifact-grounded claim with a status (positive, negative, diagnostic, uncertain, procedural) rather than as a number. The arrow returning across the top of the row is the peer's own inner loop: within a generation, a peer may revise its artifact against its own evaluated findings before the generation boundary.

The \textbf{bottom row} is \emph{global synthesis and inheritance}: it decides what the run keeps. Reading right to left, a panel of \emph{PI} roles and a \emph{Chair} (\S\ref{sec:method_synthesis}) interpret the generation's pooled findings from complementary perspectives and arbitrate between them. The \emph{frontier update} then sorts the surviving evidence into inheritance lanes---confirmed, candidate, diagnostic, and validation in the configuration used here---so that a later generation inherits not only what was measured but how far it can be trusted and what should be done with it. The \emph{next agenda} converts that into a policy for the following generation, issuing per-direction verdicts to continue, validate, stop, or explore. Where memory compression is enabled, every $\rho$ generations it distills recurring lessons into Gems, keeping long runs within a bounded context. Together, the updated frontier, agenda, and Gems form the state inherited by generation $g{+}1$, closing the loop back to the left of the figure. Running beneath both rows, the \emph{lineage trace} (\S\ref{sec:method_memory}) records research objects as they are produced---artifacts, findings, decisions, agendas, memory---and the typed relations between them; it is the run's second deliverable, reported alongside the final artifact as an account of how the outcome was reached.

\begin{table}[t]
\centering
\small
\caption{Notation used in the method; lane names and vocabularies are those of the configuration used in this work. Detailed schemas are given in Appendix~\ref{app:additional-methods}.}
\label{tab:method_notation}
\begin{tabular}{ll}
\toprule
Symbol & Meaning \\
\midrule
$g,\; G$ & generation index and number of generations \\
$\mathcal{S}_g=(\mathcal{F}_g,\mathcal{A}_g,\mathcal{G}_g,\mathcal{L}_g)$ & research state inherited into generation $g$ \\
$C$ & cohort size: number of parallel peers per generation \\
$p_i,\; i\in\{1,\dots,C\}$ & the $i$-th peer of a generation \\
$d_i$ & design contract assigned to peer $p_i$ \\
$c_i\in\mathcal{C}$ & design cell occupied by $d_i$ in the behavior space $\mathcal{C}$ \\
$a_i$ & reproducible artifact constructed by peer $p_i$ \\
$e_i=\textsc{Eval}(a_i)$ & external-evaluator outcome for artifact $a_i$ \\
$\phi$ & a finding; $\Phi_g=\bigcup_i \Phi_i$ the generation's findings \\
$\mathcal{F}=\mathcal{F}^{\mathrm{cf}}\!\cup\mathcal{F}^{\mathrm{cd}}\!\cup\mathcal{F}^{\mathrm{dg}}\!\cup\mathcal{F}^{\mathrm{vl}}$ & lanes used here: confirmed, candidate, diagnostic, validation \\
$\mu_r$ & memo emitted by PI role $r$ \\
$\mathcal{A}$ & agenda: per-direction directives and peer contracts \\
$\mathcal{G}$ & Gems: durable compressed lessons \\
$\rho$ & compression period, where Gem compression is enabled \\
$\mathcal{L}$ & lineage trace: typed records of research events \\
$a^{\star}$ & final reported artifact \\
\bottomrule
\end{tabular}
\end{table}

\begin{algorithm}[t]
\caption{\textsc{Praxist} generational research cycle. Section references point to the paragraph that defines each operator; appendix pointers for field-level schemas appear in those paragraphs.}
\label{alg:praxis}
{\small
\textbf{Input:} task with external evaluator $\textsc{Eval}(\cdot)$; cohort size $C$; generations $G$; compression period $\rho$ if Gems are enabled.\\
\textbf{Init:} $\mathcal{F}_0\!\leftarrow\!\varnothing$, $\mathcal{A}_0\!\leftarrow\!\varnothing$, $\mathcal{G}_0\!\leftarrow\!\varnothing$, $\mathcal{L}_0\!\leftarrow\!\varnothing$.\\[0.4ex]
\textbf{for} $g=0$ \textbf{to} $G-1$ \textbf{do}\\
\hspace*{1.5em}$\{d_i\}_{i=1}^{C} \leftarrow \textsc{Allocate}(\mathcal{A}_g,\mathcal{F}_g,\mathcal{G}_g)$
\hfill\textit{// design allocation: one design cell $c_i$ per peer (\S\ref{sec:method_alloc})}\\
\hspace*{1.5em}\textbf{parallel for} $i=1$ \textbf{to} $C$ \textbf{do}
\hfill\textit{// local experimentation}\\
\hspace*{3.0em}$a_i \leftarrow \textsc{Build}(d_i,\mathcal{S}_g)$
\hfill\textit{// peer constructs a reproducible artifact (\S\ref{sec:method_artifact})}\\
\hspace*{3.0em}$e_i \leftarrow \textsc{Eval}(a_i)$
\hfill\textit{// external evaluation (\S\ref{sec:method_artifact})}\\
\hspace*{3.0em}$\Phi_i \leftarrow \textsc{Interpret}(a_i,e_i,d_i)$
\hfill\textit{// artifact-to-finding extraction (\S\ref{sec:method_findings})}\\
\hspace*{1.5em}$\Phi_g \leftarrow \bigcup_{i=1}^{C}\Phi_i$\\
\hspace*{1.5em}$\{\mu_r\}_r \leftarrow \{\textsc{PI}_r(\Phi_g,\mathcal{F}_g)\}_r$
\hfill\textit{// PI panel memos on the generation's findings (\S\ref{sec:method_synthesis})}\\
\hspace*{1.5em}$\mathcal{F}_{g+1} \leftarrow \textsc{Promote}(\mathcal{F}_g,\Phi_g,\{\mu_r\}_r)$
\hfill\textit{// frontier update into the task's lanes (\S\ref{sec:method_frontier})}\\
\hspace*{1.5em}$\mathcal{A}_{g+1} \leftarrow \textsc{Chair}\bigl(\{\mu_r\}_r,\mathcal{F}_{g+1}\bigr)$
\hfill\textit{// Chair arbitration $\rightarrow$ next agenda (\S\ref{sec:method_agenda})}\\
\hspace*{1.5em}\textbf{if} $g\bmod\rho = \rho-1$ \textbf{then} $\mathcal{G}_{g+1}\leftarrow\textsc{Compress}(\mathcal{F}_{g+1},\mathcal{G}_g)$ \textbf{else} $\mathcal{G}_{g+1}\leftarrow\mathcal{G}_g$
\hfill\textit{// Gems, only if enabled (\S\ref{sec:method_memory})}\\
\hspace*{1.5em}$\mathcal{L}_{g+1} \leftarrow \mathcal{L}_g \cup \textsc{Record}(\{a_i\},\Phi_g,\mathcal{F}_{g+1},\mathcal{A}_{g+1},\mathcal{G}_{g+1})$
\hfill\textit{// lineage trace (\S\ref{sec:method_lineage})}\\
\textbf{write out} $\mathcal{F}_G,\mathcal{L}_G$, run summary; \textbf{report} $a^{\star}\leftarrow\arg\max_{a\in\mathrm{art}(\mathcal{F}^{\mathrm{cf}})}\textsc{score}(a)$ with $\mathcal{L}_G$
\hfill\textit{// selected at reporting time (\S\ref{sec:method_memory})}
}
\end{algorithm}

\subsection{Inherited State and Design Allocation}
\label{sec:method_alloc}

\paragraph{Inherited state.}
\label{sec:method_inherit}
Each generation begins from the inherited state $\mathcal{S}_g$ of Eq.~\ref{eq:state} (leftmost box of Figure~\ref{fig:praxis_method_overview}), which supplies three kinds of prior evidence. The \emph{frontier} $\mathcal{F}_g$ is the inheritable evidence eligible to influence this generation; the \emph{agenda} $\mathcal{A}_g$ is the generation policy produced by the previous synthesis step; and the \emph{Gems} $\mathcal{G}_g$ are durable lessons retained across many generations. Because a long run produces far more evidence than any single prompt context can hold, inheritance is \emph{selective}: a peer receives only the frontier entries, Gems, and parent lineage relevant to its assigned direction, the PI panel receives generation-level evidence, and the Chair receives the panel's memos (Appendix~\ref{app:gems-memory}). The unit that grounds all of this evidence is the \emph{reproducible artifact}, defined next.

\paragraph{Reproducible artifacts.}
The \textbf{reproducible artifact} $a$ is the minimal file set required to reproduce, inspect, or evaluate one attempt. In the machine-learning-engineering (MLE) task family used as our running example \citep{chan2024mlebench}, an artifact comprises the submission file, the code that produced it, and any supporting files needed for reproduction. Artifacts are what ground evaluator outcomes and downstream research claims in concrete experimental products, mirroring provenance models that connect entities, activities, and derived results \citep{groth2013prov} and ML experiment-tracking systems that record code, parameters, and metrics \citep{zaharia2018mlflow,greff2017sacred} (artifact schema and status labels: Appendix~\ref{app:artifact-evaluation-finding-records}).

\paragraph{Deep Innovation Gate (DIG).}
Before code is written, \textsc{Praxist} applies a pre-artifact \emph{innovation gate} that makes exploration intentional (second box of Figure~\ref{fig:praxis_method_overview}). Its scope is configurable: by default the gate runs as the campaign's opening allocation, and a task may enable it for every generation. The rocket campaign reported here uses the default, so its systematic pre-artifact contracts belong to the first generation, after which the Chair's agenda carries the same fields forward. The allocation operator $\{d_1,\dots,d_C\} = \textsc{Allocate}(\mathcal{A}_g,\mathcal{F}_g,\mathcal{G}_g)$ returns one \textbf{design contract} $d_i$ for each peer $p_i$. Its peer-level gate is the \textbf{Deep Innovation Gate (DIG)}: DIG requires a peer to inspect the inherited state, identify a testable mechanism, and fix the intended intervention before construction begins. The resulting contract specifies the mechanism family being tested, the intervention surface it changes, the parent lineage it extends, the evidence signature that would support or weaken it, the validation or ablation hook that makes the result interpretable, and the forbidden changes that would invalidate the test. DIG is read-only: it does not run experiments, create variants, or write result artifacts. Its purpose is to make the later artifact interpretable by letting finding extraction compare the \emph{intended} experiment against the artifact's \emph{observed} behavior. The full contract field list appears in Appendix~\ref{app:design-contracts} (Table~\ref{tab:design-contract-fields}).

\paragraph{Quantified Diversity (QD).}
The cohort-level component of \textsc{Allocate} is \textbf{Quantified Diversity (QD)}. Each contract $d_i$ occupies a \textbf{design cell} $c_i$ in a behavior space $\mathcal{C}$, where a cell is the tuple
\begin{equation}
\label{eq:cell}
c \;=\; (\text{mechanism family},\; \text{intervention surface},\; \text{intent}),
\end{equation}
and QD distributes the $C$ contracts to \emph{cover} distinct cells rather than concentrating effort on one apparent direction. Diversity is therefore represented by explicit cell coordinates and controlled by caps over mechanism family, intervention surface, and intent. The caps are hard constraints in the opening allocation, where a greedy allocator enforces them; in later generations the Chair plans the same cell coordinates under the same caps as soft targets, and the agenda validator warns on missing or unknown planning dimensions rather than rejecting the agenda. This mechanism is related to quality-diversity search, which maintains diverse high-quality candidates across a feature space \citep{mouret2015map_elites,pugh2016quality,cully2017quality}, but \textsc{Praxist} applies the idea to research design rather than to final artifacts. Within one generation, QD can allocate peers to exploit a confirmed direction, validate a fragile candidate, diagnose a failure mode, and explore a distinct mechanism family (the four design foci ``mechanism, surface, validation, diagnosis'' shown in the figure). The allocation rule and its caps are detailed in Appendix~\ref{app:design-contracts}.

\subsection{Artifact Construction, Evaluation, and Findings}
\label{sec:method_local}

\paragraph{Peers and artifact construction.}
\label{sec:method_artifact}
A \textbf{peer} $p_i$ is one parallel experimental worker: an autonomous agent that owns a single design contract $d_i$ and drives it to an evaluated artifact (third and fourth boxes of Figure~\ref{fig:praxis_method_overview}). Given its contract and the selectively inherited state, peer $p_i$ constructs an artifact $a_i = \textsc{Build}(d_i,\mathcal{S}_g)$, iterating internally---writing code, running experiments, and revising---until the artifact reaches a reportable stage (the return arrow across the top of Figure~\ref{fig:praxis_method_overview}). The $C$ peers of a generation run concurrently and coordinate only through published evidence, never by editing shared state directly; this separates local experimentation from the global allocation and promotion decisions that occur at the generation boundary. Storing and reusing experience in this way follows the broader view of agents that persist, synthesize, and retrieve memory across tasks \citep{park2023generativeagents,wang2023voyager}. Peer experiments are executed on shared hardware by a thin resource-scheduling layer that paces launches so that each generation completes enough mature evidence before its wall-clock horizon; Appendix~\ref{app:resource-scheduling} gives its specification and the supporting paired event simulation.

\paragraph{External evaluation.}
Each artifact is scored by the task's \textbf{external evaluator}, $e_i = \textsc{Eval}(a_i)$, which returns a task-grounded outcome together with a validity state and an \emph{evidence stage}. The evidence stage records how much of the evaluation an artifact has passed, on an ordered ladder from cheap sanity checks (\emph{smoke}), through partial probes (\emph{scout}), to a full scored evaluation and its replications. The core distinguishes those three levels generically; the names of the intermediate rungs are task-owned, so the rocket evaluator reports \emph{canary}, \emph{development}, and \emph{complete} where the trading evaluator reports \emph{smoke}, \emph{scout}, and three replication tiers. Evidence stage is deliberately distinct from score: a high score obtained only at a preliminary stage is treated as immature and cannot crowd out a complete result during promotion (evaluator interface, validity states, and the exact evidence-stage ladder: Appendix~\ref{app:artifact-evaluation-finding-records}).

\paragraph{Findings.}
\label{sec:method_findings}
An evaluated artifact is not yet reusable knowledge; the \textbf{finding} is the reusable unit (rightmost box of the top row). The extraction operator $\Phi_i = \textsc{Interpret}(a_i, e_i, d_i)$ converts each artifact, its outcome, and its intended design into one or more artifact-grounded claims. A finding is the tuple
\begin{equation}
\label{eq:finding}
\phi \;=\; \bigl(\text{intervention},\; \text{outcome},\; \text{evidence},\; \tau,\; m,\; \alpha\bigr),
\end{equation}
recording what was attempted, what happened, the evidence, and three labels that govern inheritance: the finding \emph{type} $\tau$, the evidence \emph{maturity} $m$, and the recommended inheritance \emph{action} $\alpha$. The type takes one of five values: \emph{positive} (an improvement or reusable mechanism), \emph{negative} (a weakened assumption or intervention), \emph{diagnostic} (a constraint, failure mode, or invalidity condition), \emph{uncertain} (evidence worth keeping pending further checks), or \emph{procedural} (a constraint on how future experiments must run). The maturity $m$ inherits the artifact's evidence stage, so interpretation confidence stays tied to how thoroughly it was evaluated. The action $\alpha\in\{\text{reuse},\text{validate},\text{avoid},\text{diagnose},\text{preserve},\text{archive}\}$ is a recommendation for what the next generation should do with it. Both $\tau$ and $\alpha$ name the operational roles the method is defined over; the implementation files a finding under a task-facing record enum (\texttt{result}, \texttt{hypothesis}, \texttt{insight}, \texttt{challenge}, \texttt{error}) and carries the inheritance recommendation in next-step-intent metadata rather than in a single $\alpha$ column (field and vocabulary tables: Appendix~\ref{app:artifact-evaluation-finding-records}).

\paragraph{Failures as first-class evidence.}
Negative and diagnostic findings are first-class evidence, not discarded failures. A failed artifact can reveal an invalid assumption, an evaluator constraint, a data-processing error, or a fragile score, and encoding this as a finding lets later generations inherit the failure as a constraint rather than rediscovering it independently. This treatment of feedback as persistent, reusable state generalizes verbal-reflection and self-refinement mechanisms in language agents \citep{shinn2023reflexion,madaan2023self_refine} from within-episode iteration to cross-generation inheritance. The pooled findings of the generation are $\Phi_g=\bigcup_{i=1}^{C}\Phi_i$.

\subsection{Global Synthesis: Frontier Update and Next Agenda}
\label{sec:method_synthesis}

\paragraph{PI/Chair synthesis.}
The bottom row of Figure~\ref{fig:praxis_method_overview} turns the pooled findings $\Phi_g$ into the next generation's inheritable state. It begins with \textbf{PI/Chair synthesis}, which converts peer-local evidence into generation-level research policy. A panel of \emph{Principal-Investigator} (PI) roles independently interprets $\Phi_g$ from complementary perspectives---a \emph{Builder} that assembles the strongest evidence-backed mainline, a \emph{Skeptic} that audits unsupported or fragile claims, and a \emph{Portfolio} role that balances effort across approach families; in \emph{high-stakes mode}, used when final claims or reproducibility risks require extra scrutiny, the panel adds an \emph{External-validity} role that checks reproducibility and evidence boundaries---each role emitting a memo $\mu_r = \textsc{PI}_r(\Phi_g,\mathcal{F}_g)$, after which a \emph{Chair} merges the memos with the updated frontier into a single agenda, $\mathcal{A}_{g+1} = \textsc{Chair}(\{\mu_r\}_r,\mathcal{F}_{g+1})$. Role specialization of this kind is widely used in multi-agent LLM collaboration \citep{li2023camel,wu2023autogen,hong2023metagpt}; \textsc{Praxist} applies it specifically to research governance, so that scores, reliability, and research utility are weighed against one another before inheritance. The panel proceeds through an evidence-freeze step, parallel independent memos, and an anonymized cross-review, after which the Chair arbitrates (role definitions, panel modes, and round structure: Appendix~\ref{app:pi-chair-synthesis}).

\paragraph{Frontier update.}
\label{sec:method_frontier}
Synthesis drives the \textbf{frontier update} (the ``Frontier Update'' box of Figure~\ref{fig:praxis_method_overview}), the operation $\mathcal{F}_{g+1} = \textsc{Promote}(\mathcal{F}_g,\Phi_g,\{\mu_r\}_r)$ that decides which evidence becomes inheritable and in what capacity. In the configuration used throughout this work, the frontier is partitioned into four \emph{lanes} by operational role,
\begin{equation}
\label{eq:lanes}
\mathcal{F} \;=\; \mathcal{F}^{\mathrm{cf}} \,\cup\, \mathcal{F}^{\mathrm{cd}} \,\cup\, \mathcal{F}^{\mathrm{dg}} \,\cup\, \mathcal{F}^{\mathrm{vl}},
\end{equation}
namely \emph{confirmed} ($\mathcal{F}^{\mathrm{cf}}$): mature enough to serve as a parent or constraint; \emph{candidate} ($\mathcal{F}^{\mathrm{cd}}$): promising but immature evidence; \emph{diagnostic} ($\mathcal{F}^{\mathrm{dg}}$): failures, controls, and failure modes that shape exploration; and \emph{validation} ($\mathcal{F}^{\mathrm{vl}}$): evidence scheduled for reproduction, ablation, or checking before promotion. Separating the frontier this way lets \textsc{Praxist} inherit not just the best score but also reliability, diagnostic value, and validation priority. Lanes are task-defined rather than built into the system: a task realization declares whichever lanes it needs---the trading campaign uses \emph{confirmed alpha}, \emph{alpha incubator}, \emph{benchmark floor}, and \emph{diagnostic control}---and a task that declares none falls back to a single-metric frontier. Promotion admits, per lane, the top findings under that lane's criterion, gating out immature evidence (smoke/scout stages) and non-promotable results (promotion criteria, per-lane caps, and gating rules: Appendix~\ref{app:frontier-agenda-decisions}).

\paragraph{Next agenda.}
\label{sec:method_agenda}
The frontier and the panel memos yield the \textbf{next agenda} $\mathcal{A}_{g+1}$ (the ``Next Agenda'' box), the control object for the next generation. It labels each inherited direction with one of four dispositions---\emph{continue} a confirmed direction, \emph{stop} a weakened one, \emph{validate} a fragile candidate, or \emph{explore} an underexplored cell---and attaches, for each of the $C$ peers, a \emph{peer contract} that specifies its role, hypothesis, success signal, and forbidden actions. As with the finding vocabulary, the four dispositions are method-level roles; the emitted agenda records them through a finer next-step-intent field that distinguishes, for example, repairing a failure mode from ablating one. $\textsc{Allocate}$ (\S\ref{sec:method_alloc}) consumes the agenda at the start of generation $g{+}1$, closing the loop between global synthesis and local experimentation (agenda schema and lane-to-disposition mapping: Appendix~\ref{app:frontier-agenda-decisions}).

\subsection{Memory, Lineage, and Final Output}
\label{sec:method_memory}

\paragraph{Memory compression and Gems.}
Frontier entries carry the operational status of \emph{current} evidence, but a long campaign also accumulates lessons that should survive after their originating evidence is archived. Where a campaign enables it, every $\rho$ generations \textsc{Praxist} performs a \textbf{memory compression} $\mathcal{G}_{g+1} = \textsc{Compress}(\mathcal{F}_{g+1},\mathcal{G}_g)$ that distills the balanced frontier into \textbf{Gems} (the ``Memory Compression'' box). A Gem is a compact, durable lesson---a validated mechanism, a rejected assumption, a recurring failure mode, or a procedural constraint---retained across the reset boundary so that later peers keep the benefit of past evidence without its bulk. Compression is lane-balanced (preserving control and diagnostic lessons, not only high scorers) and bounded (only a small active Gem set is kept). The facility is off by default: among the case studies reported here only the trading campaign runs it, with a period of six generations and at most four active Gems. This extends experience-storage-and-retrieval memory patterns \citep{park2023generativeagents,wang2023voyager,shinn2023reflexion} to evaluator-grounded campaigns (compression trigger, selection policy, and Gem schema: Appendix~\ref{app:gems-memory}).

\paragraph{Lineage trace.}
\label{sec:method_lineage}
Throughout the cycle, research events are appended to the \textbf{lineage trace} (the ``Lineage Trace'' band along the bottom of Figure~\ref{fig:praxis_method_overview}):
\[
\mathcal{L}_{g+1} \;=\; \mathcal{L}_g \,\cup\, \textsc{Record}\bigl(\{a_i\},\Phi_g,\mathcal{F}_{g+1},\mathcal{A}_{g+1},\mathcal{G}_{g+1}\bigr).
\]
The lineage records the research objects defined above---artifacts, findings, synthesis decisions, agendas, and memory (Gem) updates---together with typed relations among them such as \emph{derived-from}, \emph{supports}, \emph{challenges}, and \emph{updates}. It is materialized as several correlated ledgers that share identifiers---an edge list over findings, an artifact index, an event trajectory, and per-generation frontier, agenda, and Gem state---rather than as one object graph, and a chain is traversed by following those shared identifiers across them. It is accumulated \emph{during} the run rather than reconstructed after the fact: each artifact points to the parents, contract, and findings it descended from, and each promotion or agenda decision records why evidence became inheritable. During execution the lineage determines what later peers inherit; after execution it is the structured account of the research trajectory (node and edge vocabulary: Appendix~\ref{app:lineage-trace}).

\paragraph{Final output.}
After $G$ generations, the outcome of a \textsc{Praxist} campaign is reported as a final artifact together with its lineage,
\begin{equation}
\label{eq:output}
a^{\star} \;=\; \arg\max_{a\,\in\,\mathrm{art}(\mathcal{F}^{\mathrm{cf}})} \textsc{score}(a), \qquad \text{output} = (a^{\star},\,\mathcal{L}_G),
\end{equation}
where $\mathrm{art}(\mathcal{F}^{\mathrm{cf}})$ denotes the artifacts that ground the confirmed-lane findings: under the default rule the final artifact is the best-scoring member of the confirmed frontier. Eq.~\ref{eq:output} states that rule, not an automatic runtime call. The loop itself terminates by writing out the frontier, the lineage ledgers, and a run summary; the selection is applied over those outputs at reporting time, and each case study below states which artifact it selected and under which rule---the Rocket and Quant studies report an artifact selected post-run from outside the confirmed lane, and say so where they report it. The output is thus not only an evaluated artifact but the \emph{solution lineage} that explains how it was obtained---which evidence supported it, which failures constrained it, which candidates required validation, and which agendas shaped its construction---providing reusable state for future researchers or autonomous systems.

%% file: sections/3_experiments.tex
\section{Experiments} \label{sec:experiments}

% ============================================================
\subsection{Overview}
\label{sec:exp_overview}

Our evaluation has two parts. The first is a large-scale comparative benchmark: we run \textsc{Praxist} on MLE-bench \citep{chan2024mlebench}, a suite of 75 Kaggle-derived machine-learning-engineering competitions with an external grading harness, and compare it head-to-head against a Claude Code baseline that we run ourselves under an identical protocol (\Secref{sec:exp_mle}). MLE-bench is our primary evaluation because it is the setting in which directly comparable cross-system claims are possible: every system faces the same tasks, the same data, and the same medal thresholds.

The second part probes generality beyond competition-style ML engineering through five in-depth case studies on harder, more open-ended R\&D problems: rocket design (\Secref{sec:exp_rocket}), quantitative trading (\Secref{sec:exp_quant}), LiDAR-inertial-visual SLAM (\Secref{sec:exp_slam}), tokamak magnetic control for fusion (\Secref{sec:exp_fusion}), and dynamic job-shop rescheduling (\Secref{sec:exp_jobshop}). We emphasize that the case studies are not standardized cross-system comparisons in the sense of MLE-bench: each is scored by its own domain-native evaluator against a task-grounded baseline, and we analyze both the resulting artifact and its discovery process. The Rocket study additionally reports an external autonomous code optimizer run on the same task and the same success metric, an outside reference point rather than a matched-model or matched-compute comparison. In the Rocket study, \textsc{Praxist} constructs a deterministic first-contact landing controller that reaches $12{,}288/12{,}288$ ($100\%$) on the frozen protocol, against $17.12\%$ for Weco, an autonomous code optimizer run on the same task and objective from the same $4.03\%$ starting point; a separate post-run all-row audit of the three fixed source banks records $40{,}959$, $40{,}959$, and $40{,}960$ successes out of $40{,}960$ rows per bank. In the Quant study, it discovers a recurrent, execution-aware trading policy whose 53\% walk-forward CAGR is 2.3 times the 23\% of its paired all-eligible equal-weight baseline.
In the SLAM study, it shows that state-of-the-art LIVO systems over-spend visual computation on redundant frames and repeated map observations. Replacing the uniform visual-update policy with an observability-aware scheduler and a geometry-aware map-admission gate cuts the evaluator-captured visual-path processing time by roughly 72\% across fourteen sequences at no cost in trajectory accuracy; the two arms stamp their poses differently, so \Secref{sec:exp_slam} reports the accuracy columns side by side rather than as a gain.
% In the SLAM study, it finds that a state-of-the-art odometry system spends visual updates on redundant frames, and replaces that uniform policy with a state-dependent schedule that cuts mean full-sequence trajectory error by 45.4\% across all fourteen benchmark sequences while cutting per-frame runtime by roughly 72\%. 
In the Fusion study, it synthesizes a tokamak magnetic-control law that achieves higher aggregate survival and lower common-horizon tracking error than a task-native controller based on the plasma control system deployed on the MAST-U device, while that baseline keeps the edge on the benchmark's original full-horizon metric and on completion rate. In the Job-Shop study, it re-engineers how a frozen neural rescheduling policy estimates durations, admits plans, and repairs them after disruptions, raising the mean makespan improvement over a dispatching rule from the incumbent's 1.38\% to 3.38\% on a fixed 50-episode calibration set without retraining the network. Appendix~\ref{app:experiment-setups} records the execution layer for all six studies: the repetitions behind each headline number, the hardware/time envelope, and the \textsc{Praxist} campaign setup.

% ============================================================
\subsection{MLE-bench}
\label{sec:exp_mle}

\subsubsection{Experimental Setup}
\label{sec:exp_mle_setup}

\textbf{Benchmark.} MLE-bench \citep{chan2024mlebench} consists of 75 Kaggle competitions spanning tabular modeling, computer vision, natural-language processing, time-series forecasting, and signal-processing tasks. An agent must produce a valid submission file for each competition; submissions are scored by the official grading harness against the competition's private leaderboard, and a run earns a bronze, silver, or gold medal by clearing the corresponding human-leaderboard threshold. Tasks are stratified by complexity into a \emph{Low} tier (22 tasks; this tier constitutes the MLE-bench Lite subset), a \emph{Medium} tier (38 tasks), and a \emph{High} tier (15 tasks). We evaluate on the full 75-task set and report results per tier. Medals are determined by the official MLE-bench grading harness from each submission's private-leaderboard score, and a task contributes a medal only if its recorded score clears the corresponding MLE-bench threshold. Entries without a benchmark-available final score are kept in the denominator and counted as no medal. Every row we report is a local 75-task evaluation graded by that same harness on the same hardware pool, which makes the rows directly comparable to one another; we therefore do not place them beside the public MLE-bench leaderboard, whose entries are produced under different run conventions and reported as multi-seed run-group averages. Each arm is one full 75-task sweep, so the medal rates are single-sweep outcomes rather than variance estimates over repeated sweeps. These results nevertheless cover all 75 benchmark competitions, providing broad evidence about cross-task reliability under the stated reporting protocol.

\textbf{Setup.} We report one finalized result per task from the release ledger. That ledger records an explicit integrity adjudication for each of the 75 tasks: every reported \textsc{Praxist} result is the highest-scoring official attempt that passes the adjudication, with the selected submission verified by SHA-256 against the run journal, and attempts drawn from a contaminated lineage are excluded outright rather than down-weighted (90{,}423 attempts were rejected across the campaign). On 9 of the 75 tasks the whole lineage behind the previously best attempt was excluded and a clean actor's submission was substituted in its place, which lowers the medal on four of those nine, and 6 further tasks were re-reviewed and cleared; Appendix~\ref{app:mle-performance-overview} reports the per-task statuses. This adjudication is internal to the \textsc{Praxist} ledger and is separate from the benchmark's own post-run screen for severe-cheating violations, which applies to the Claude Code baseline described below. \textsc{Praxist} runs deepseek-v4-pro with a 1M-token context window as its research-agent model on all 75 competitions and evaluates each competition using the procedure described in \Secref{sec:method}. We compare it against a Claude Code baseline swept over all 75 tasks locally under the same grading harness. The baseline runs \textbf{Claude Opus 4.8} with maximum thinking budget; after the benchmark's post-run screen for severe-cheating violations, its finalized ledger retains 70 accepted scores, marks five tasks not scored, and records a dollar-equivalent cost of US\$38{,}370 (Appendix~\ref{app:mle-per-task} gives its per-task results). Both arms ran on the same pool of H100 80GB GPUs, with each experiment scheduled as a single-GPU job and up to eight experiments in flight per task; 70 tasks ran under a 24-hour per-task wall cap and the remaining five under a 36-hour cap. Appendix~\ref{app:config} reports the generation, cohort, and budget settings of the single \textsc{Praxist} campaign, and Appendix~\ref{app:setup-mle} records the execution environment and the campaign totals behind the finalized ledger.

\subsubsection{Main Results}
\label{sec:exp_mle_main}

Table~\ref{tab:mle_main} reports the base LLM for each arm, Any Medal rates per complexity tier, and all-task medal composition. The audited Claude Code + Opus 4.8 sweep records 55 accepted medals across the 75-task suite (73.3\%), with five tasks not scored. \textsc{Praxist} exceeds the Opus 4.8 sweep on every tier---90.9\% versus 81.8\% on Low, 81.6\% versus 76.3\% on Medium, and 60.0\% versus 53.3\% on High---and has the larger gold count in every tier, 49 versus 34 overall, with 19 of its 20 Low-tier medals gold. The available dollar-equivalent ledgers add a separate efficiency contrast: the recorded \textsc{Praxist} 75-task spend is approximately US\$3{,}054, whereas the Opus 4.8 sweep records US\$38{,}370, roughly an order of magnitude apart on the same task suite.

\begin{table}[t]
\centering
\footnotesize
\caption{MLE-bench Any Medal rate (\%) by complexity tier and base LLM, with all-task medal composition. Bold marks the best value in each column.}
\label{tab:mle_main}
\setlength{\tabcolsep}{2.4pt}
\renewcommand{\arraystretch}{1.12}
\begin{tabular*}{0.98\textwidth}{@{\extracolsep{\fill}}llccccc@{}}
\toprule
& & \multicolumn{4}{c}{Any Medal (\%)} & \multicolumn{1}{c}{All-task} \\
\cmidrule(lr){3-6}
Agent & Base LLM & Low (22) & Medium (38) & High (15) & All (75) & G/S/B \\
\midrule
Claude Code & Claude Opus 4.8 & 81.8 & 76.3 & 53.3 & 73.3 & 34/16/5 \\
\textsc{Praxist} (ours) & deepseek-v4-pro & \textbf{90.9} & \textbf{81.6} & \textbf{60.0} & \textbf{80.0} & \textbf{49}/10/1 \\
\bottomrule
\end{tabular*}
\end{table}

The medal composition is the most distinctive feature of the \textsc{Praxist} result. Of \textsc{Praxist}'s 60 medals, 49 (81.7\%) are gold, compared with 34 of 55 accepted medals (61.8\%) for Claude Code on Opus 4.8. Because both arms are single locally measured runs rather than official multi-seed leaderboard rows, the margins should be interpreted as benchmark-wide outcomes under the stated reporting protocol rather than seed-averaged estimates of a distribution over repeated runs.

\subsubsection{Analysis by Task Category}
\label{sec:exp_mle_analysis}

Beyond the per-tier aggregates, we compare \textsc{Praxist} head-to-head against the Claude Code + Opus~4.8 baseline competition by competition. Table~\ref{tab:mle_task_main} shows this comparison on 19 competitions drawn from the Medium- and High-complexity tiers, where several category-level differences are visible. The selection spans both categories where \textsc{Praxist} improves the medal outcome and categories where Claude Code has the stronger score; the complete 75-task comparison, including the Low tier, is given in Appendix~\ref{app:mle-per-task}.

\begin{table}[t]
\centering
\small
\begingroup
\setlength{\fboxsep}{0.5pt}
\caption{Per-task comparison of the Claude Code + Opus~4.8 baseline and \textsc{Praxist} on 19 representative Medium- and High-tier MLE-bench competitions. Shading encodes the medal (\colorbox{medalgold}{gold}, \colorbox{medalsilver}{silver}, \colorbox{medalbronze}{bronze}; unshaded = none); bold marks the better score in the metric's direction. The \# column indexes the full 75-task listing of Appendix~\ref{app:mle-per-task}.}
\label{tab:mle_task_main}
\endgroup
\setlength{\tabcolsep}{5pt}
\resizebox{0.98\textwidth}{!}{%
\begin{tabular}{lcllcc}
\toprule
Category & \# & Task & Metric & Claude Code & \textsc{Praxist} (ours) \\
\midrule
\multirow{5}{*}{Image Classification}
	 & 2 & \texttt{alaska2-image-steganalysis} & weighted AUROC $\uparrow$ & \cellcolor{medalsilver}\textbf{0.9260} & 0.9140 \\
	 & 4 & \texttt{cassava-leaf-disease-classification} & accuracy $\uparrow$ & \cellcolor{medalsilver}0.9010 & \cellcolor{medalgold}\textbf{0.9021} \\
	 & 17 & \texttt{kuzushiji-recognition} & F1 $\uparrow$ & \cellcolor{medalgold}0.9714 & \cellcolor{medalgold}\textbf{0.9741} \\
	 & 23 & \texttt{rsna-breast-cancer-detection} & probabilistic F1 $\uparrow$ & \cellcolor{medalsilver}\textbf{0.4719} & 0.2752 \\
	 & 25 & \texttt{statoil-iceberg-classifier-challenge} & log loss $\downarrow$ & \cellcolor{medalbronze}0.1451 & \cellcolor{medalsilver}\textbf{0.1271} \\
\midrule
\multirow{2}{*}{Image (Other)}
	 & 40 & \texttt{petfinder-pawpularity-score} & RMSE $\downarrow$ & \cellcolor{medalgold}\textbf{16.77} & \cellcolor{medalgold}16.86 \\
	 & 41 & \texttt{rsna-miccai-brain-tumor-radiogenomic-classification} & AUROC $\uparrow$ & 0.5341 & \cellcolor{medalgold}\textbf{0.6588} \\
\midrule
\multirow{6}{*}{Text Classification}
	 & 29 & \texttt{AI4Code} & Kendall $\tau$ $\uparrow$ & \cellcolor{medalsilver}\textbf{0.8606} & 0.8227 \\
	 & 31 & \texttt{facebook-recruiting-iii-keyword-extraction} & micro F1 $\uparrow$ & \cellcolor{medalsilver}0.7851 & \cellcolor{medalgold}\textbf{0.7959} \\
	 & 33 & \texttt{jigsaw-unintended-bias-in-toxicity-classification} & bias-weighted AUC $\uparrow$ & 0.8635 & \textbf{0.8640} \\
	 & 34 & \texttt{learning-agency-lab-automated-essay-scoring-2} & quadratic kappa $\uparrow$ & 0.8312 & \cellcolor{medalgold}\textbf{0.8384} \\
 & 35 & \texttt{lmsys-chatbot-arena} & log loss $\downarrow$ & \cellcolor{medalgold}\textbf{0.8718} & \cellcolor{medalgold}0.9812 \\
 & 38 & \texttt{tweet-sentiment-extraction} & Jaccard $\uparrow$ & \cellcolor{medalsilver}0.7225 & \cellcolor{medalsilver}\textbf{0.7239} \\
\midrule
Training LLMs
 & 58 & \texttt{chaii-hindi-and-tamil-question-answering} & word Jaccard $\uparrow$ & \cellcolor{medalsilver}0.7585 & \cellcolor{medalgold}\textbf{0.8894} \\
\midrule
Forecasting
	 & 60 & \texttt{osic-pulmonary-fibrosis-progression} & Laplace log-lik. $\uparrow$ & -7.036 & \cellcolor{medalgold}\textbf{-5.6407} \\
\midrule
Tabular
	 & 48 & \texttt{icecube-neutrinos-in-deep-ice} & angular error $\downarrow$ & \cellcolor{medalsilver}\textbf{1.005} & 1.0563 \\
\midrule
\multirow{2}{*}{Image Segmentation}
	 & 64 & \texttt{hubmap-kidney-segmentation} & Dice $\uparrow$ & \cellcolor{medalgold}0.9487 & \cellcolor{medalgold}\textbf{0.9488} \\
	 & 67 & \texttt{uw-madison-gi-tract-image-segmentation} & Dice--Hausdorff $\uparrow$ & 0.6274 & \cellcolor{medalsilver}\textbf{0.8722} \\
\midrule
Audio Classification
 & 42 & \texttt{freesound-audio-tagging-2019} & LRAP $\uparrow$ & \cellcolor{medalgold}\textbf{0.7471} & \cellcolor{medalgold}0.7445 \\
\bottomrule
\end{tabular}}
\end{table}

\textbf{Where \textsc{Praxist} improves the medal outcome.} In the selected 19-task subset, \textsc{Praxist} records the better score on 12 tasks and the stronger gold count: 11 gold outcomes versus 5 for Claude Code. Each arm medals on 14 of the 19, but the medal is strictly better for \textsc{Praxist} on eight tasks and strictly better for the baseline on four, so the difference is concentrated in medal grade rather than medal count. The clearest gains are conversions from no medal at all: brain-tumor radiogenomic classification (\#41) moves from the baseline's 0.5341 AUROC to 0.6588, automated essay scoring (\#34) from 0.8312 to 0.8384 quadratic kappa, and pulmonary-fibrosis progression (\#60) from $-7.04$ to $-5.64$ Laplace log-likelihood, each crossing from no medal to gold. A second family upgrades a baseline that already medals: on cassava leaf disease (\#4), keyword extraction (\#31), and Hindi/Tamil question answering (\#58, where word Jaccard rises from 0.7585 to 0.8894), Claude Code earns silver and \textsc{Praxist} reaches gold, while GI-tract segmentation (\#67) rises from no medal to silver and iceberg classification (\#25) from bronze to silver. This is the per-task analogue of the medal-composition pattern in \Secref{sec:exp_mle_main}: the frontier can retain strong parents and schedule additional attempts after a first medal-clearing artifact has been found.

\textbf{Where Claude Code remains stronger.} Claude Code has the better score on the remaining seven of the selected tasks, and the raw-score picture over the full suite is closer still. On the 70 tasks where both arms record a score, the baseline has the better raw score on 36 and \textsc{Praxist} on 33, with one tie; \textsc{Praxist} additionally returns a score on the five tasks the baseline leaves unscored. Its benchmark advantage therefore lies in which side of a leaderboard threshold a run lands on, not in winning the majority of head-to-head score comparisons. Four rows of Table~\ref{tab:mle_task_main} cost it a medal outright: image steganalysis (\#2), notebook-cell ordering (\#29), and IceCube event reconstruction (\#48) each leave a baseline silver unmatched, and screening mammography (\#23) falls from the baseline's 0.4719 probabilistic F1 to 0.2752. The full table in Appendix~\ref{app:mle-per-task} shows the same pattern more broadly for several large-label-space recognition tasks. These competitions are often compute- and pipeline-intensive, so fewer complete training/evaluation cycles may fit inside a fixed campaign budget. On tasks where both systems clear gold---audio tagging (\#42), PetFinder Pawpularity (\#40), and LMSYS Chatbot Arena (\#35)---residual score differences remain informative for score-level analysis but do not change the benchmark medal outcome.

\textbf{Cost accounting.} For the two arms with finalized model-spend ledgers, \textsc{Praxist} records a 75-task cost of about US\$3{,}054, converted from CNY~20{,}695 using the Federal Reserve H.10 exchange rate of 6.7766 CNY per US dollar on 10 July 2026.\footnote{\url{https://www.federalreserve.gov/releases/h10/hist/dat00_ch.htm}} The finalized Claude Code + Opus 4.8 ledger records US\$38{,}370 for the same suite, twelve times the \textsc{Praxist} figure. Cost figures provide resource context only; medals in Tables~\ref{tab:mle_main} and~\ref{tab:mle_task_main} are determined by task scores and MLE-bench thresholds rather than cost.

% ============================================================
% Keep the preceding subsection's floats from spilling into this case study.
\FloatBarrier
\subsection{Case Study: Rocket (Reusable Rocket Landing)}
\label{sec:exp_rocket}

\begin{figure}[!ht]
    \centering
    \includegraphics[width=0.98\textwidth]{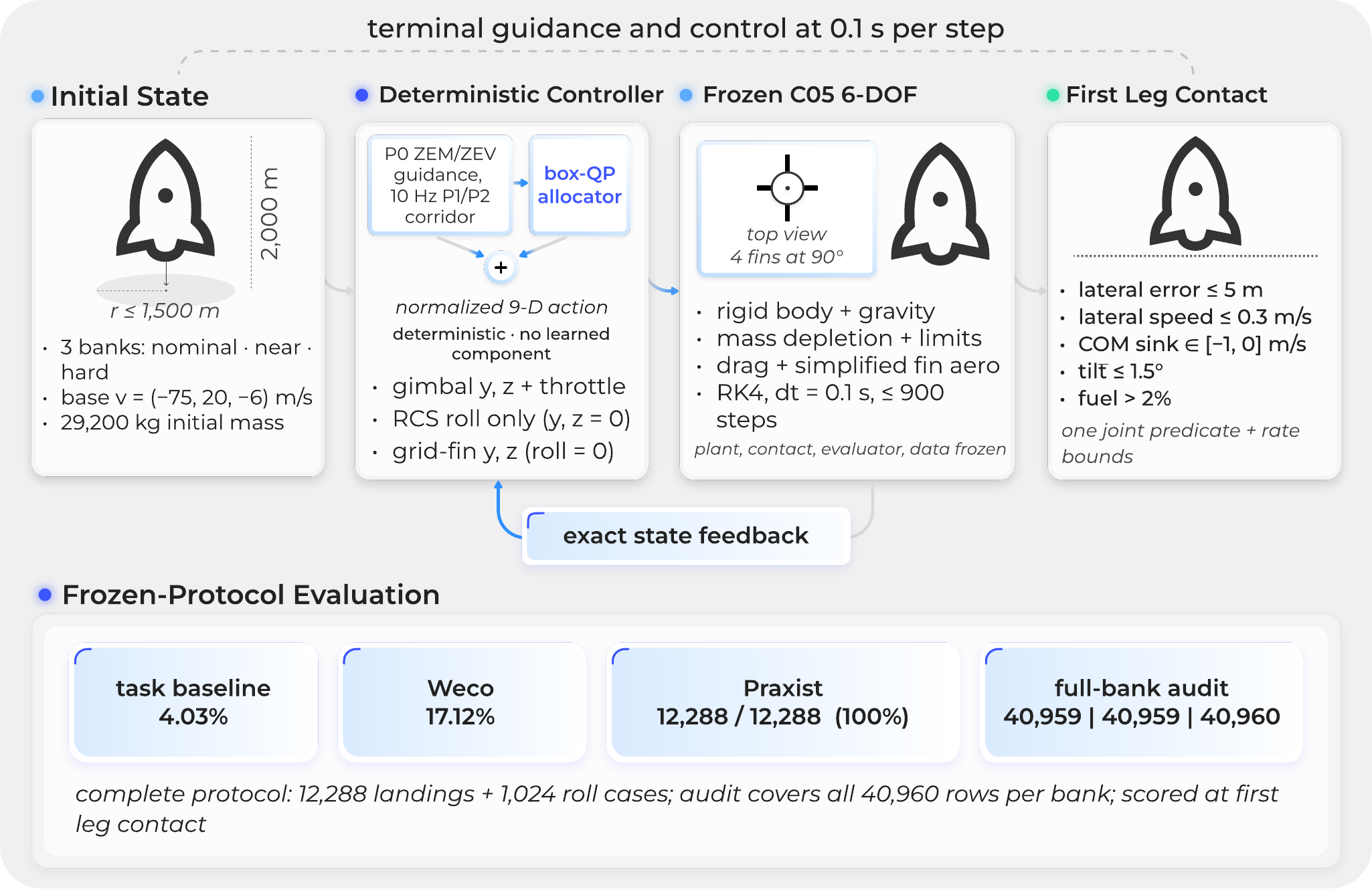}
    \caption{Rocket task overview. Initial states are drawn from three fixed banks---a nominal area-uniform disk, a row-paired slower near-OOD bank, and a faster hard-OOD outer annulus---and are flown by a deterministic hybrid controller: rolling ZEM/ZEV guidance in phase P0, fuel-commit and phase guards, and a P1/P2 terminal-descent corridor. Geometric attitude control feeds a closed-form pitch/yaw box QP that splits each torque demand between gimbal and grid fins, with the RCS restricted to roll. The frozen C05 6DoF plant advances under RK4 at $0.1$-s steps, and every trajectory is scored once, at the interpolated first landing-leg contact, by a single joint success predicate. The evidence strip records results on the shared task and objective---the $4.03\%$ starting artifact, Weco's reported $17.12\%$, and the selected controller at $12{,}288/12{,}288$---and, separately, its full-bank audit ($40{,}959$, $40{,}959$, and $40{,}960$ successes per $40{,}960$-row bank).}
    \label{fig:rocket_task_overview}
\end{figure}

\textbf{Problem setup.} The task is terminal guidance and control for a reusable, vertically landing rocket (Figure~\ref{fig:rocket_task_overview}). The vehicle is the frozen Swordfish C05 six-degree-of-freedom rigid-body plant, starting at $2{,}000\,$m with $22{,}200\,$kg of dry mass and $7{,}000\,$kg of main propellant, or $29{,}200\,$kg of initial total mass, and advancing at $0.1\,$s steps under exact state feedback. The controller commands two gimbal deflections, throttle, three reaction-control-system (RCS) channels, and three grid-fin channels, but the contract hard-locks the pitch and yaw RCS channels and the grid-fin roll channel to zero, so RCS is restricted to roll while pitch and yaw remain actuated by the gimbal and the grid fins, whose effectiveness depends strongly on geometry and flight regime \citep{liu2024gridfins}. Plant, contact model, integrator, evaluator, and data are frozen for the whole campaign: a search may change only the controller, and cannot buy apparent gains by loosening the physics or moving the scoring endpoint. Appendix~\ref{app:setup-rocket} gives the full protocol.

\textbf{Evaluation protocol.} Initial states come from three fixed banks that share a base velocity of $(-75,20,-6)\,\mathrm{m\,s^{-1}}$: a nominal area-uniform disk of radius $1{,}500\,$m, a near-OOD bank pairing those positions row by row with velocities scaled by $s\sim U[0.92,0.98)$, and a hard-OOD bank drawing an annulus at $1{,}500\le r<1{,}650\,$m with $s\sim U[1.02,1.08)$. The banks are fixed and adaptively reused rather than held out. Success is one joint task-native predicate evaluated at the interpolated state where the first landing leg tips the ground---bounding lateral error, sink speed, lateral speed, tilt, angular rates, and remaining propellant simultaneously---so post-contact spring--damper response cannot improve the scored state. Two evidence scopes stay separate throughout: a complete private validation of $4{,}096$ trajectories per bank, hence $12{,}288$ landing trajectories plus $1{,}024$ roll cases, used for selection; and a separate post-run audit over every row of all three banks.

\textbf{Baseline.} Our point of comparison is another autonomous research system, not merely the artifact the task ships with. We ran Weco's LLM-driven code optimizer---built on the AIDE tree search over code \citep{jiang2025aide}---on the same task package,\footnote{\url{https://dashboard.weco.ai/share/xZJAbpYNN7aJCvyiZuuZFsp_hzQdPqTy}} with the same \texttt{landing\_success\_rate} objective and the same starting artifact, which scores $495/12{,}288$ trajectories, or $4.03\%$, under our complete protocol. Weco evaluated $793$ candidate controllers in $2$\,h $37$\,min and reached a best score of $17.12\%$, a $4.25\times$ relative gain that nonetheless leaves the task unsolved \citep{weco2026rocketrun}. That budget is not small: its $793$ evaluations amount to roughly $800$ optimization steps, which in \textsc{Praxist}'s generational structure would correspond to about $100$ generations running eight parallel candidate experiments each---an illustration of scale, not a description of the rocket run, which committed $12$ generations of $16$ peers. Recorded model spend runs the opposite way from the scores: the \textsc{Praxist} run cost US\$196.05 against US\$1{,}009.66, over its own $697$ completed evaluation jobs, so the higher success rate was not bought with a larger budget. Table~\ref{tab:rocket-weco} places both systems against the shared starting artifact. Weco's $17.12\%$ is the best score reported by its run dashboard on that task and objective; \textsc{Praxist}'s $100\%$ is measured by the frozen $12{,}288$-trajectory protocol reported here. The systems use different underlying language models and wall-clock budgets, and Weco's editable search surface additionally included the task-variant manifest whereas the \textsc{Praxist} run could edit the controller alone, so Weco serves as an external autonomous-optimizer reference rather than a controlled cross-model comparison.

\textbf{Discovered controller.} The selected artifact is deterministic throughout---no neural network, reinforcement learning, learned residual, or training seed appears anywhere in it. Rolling zero-effort-miss/zero-effort-velocity guidance \citep{guo2013zemzev} sets the descent in phase P0, a fuel-commit governor and phase guards handle the transition, a terminal descent corridor carries the vehicle to first leg contact, and geometric attitude control \citep{lee2010geometric} drives the airframe while a damped roll loop absorbs the roll channel. The newest method contribution, committed at generation~11, is a closed-form box-constrained allocator that splits each pitch/yaw torque request between gimbal and grid fins by solving $\min\,(a\,u_f+b\,u_g-d)^2+w\,u_f^2$ under amplitude and rate boxes, with only the grid deflection regularized so the cost is gimbal-primary \citep{bodson2002allocation,johansen2013allocation}. It is not an iterative solver: the implementation enumerates clipped KKT candidates and returns the minimum-cost feasible point. The evaluated instance is that controller's generation-12 configuration, byte-identical in code and differing in a single value, which is the run's current complete-evidence champion, selected post-run rather than promoted as a committed frontier artifact.

\begin{table}[!htb]
\centering
\small
\caption{Rocket case study: landing success of the shared starting artifact, the Weco
autonomous-optimizer baseline \citep{jiang2025aide,weco2026rocketrun}, and \textsc{Praxist}
on the same task package and \texttt{landing\_success\_rate} objective. Relative gains are
taken against the starting artifact, which is the given controller and so has no search
budget of its own. Evaluation counts are each system's own unit of work: candidate
controllers scored for Weco, completed evaluation jobs for \textsc{Praxist}.}
\label{tab:rocket-weco}
\setlength{\tabcolsep}{6pt}
\begin{tabular}{@{}lrrrr@{}}
\toprule
System & Landing success & Relative gain & Evaluations & Recorded spend \\
\midrule
Starting artifact & $4.03\%$ & --- & --- & --- \\
Weco & $17.12\%$ & $4.25\times$ & $793$ & US\$1{,}009.66 \\
\textsc{Praxist} & $\mathbf{100\%}$ & $24.8\times$ & $697$ & US\$196.05 \\
\bottomrule
\end{tabular}
\end{table}

\textbf{Results.} Under the frozen complete protocol, the selected controller lands $12{,}288/12{,}288$ trajectories ($100\%$), compared with the shared starting artifact's $4.03\%$ and Weco's reported $17.12\%$ on the shared task and objective, with a descriptive Wilson $95\%$ lower bound of $99.9687\%$ on that fixed set (Table~\ref{tab:rocket-matched-protocol}). Per-bank rates rise from $5.6641\%$, $6.4209\%$, and $0\%$ to $100\%$ throughout, and the worst radius bin from $0\%$ to $100\%$. The improvement is one of contact quality rather than contact occurrence: the baseline already reaches first contact in every trajectory, and what changes is the state in which that contact happens. The $95$th-percentile sink speed falls from $66.3928$ to $0.32636\,\mathrm{m\,s^{-1}}$, lateral speed from $1.40592$ to $0.04705\,\mathrm{m\,s^{-1}}$, and tilt from $4.56062^\circ$ to $0.35184^\circ$, while fuel depletion drops from $88.6393\%$ of trajectories to none. Two metrics move adversely and we retain them: gimbal total variation rises $80.43\%$ and roll-to-pitch/yaw coupling $87.55\%$, consistent with moving pitch/yaw activity onto the gimbal.

A separate post-run evaluation over all $40{,}960$ rows of each bank records $40{,}959$, $40{,}959$, and $40{,}960$ successes (Table~\ref{tab:rocket-full-bank-audit}). We report these per bank because the three banks are differently defined and are not one IID reliability sample. Both failures violated only the lateral-speed conjunct and lie in the $[0,450)\,$m initial-radius bin; they are residual-motion failures rather than impacts or fuel exhaustion. The hard bank's $100\%$ count should not be read as broad margin, since its minimum successful fuel reserve is only $2.8908\%$, just $0.8908$ points above the gate.

\textbf{Lineage and claim boundary.} The gain accumulated along the parent chain rather than arriving in one step: a fuel-commit governor and settled-release gate first, then P0 attitude authority and a low-altitude bandwidth schedule, then the initial-radius and slew branch with a sink guard, which together closed the set at $100\%$ by generation~9, and only then the box-QP allocator at generation~11 (Table~\ref{tab:rocket-lineage}). This is cumulative direction-of-progress evidence, not an additive causal decomposition. The allocator's own attribution comes from a one-key ablation: it lowers sink and lateral-speed P95 and removes the residual gimbal saturation, but its immediate parent was already at $100\%$, so it did not cause the success gain. That gain belongs to the accumulated lineage, whose mechanisms were established, validated, and recombined across generations. The result is bounded by a frozen low-order simulator, by fixed banks that are adaptively reused and so carry selection-overfitting risk, by exact state feedback, and by scoring that stops at first contact; nothing is claimed about post-contact dwell, bounce, or leg loads, and the measured rates are not real-world landing reliability. Appendix~\ref{app:rocket-v12-validation} gives the full protocol, matched results, audit, ablation, and lineage.

% ============================================================
% Keep the preceding subsection's floats from spilling into this case study.
\FloatBarrier
\subsection{Case Study: Quantitative Finance (Quant)}
\label{sec:exp_quant}

\begin{figure}[!ht]
    \centering
    \includegraphics[width=0.98\textwidth]{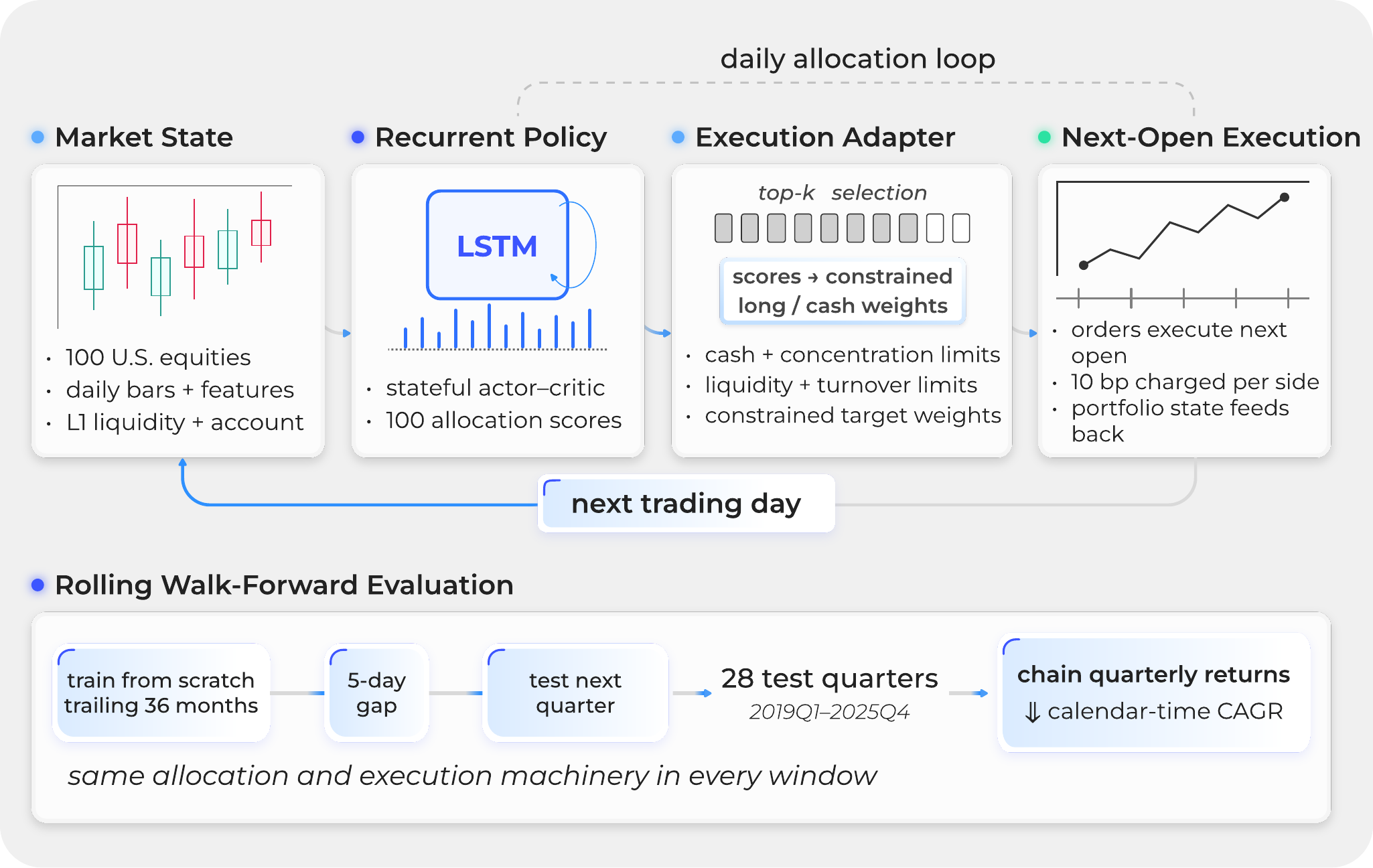}
    \caption{Quant task overview. Daily market, liquidity, and account state are mapped to 100 allocation scores; a deterministic execution adapter converts those scores into constrained long/cash weights, and orders execute at the next open with per-side costs. The evaluator retrains the policy on each trailing 36-month window, leaves a five-trading-day gap, tests the next quarter, and chains all 28 quarterly returns into calendar-time CAGR.}
    \label{fig:quant_task_overview}
\end{figure}

\textbf{Problem setup.} The task is daily stock allocation (Figure~\ref{fig:quant_task_overview}): every trading day, decide what share of the portfolio to hold in each of a fixed 100-stock, multi-sector U.S. universe (technology and semiconductors, health care, transport and logistics, materials, REITs, and large banks) and what share to leave in cash. Positions are long or flat, never short. Inputs are task-local Alpaca SIP split-adjusted daily bars, expanded per-symbol features, daily top-of-book L1 features, and the current state of the account, including its cash balance and open positions.

\textbf{Evaluation protocol and baseline.} A policy is never tested on data it was trained on. Evaluation walks forward through the 28 consecutive quarters from 2019Q1 to 2025Q4: for each quarter, a model is trained from scratch on the preceding 36 months, a five-trading-day gap is left between the end of that training window and the start of trading, and that quarter is then traded once. The environment also excludes target columns and rejects forward-looking feature names, so the protocol is structurally causal by construction; we did not additionally audit every engineered feature column for leakage. Orders execute at the next day's open with 10 basis points charged per side, which is more conservative than a frictionless close-to-close backtest. The headline metric chains all 28 quarterly returns into a calendar-time compound annual growth rate (CAGR). The \textbf{baseline} is an all-eligible equal-weight buy-and-hold comparator over the same fixed universe. In each independently cash-reset window, it invests 98\% of the account equally across every protocol-eligible stock at the next open using fractional shares, charges the same 10 basis-point buy cost, and holds those shares unchanged through the final close. Chaining the 28 window endpoints gives a 22.80\% paired-evaluation CAGR over the same period. Every window retrains from scratch, so a full five-seed evaluation of one candidate is 145 independent training runs; that is the campaign's top evaluation tier, and the artifact reported below was evaluated at the first tier, a single seed over 29 cells. Appendix~\ref{app:setup-quant} gives the evaluation tiers, the per-cell training cost, and the campaign configuration.

\textbf{Discovered artifact.} The artifact reported here is a recurrent, execution-aware policy: a flat LSTM actor--critic \citep{hochreiter1997lstm} trained with PPO \citep{schulman2017ppo}, bounded replay of recent recurrent rollouts, supervised cross-sectional warm-up objectives, and a deterministic execution adapter that converts the 100 policy scores into long/cash weights under top-$k$ selection, cash, concentration, liquidity, and turnover constraints.
Its lineage traces to an earlier LSTM--PPO parent with recent-rollout replay and behavior-cloning warm-up; a later repair step added effective-number regularization and maximum-weight penalties to that parent while keeping its return. We select it post hoc, as the highest walk-forward CAGR recorded anywhere in the campaign; it is not the campaign's own promoted artifact. Under the task's promotion rule it sits in the incubator lane---first-tier evidence on one seed, with three hard constraint violations---while the highest confirmed-lane result is a different, generation-19 cross-sectional attention policy that completed the full five-seed tier cleanly at a lower CAGR. The numbers below therefore characterize a case-study artifact chosen by the reported metric, not an artifact that cleared the campaign's clean-promotion gate.

\textbf{Results.} Table~\ref{tab:quant_results} reports the walk-forward comparison.
Over the 28 windows the discovered policy compounds to a 1{,}864.5\% cumulative return, a \textbf{53\% calendar-time CAGR versus 23\% for the paired baseline}---a 2.3-fold ratio in growth rate and an advantage of roughly 30 percentage points. The policy is positive in 26 of 28 quarters, with a 1.56 quarterly zero-rate Sharpe ratio, a 12.75\% mean and 33.48\% worst quarterly-window maximum drawdown, and beats its baseline in every calendar year of the test period. On a strictly training-excluded 2026 validation (model trained only through 23 December 2025, evaluated 2 January--21 May 2026), the policy returns 21.85\% with a 10.97\% maximum drawdown and a 1.73 daily zero-rate Sharpe. The result is also robust to execution costs: under an additional 50 basis points per executed side, the mean quarterly return declines from 12.28\% to 7.09\% but remains positive.

\begin{table}[!htb]
\centering
\small
\caption{Quant case study: rolling walk-forward results of the discovered policy versus the paired all-eligible equal-weight baseline (2019Q1--2025Q4, 28 quarterly windows). Calendar-year rows compound that year's four independently cash-reset quarterly windows; worst quarter and positive quarters refer to the \textsc{Praxist} policy.}
\label{tab:quant_results}
\begin{tabular}{lrrrc}
\toprule
Year & \textsc{Praxist} policy & Baseline & Worst quarter & Positive quarters \\
\midrule
2019 & 64.17\% & 31.20\% & 5.59\% & 4/4 \\
2020 & 194.51\% & 53.34\% & 7.34\% & 4/4 \\
2021 & 47.59\% & 31.76\% & 0.81\% & 4/4 \\
2022 & $-$19.66\% & $-$22.88\% & $-$29.14\% & 3/4 \\
2023 & 62.24\% & 34.41\% & 5.18\% & 4/4 \\
2024 & 39.96\% & 19.98\% & 3.58\% & 4/4 \\
2025 & 50.90\% & 27.63\% & $-$7.21\% & 3/4 \\
\midrule
Cumulative return (2019--2025) & 1{,}864.5\% & 320.8\% & & \\
Calendar-time CAGR & \textbf{53.07\%} & 22.80\% & & \\
\bottomrule
\end{tabular}
\end{table}

\textbf{Qualitative analysis.} The discovery trajectory illustrates the generational dynamics of \Secref{sec:method}. The best-so-far CAGR rises through intermediate recurrent-policy variants before reaching 53\%, and progress is deliberately non-monotonic at the attempt level: later generations spend much of their effort on exploration, ablation, and validation rather than on chasing the incumbent's score.
Subsequent candidates include falsification probes, positive controls, and cost-penalty variants that never exceed the incumbent but harden the evidence around it. The selected mechanism is coherent rather than accidental: the frontier retained the recurrent-PPO parent for its return, a diagnostic finding flagged its concentration risk, and the repair intervention targeted diversification directly---whose effect is visible in the reported policy's realized statistics (8.01 mean effective names, 21.69\% maximum mean single-name weight, 7.95\% mean daily turnover, 9.57\% mean cash). The repair moved concentration in the intended direction but fell short of its own pre-registered success signals of more than ten effective names and a maximum weight below 0.15, and the strong setting was retained for its return rather than for meeting those targets.

% ============================================================
% No \FloatBarrier before this subsection: Table 5 is already placed by this
% point, and the barrier would clear the page and leave most of it blank.
\subsection{Case Study: SLAM}
\label{sec:exp_slam}

\begin{figure}[!ht]
    \centering
    \includegraphics[width=0.98\textwidth]{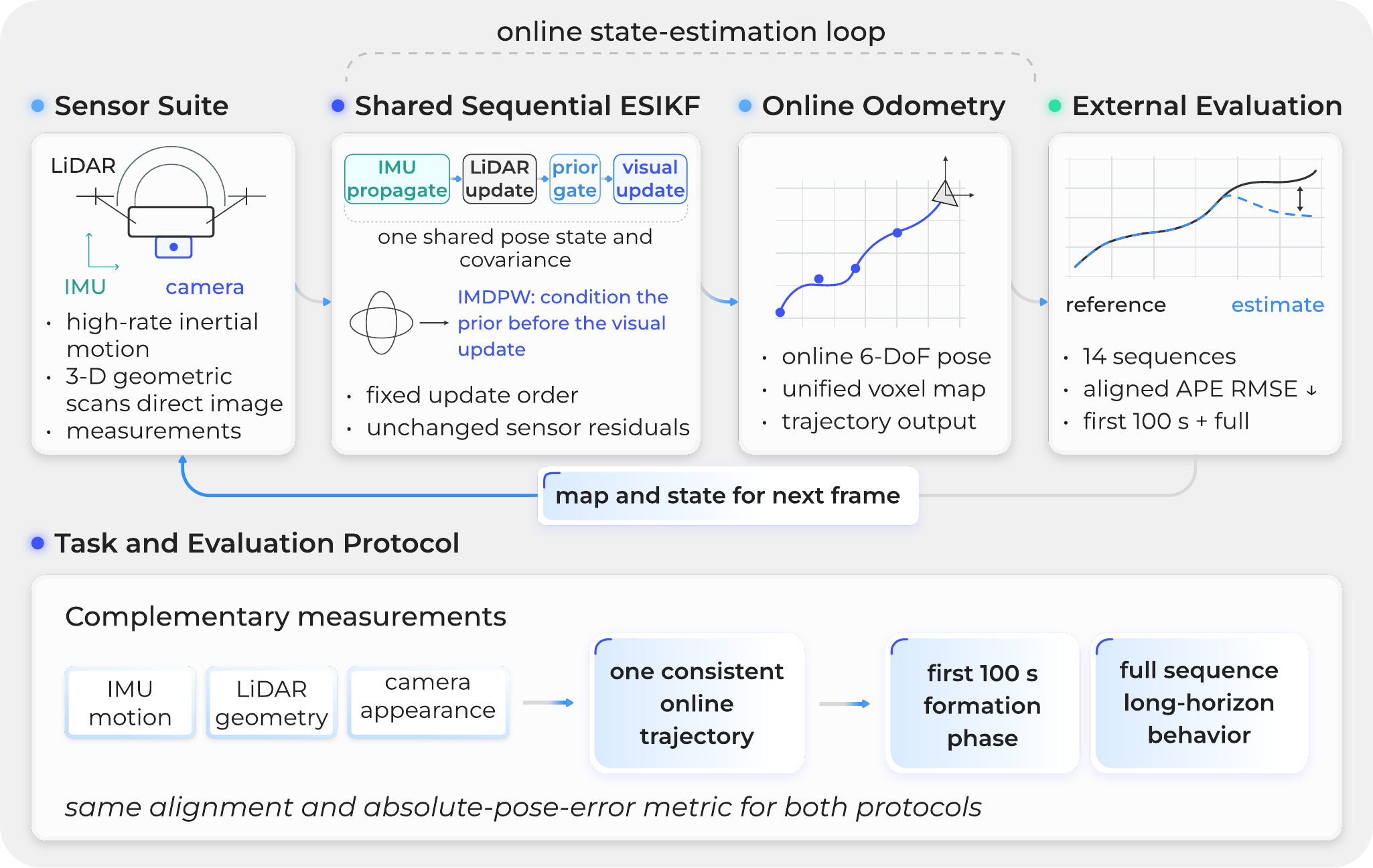}
    \caption{SLAM task overview. Synchronized IMU, LiDAR, and camera measurements update one shared ESIKF state in sequence, producing an online 6-DoF trajectory and map. \textsc{CovSched}, the blue gate, intervenes only at the visual pathway: it schedules sparse-direct visual updates by marginal value and admits map points only when they add new geometry, leaving the update order and the sensor residuals unchanged. Evaluation aligns the estimated and reference trajectories and reports full-sequence APE RMSE.}
    % The blue gate marks where \textsc{CovSched} intervenes: it runs the sparse-direct visual update only on the frames it judges informative, leaving the update order and the sensor residuals untouched. Evaluation aligns the estimated and reference trajectories and reports full-sequence APE RMSE.}
    \label{fig:slam_task_overview}
\end{figure}

\begin{table}[!htb]
\centering
\small
\caption{SLAM case study: full-sequence APE RMSE (m) of the FAST-LIVO2 baseline versus the discovered \textsc{CovSched} module on all fourteen NTU-VIRAL sequences. \emph{Coverage} is the ratio of \textsc{CovSched} to baseline ground-truth-associated samples, and the last column the reduction in the evaluator-captured amortized visual-path (VIO) processing time; both mean-row entries there are macro averages of per-sequence values, not ratios of means. Because the two arms stamp their poses differently, the APE columns are reported side by side with no relative-improvement column; \Secref{sec:exp_slam} bounds the difference between them.}
\label{tab:slam_results}
\setlength{\tabcolsep}{5pt}
\resizebox{0.98\textwidth}{!}{%
\begin{tabular}{lcccc}
\toprule
Sequence & FAST-LIVO2 RMSE & \textsc{CovSched} RMSE & Coverage & Evaluator-captured VIO time \\
\midrule
\texttt{eee\_01}  & 0.0582 & 0.0279 & 1.03 & 77.8\% \\
\texttt{eee\_02}  & 0.0602 & 0.0408 & 0.90 & 75.6\% \\
\texttt{eee\_03}  & 0.0725 & 0.0322 & 0.79 & 70.2\% \\
\texttt{nya\_01}  & 0.0537 & 0.0308 & 0.99 & 73.6\% \\
\texttt{nya\_02}  & 0.0856 & 0.0353 & 1.76 & 75.6\% \\
\texttt{nya\_03}  & 0.0848 & 0.0339 & 0.99 & 75.9\% \\
\texttt{rtp\_01}  & 0.0959 & 0.0525 & 1.23 & 81.6\% \\
\texttt{sbs\_01}  & 0.0524 & 0.0300 & 0.98 & 73.7\% \\
\texttt{sbs\_02}  & 0.0567 & 0.0341 & 0.99 & 75.6\% \\
\texttt{sbs\_03}  & 0.0614 & 0.0308 & 0.98 & 74.1\% \\
\texttt{spms\_01} & 0.1795 & 0.0925 & 0.93 & 74.8\% \\
\texttt{spms\_02} & 0.2407 & 0.1132 & 0.92 & 43.2\% \\
\texttt{spms\_03} & 0.1118 & 0.0500 & 1.15 & 70.1\% \\
\texttt{tnp\_03}  & 0.0978 & 0.0979 & 0.84 & 71.6\% \\
\midrule
Mean (14 seq.) & 0.0937 & \textbf{0.0501} & 1.03 & 72.4\% \\
\bottomrule
\end{tabular}}
\end{table}

\textbf{Problem setup.}
We consider online LiDAR--inertial--visual odometry (LIVO), where a moving platform estimates its six-degree-of-freedom pose by sequentially fusing inertial, LiDAR, and camera measurements into a shared state estimate (Figure~\ref{fig:slam_task_overview}). Our starting point is FAST-LIVO2~\citep{zheng2024fastlivo2}, a state-of-the-art tightly coupled LIVO system built around a shared error-state iterated Kalman filter (ESIKF). FAST-LIVO2 applies a fixed estimation sequence: inertial propagation, LiDAR point-to-plane update, and sparse-direct visual update. This design gives a strong reference pipeline, but it also imposes a uniform visual-computation policy: the visual update is attempted on each incoming camera frame, regardless of whether that frame provides substantial marginal information beyond the current LiDAR-constrained state and visual map.
We use aligned absolute pose error (APE) RMSE as the primary accuracy metric, computed against Leica ground truth under the evaluator's rigid Umeyama alignment with $0.02$\,s association, on the UAV-borne NTU-VIRAL benchmark~\citep{nguyen2022ntuviral}. The baseline is the stock FAST-LIVO2 implementation as built for this campaign---close to, but not byte-identical to, that build with \textsc{CovSched} disabled---and it achieves a mean full-sequence APE RMSE of $0.0937\,\mathrm{m}$ across the evaluated sequences. The campaign is deterministic and carries no training seed, so we report no repetition variance. Appendix~\ref{app:setup-slam} gives the full evaluation protocol, the timing environment, the campaign configuration, and the run-acceptance caveat that qualifies which of the campaign's repeated replays reach the table.

\textbf{Discovered mechanism.}
The mechanism selected by \textsc{Praxist}, \textsc{CovSched}, treats the sparse-direct visual update as a resource to be allocated rather than as an unconditional step in the estimation loop. It adds two visual-pathway controls to the otherwise unchanged FAST-LIVO2 pipeline: a frame-level marginal-value scheduler that decides whether an incoming camera frame triggers a sparse-direct visual update, and a deduplicating map-admission policy that limits redundant map growth by suppressing candidate landmarks which re-observe geometry already represented in the sparse map. IMU propagation, the LiDAR point-to-plane update, and every residual definition are left unchanged, so no estimator is retuned. The two builds are not, however, identical outside the visual pathway: the \textsc{CovSched} build publishes odometry stamped with the LiDAR measurement time, whereas unmodified FAST-LIVO2 stamps its output with wall-clock publication time. Because APE associates estimated and reference poses inside a $20$\,ms window, that difference is itself an accuracy factor, and we quantify it below.

In the champion configuration the scheduler (\texttt{PVTR\_MODE=8}) keys on a LiDAR translation-observability signal $o_{\min} = \lambda_{\min}(\sum_i \mathbf{n}_i \mathbf{n}_i^\top)$, where $\mathbf{n}_i$ is the normal of a LiDAR point-to-plane correspondence. Large $o_{\min}$ indicates that the LiDAR geometry already constrains the local translation subspace well, so the marginal value of a visual update is low; small $o_{\min}$ indicates a weakly constrained translation direction, where visual information is most valuable. \textsc{CovSched} combines this instantaneous redundancy signal with a persistent low-observability dwell counter and integrates the resulting demand through a saturating skip budget with a hard cap of three consecutively skipped frames, which is what prevents permanent visual starvation. The recorded schedules are strongly protective of the frames that need vision: on \texttt{nya\_03}, high-$o_{\min}$ frames are skipped $74.9\%$ of the time against $0.3\%$ for low-$o_{\min}$ frames. The second component (\texttt{VMAP\_DEDUP=1}) discards a candidate visual point as a re-observation rather than admitting it as new coverage when it lies within $0.08$\,m of an existing map point in the same voxel with a normal deviation below $15^\circ$. \textsc{CovSched} thus governs both when visual computation is spent and which visual observations are allowed to grow the map; Appendix~\ref{app:setup-slam} records the parameter settings and two qualifications they carry.

\textbf{Results.}
Table~\ref{tab:slam_results} reports the full-sequence comparison against FAST-LIVO2. Across the fourteen NTU-VIRAL sequences the two arms record mean APE RMSE of $0.0937\,\mathrm{m}$ and $0.0501\,\mathrm{m}$ respectively, with the \textsc{CovSched} value lower on thirteen sequences and effectively tied on \texttt{tnp\_03} ($0.0978\,\mathrm{m}$ versus $0.0979\,\mathrm{m}$). We deliberately do not report that gap as an accuracy improvement. The two arms stamp their published poses under different conventions, and re-associating them under a single rule removes the gap almost entirely; the caveats below give the measurement and the bound it implies. The accuracy claim we do make is the weaker one: on this benchmark, spending far less visual computation does not cost trajectory accuracy.

The main measured efficiency gain is on the visual-processing path. Across the same fourteen runs, the evaluator-captured amortized VIO processing time decreases by $72.4\%$ on average, with a median reduction of $74.4\%$; every sequence shows a reduction, ranging from $43.2\%$ on \texttt{spms\_02} to $81.6\%$ on \texttt{rtp\_01}. Because FAST-LIVO2's LiDAR mapping and visual processing are reported by separate asynchronous timing streams, we do not interpret this number as a $72.4\%$ reduction in end-to-end SLAM latency, throughput time, or CPU load. The LiDAR-side timing is essentially unchanged. As a conservative aggregate proxy, summing the internally reported LIO and VIO thread-wall workloads gives a $22.1\%$ reduction. The supported resource claim is therefore that \textsc{CovSched} substantially reduces the measured visual-path workload.

% \textbf{Results.}
% Table~\ref{tab:slam_results} reports the full-sequence comparison against unmodified FAST-LIVO2. Across the fourteen NTU-VIRAL sequences, \textsc{CovSched} reduces mean APE RMSE from $0.0937\,\mathrm{m}$ to $0.0501\,\mathrm{m}$. The mean per-sequence relative improvement is $45.4\%$, with a median of $49.2\%$. \textsc{CovSched} improves thirteen of the fourteen sequences and is effectively neutral on \texttt{tnp\_03}, which moves from $0.0978\,\mathrm{m}$ to $0.0979\,\mathrm{m}$. The same runs also show a large resource effect: mean per-LiDAR-frame processing time falls by $72.4\%$ on average (median $74.4\%$), and runtime improves on every sequence, from $43.2\%$ on \texttt{spms\_02} to $81.6\%$ on \texttt{rtp\_01}.

These results support the resource-allocation reading of the mechanism: \textsc{CovSched} does not trade accuracy for speed by uniformly thinning the visual stream, but conditions visual computation on state-dependent marginal value and limits redundant sparse-map growth. The saving appears on every sequence, whereas the residual APE difference does not track task difficulty ($r=0.13$ between baseline APE and the per-sequence gap). That is the pattern the timestamp analysis below predicts, and it is consistent with the hypothesis motivating the mechanism---that many baseline visual updates are redundant given the current LiDAR-constrained state and visual map---while leaving the accuracy side of the comparison uncommitted.

The claim is nevertheless bounded, most consequentially by a timestamp confound. The two arms of Table~\ref{tab:slam_results} do not share a pose-timestamp convention: \textsc{CovSched} publishes measurement-time stamps and the unmodified baseline publishes wall-clock stamps, and a same-binary control that toggles only that setting on a fixed trajectory moves APE by a factor of $1.74$--$1.80$. Re-associating the baseline poses under the \textsc{CovSched} rule---possible without interpolation on the ten sequences where both arms emit the same number of poses---leaves the mean relative APE change at $-0.09\%$ rather than the $+45\%$ the raw columns suggest. The gap between the two accuracy columns is therefore dominated by the timestamp and association convention, which is why we report no improvement column and make no accuracy-gain claim. Three further caveats, detailed in Appendix~\ref{app:setup-slam}, qualify the rest: the ground-truth-associated pose sets differ between the arms, so the comparison must be read together with coverage ($0.79\times$ to $1.76\times$); relative pose error over short $1$--$30$\,s windows is mixed rather than uniformly improved; and the component controls are too thin, and too confounded with the map-admission filter, to attribute the result to either half of the mechanism. Establishing the mechanism's true effect on accuracy and supplying the missing factorial both require the same outstanding experiment---a paired re-run of all fourteen sequences with both arms built from the same source and stamped the same way. The current evidence therefore supports \textsc{CovSched} as a structured, bounded, map-aware visual-resource policy, without assigning the improvement to either component alone.

\textbf{Qualitative analysis.}
\textsc{CovSched} is a change in visual-processing policy rather than in sensor modeling: FAST-LIVO2 implicitly assumes that every incoming visual frame should enter the sparse-direct update, and \textsc{CovSched} makes that decision conditional on the current state of the LiDAR geometry and the visual map. The distinction also shaped the evidence package. Because the estimated trajectories are visually close under a direct overlay, the run reports distributional APE, executed and skipped visual-update schedules, visual-map growth, runtime, and coverage rather than relying on trajectory plots, and---following the evidence contracts of \Secref{sec:method}---it declines to claim a duplicate-merge statistic that its instrumentation never recorded, and preserves the coverage ratios and rate-matched skipping control that qualify the comparison. That restraint matters because a visual-resource mechanism is easy to overclaim: it may improve APE by reducing computation, by suppressing harmful redundant updates, by changing the evaluated pose set, or by some combination of the three. The current evidence supports \textsc{CovSched} as a strong structured resource policy, and the outstanding factorial would sharpen the causal attribution.

\FloatBarrier
\subsection{Case Study: Fusion (Tokamak Magnetic Control)}
\label{sec:exp_fusion}

\begin{figure}[!ht]
    \centering
    \includegraphics[width=0.98\textwidth]{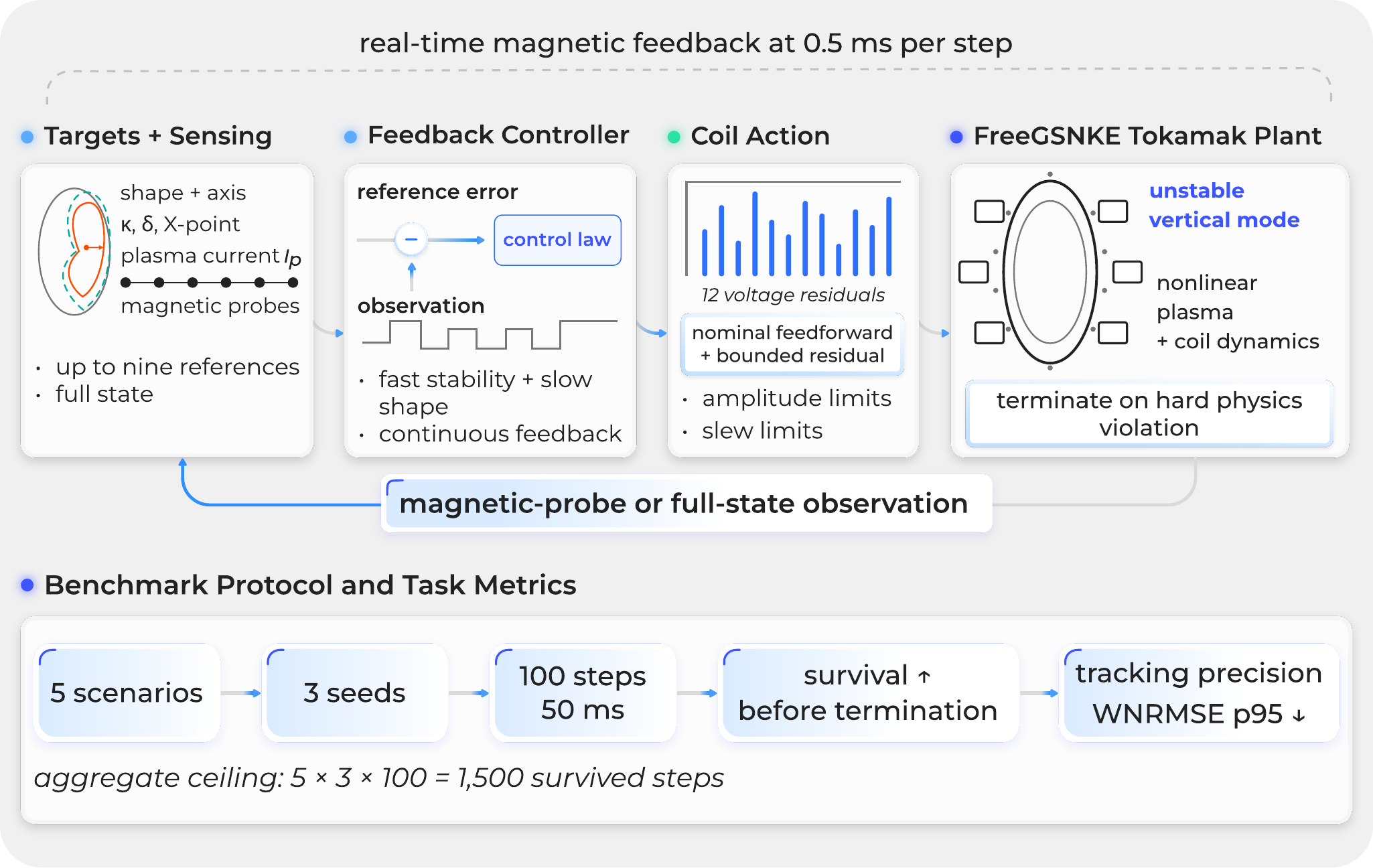}
    \caption{Fusion task overview. At every 0.5-ms step the controller compares plasma targets with the scenario's observation and adds 12 bounded voltage residuals to nominal coil feedforward; the two closed-loop controllers evaluated here read the target values from the harness rather than from the probe observation, as noted in the text. FreeGSNKE advances the nonlinear tokamak state and terminates on hard physics violations; the benchmark scores 15 episodes by aggregate survival and tracking precision, which we report on both the benchmark's full horizon and the common horizon defined in the text.}
    \label{fig:fusion_task_overview}
\end{figure}

\textbf{Problem setup.} The task is real-time magnetic control of a tokamak: holding a ring of plasma in the right shape and place inside a MAST-U-like spherical fusion device (Figure~\ref{fig:fusion_task_overview}). The plasma is simulated by FreeGSNKE, a time-dependent free-boundary Grad--Shafranov solver \citep{amorisco2024freegsnke}, wrapped as a reinforcement-learning environment. Every 0.5\,ms it adjusts 12 magnetic-coil voltages, added to the scenario's nominal schedule under hard amplitude and slew limits, and must hold up to nine targets on their references: the plasma boundary's radial extents, the position of its magnetic axis, its elongation $\kappa$ and triangularity $\delta$, the lower X-point position, and the plasma current $I_p$. Any hard physics violation ends the episode at once: the plasma drifting out of position, losing its current, exceeding a coil limit, touching the wall, or the solver diverging. The plant is genuinely unstable: left uncontrolled the plasma column runs away vertically, at open-loop growth rates matching those measured in real MAST-U vertical-displacement experiments. The scenario's nominal voltage schedule nonetheless carries an unactuated plasma through part of the horizon, which is why the zero-feedback floor in Table~\ref{tab:fusion_results} sits well above zero rather than near it.\footnote{The simulator derives growth rates of 417--470\,rad/s for the four perturbed scenarios---a vertical-displacement doubling time of ${\approx}1.5$\,ms, about three control periods---comfortably inside the 380--500\,rad/s range measured on the MAST-U device itself \citep{lvovskiy2025framework}.}

The benchmark protocol is 5 scenarios $\times$ 3 random initializations $\times$ a 100-step (50-ms) horizon, so a controller that never fails survives 1{,}500 steps in total. Four scenarios build the observation from what magnetic diagnostics would see---$32$ flux loops and $32$ pickup coils, corrupted by Gaussian noise, together with the $12$ coil currents, the previous $12$ voltage commands, and the target references, stacked over two frames and delayed by one step, for a $182$--$194$-dimensional vector---and add plasma and actuator perturbations; one certification scenario gives the full simulator state, unperturbed. There are two task-native \textbf{baselines}. The first is a zero-feedback controller, the task's performance floor: the plasma simply coasts on the scenario's nominal voltages. The second---the demanding one---is a PCS-style controller based on the plasma-control architecture deployed on MAST-U in its 2022--2023 experimental campaign \citep{anand2024mastu}: damped pseudo-inverse virtual circuits for slow shape control, an independent vertical-stabilization PD loop, and an ohmic plasma-current controller. The benchmark implementation uses FreeGSNKE's one-step voltage sensitivities and a virtual-circuit gain of $g_{\mathrm{vc}}=0.3$ selected by an offline stability scan. Two protocol boundaries qualify every number below. First, both closed-loop controllers are privileged-state controllers: the evaluation harness passes the simulator's true target values to the controller on every step, and both read their tracking errors from that channel rather than from the noisy probe observation the scenarios construct, which neither controller consumes---we accordingly refer to those four scenarios below by their perturbations rather than their observation model. The comparison between them is therefore internally matched, but it does not establish that either would work from probe measurements alone; a probe-only re-evaluation behind a state estimator is outstanding work. Second, the PCS-style baseline is an architectural reconstruction rather than a port of the deployed controller: its slow loop is proportional instead of PI, it uses one-step voltage sensitivities in place of a steady-state current Jacobian, it stabilizes vertically through the fast \texttt{PX} coil instead of the antisymmetric \texttt{P6} pair, and its vertical gains, $I_p$ block, shaping-coil counter-drain, and waypoint warm-up are inherited from the \textsc{Praxist} controller. It is a strong in-simulator baseline, not an independent reproduction of MAST-U's plasma control system. Appendix~\ref{app:setup-fusion} records the evaluator's runner and reset lifecycle, to which every score below is bound, and the campaign configuration.

\textbf{Discovered artifact.} The selected artifact, \texttt{HybridJacobianPDV1}, is a sparse Jacobian-sign PD controller: a superposition of four functionally distinct channels---a full-error vertical PD on the dedicated fast coil with a fixed $-0.30$ counter-drain onto three shaping coils, a waypoint-ramped radial-position proportional loop, an ohmic $I_p$ loop on the central solenoid, and six offline system-identified unit-norm Jacobian sign vectors that route each shape axis to the three coils identified as dominating it---under a global action clip and slew limit. The channels are decoupled by design intent, not by actuator: the vertical, radial, and shaping channels all act on \texttt{P4}--\texttt{P6}, and their contributions are summed into one voltage vector before the shared clip and slew limit. Architecturally, the artifact belongs to the same inverse-Jacobian PD family as the deployed PCS; it differs in replacing the dense damped pseudo-inverse feedback matrix with sparse sign vectors, and in closing the loop on seven geometric target axes rather than four (explicitly adding $\kappa$, $\delta$, and the magnetic-axis radial position, which the deployed PCS does not feed back). Counting the dedicated vertical and $I_p$ loops as well, the two controllers close nine and six axes respectively.

\textbf{Results.} Table~\ref{tab:fusion_results} reports the benchmark comparison on both precision metrics, and the verdict is split rather than uniform. \textsc{Praxist} attains the highest aggregate survival (1{,}264 of 1{,}500 steps, versus 1{,}222 for the PCS-style controller and 1{,}039 for zero feedback) and the lowest common-horizon tracking error (WNRMSE $p_{95}$ of 2.86 versus 2.99 and 4.89), and it has the lowest common-horizon error in all four perturbed scenarios. The PCS-style controller completes more episodes (11 of 15 against 10) and holds the lower error on the benchmark's original full-horizon metric (aggregate $p_{95}$ of 4.42 versus 4.65), leading there on three of the four perturbed scenarios as well. Because a controller that does nothing already survives 1{,}039 steps, survival margins are best read against the 461 steps of headroom that remain: \textsc{Praxist} recovers 48.8\% of that headroom and the PCS-style controller 39.7\%, a gap of 9.1 points that the raw ratios compress into a 22\% improvement over the open-loop floor and a 3.4\% improvement over the baseline.

Which precision metric to read is itself a finding of this case study. The full-horizon percentile is \emph{not} monotone in controller quality, because longer-surviving controllers contribute more near-failure, high-error steps to the sample pool, and its aggregate values here show the distortion directly: it separates the two closed-loop controllers from zero feedback by only 10\% and 5\% even though they survive 18\% and 22\% longer, so on that metric feedback registers as nearly worthless on a plant that is unstable without it. Recomputing precision on a per-seed common horizon capped at the zero-feedback baseline's survival length removes the confound and restores the physically expected ordering, at the price of discarding the late-horizon steps where the two closed-loop controllers differ most---which is why we report both rather than substituting one for the other. Neither is the benchmark's verdict: its own classification thresholds are stricter---$\leq 1.0$ for \emph{acceptable} and $\leq 0.5$ for \emph{solved}, and both additionally require a completed horizon---so none of the three controllers is classified as passing under the official rule, and both precision columns are descriptive comparisons reported separately from that classification.

On the common horizon both closed-loop controllers cut per-step tracking error by roughly 40\% relative to zero feedback; relative to the PCS-style controller, \textsc{Praxist} improves aggregate survival by 3.4\% and common-horizon precision by 4.4\%, margins that Appendix~\ref{app:setup-fusion} places against the reproduction spread of a single controller under the same protocol. The clearest qualitative difference appears on the certification scenario under the full horizon: there, explicit feedback on $\kappa$, $\delta$, and the axis position keeps late-horizon tracking bounded (full-horizon $p_{95}$ of 3.76 versus 13.46 for the PCS-style controller), the one place where the two horizons disagree by an order of magnitude rather than by a few percent.

\begin{table}[!htb]
\centering
\small
\caption{Fusion case study: MAST-U magnetic-control benchmark (5 scenarios $\times$ 3 seeds $\times$ 100-step horizon; 1{,}500 aggregate steps maximum). Survival counts steps before a hard physics termination; Compl.\ is the fraction of the 15 episodes completing the full horizon; WNRMSE $p_{95}$ is the 95th percentile per-step weighted normalized tracking error (lower is better). The six common-horizon columns cap each episode at the zero-feedback baseline's survival length for that scenario and seed; the last column is the benchmark's original full-horizon aggregate, and Table~\ref{tab:fusion_full_horizon} gives that metric per scenario. Each \emph{all} entry is pooled over every scored step rather than averaged across the scenario columns. Scenario columns abbreviate \texttt{main\_shape}, \texttt{axis\_ip}, \texttt{ip\_shape}, \texttt{xpoint}, and \texttt{cert\_shape}. Best value per column in bold. Both closed-loop controllers read their target errors from the harness rather than from the probe observation, so the table is an internally matched comparison inside a privileged-state harness, as stated in the text.}
\label{tab:fusion_results}
\setlength{\tabcolsep}{4pt}
\begin{tabular}{lccccccccc}
\toprule
& & & \multicolumn{6}{c}{Common-horizon WNRMSE $p_{95}$ ($\downarrow$)} & Full-hor. \\
\cmidrule(lr){4-9}\cmidrule(lr){10-10}
Controller & Survival (/1500) & Compl. & main & axis & ip & xpt & cert & all & all \\
\midrule
Zero-feedback baseline & 1039 & 0.267 & 4.77 & 6.10 & 5.12 & 3.85 & 4.74 & 4.89 & 4.89 \\
MAST-U PCS-style & 1222 & \textbf{0.733} & 2.63 & 3.54 & 2.63 & 1.38 & \textbf{2.76} & 2.99 & \textbf{4.42} \\
\textsc{Praxist} (ours) & \textbf{1264} & 0.667 & \textbf{2.55} & \textbf{3.48} & \textbf{2.43} & \textbf{1.21} & 2.81 & \textbf{2.86} & 4.65 \\
\bottomrule
\end{tabular}
\end{table}

\textbf{Qualitative analysis.} The value of this case study lies less in the margin than in the discovery dynamics. The champion was synthesized by \emph{composing} mechanisms discovered independently: a full-error vertical PD with coil counter-drain, its high-derivative-gain variant, simulator-identified Jacobian sign vectors, and a shared waypoint warm-up for the slow channels. The run's own probes are directional rather than single-factor. Replacing the counter-drain collapses the plasma at step 82, and extending the delayed waypoint ramp to the vertical channel collapses it at step 85; both variants change the coil allocation at the same time as the mechanism under test, and both are certification-only, single-seed probes, so they order the variants without isolating one cause. The sign vectors are the clearest case of the run correcting itself. An early comparison of a hand-tuned controller against a system-identified one showed 57 steps against a full horizon, which reads as evidence that identification matters; but those two controllers also differed in gains, coil maps, deadbands, and smoothing. A later matched control that held the trajectory and gains fixed and randomly flipped half of the identified signs scored 297 against 298 aggregate steps, with identical results on three of the probe scenarios and certification error of 3.49 against 3.48. The run therefore recorded the identified signs as non-essential, and we report that as the standing conclusion. Subsequent experiments in the reported lineage tested observation-quality gating, integral loops, coil specialization, mechanism substitutions, per-coil quadratic programs, and spectral decoupling without finding a stronger controller, and the survival-versus-certification-precision Pareto frontier stayed around this artifact. That statement is bounded to the seven generations behind the reported result: a later continuation of the campaign, reported alongside it, produced controllers with substantially higher aggregate survival (up to 1{,}481 of 1{,}500) at worse full-horizon precision, so the frontier has since acquired new survival-heavy trade-off points rather than remaining a single point.

% ============================================================
% Keep the preceding subsection's floats from spilling into this case study.
\FloatBarrier
\subsection{Case Study: Job-Shop (Dynamic Flexible Job-Shop Rescheduling)}
\label{sec:exp_jobshop}

\begin{figure}[!ht]
    \centering
    \includegraphics[width=0.98\textwidth]{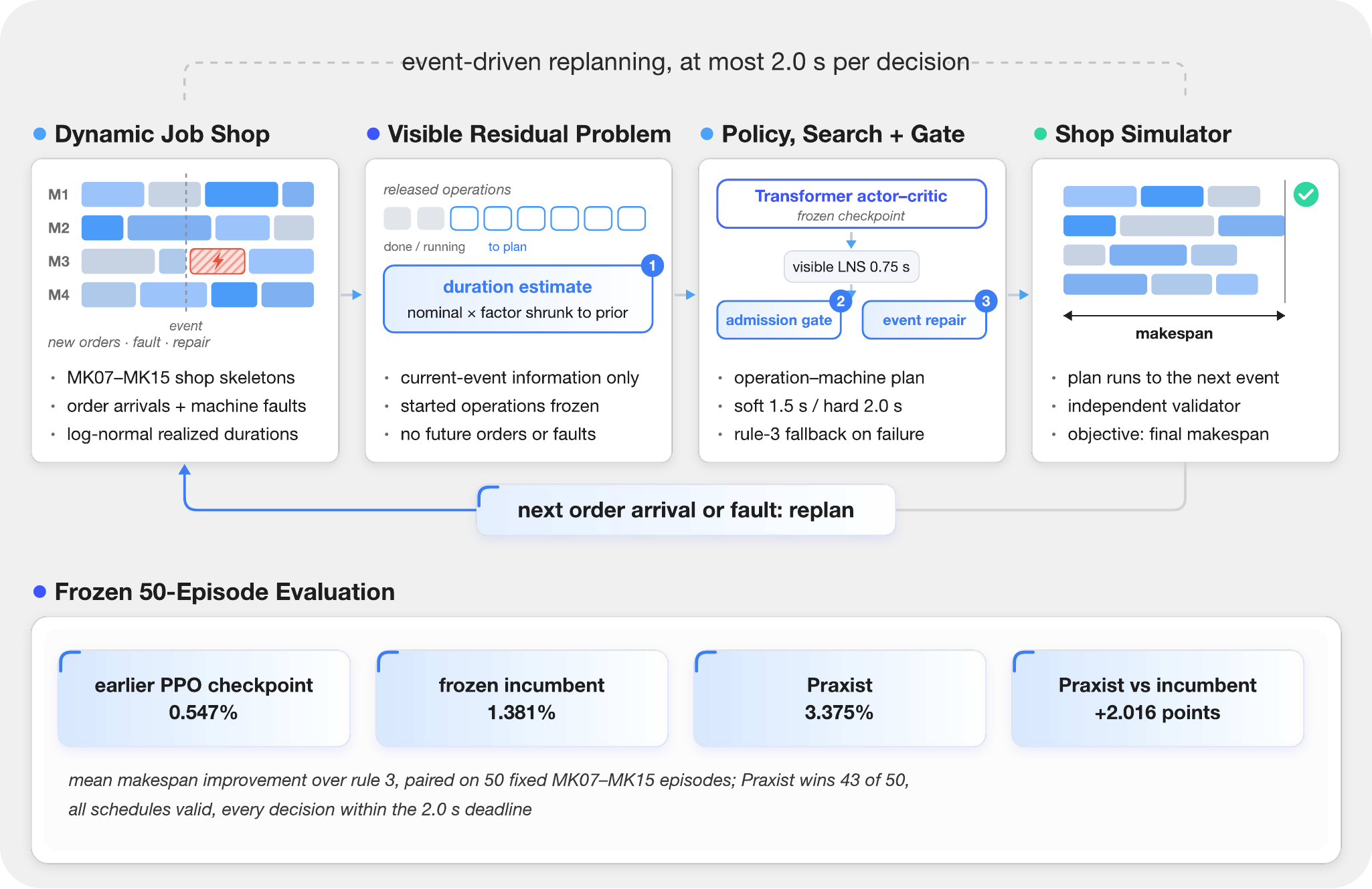}
    \caption{Job-Shop task overview. Each external event---a batch of new orders, a machine fault, or a repair---triggers a replan from the visible residual problem: released, unstarted operations with duration estimates built from nominal times and completed work only. A frozen Transformer actor--critic proposes a complete operation--machine plan, visible LNS refines it, and an admission gate decides whether to deploy it or keep rule~3; the shop simulator then plays the plan forward to the next event, and an independent validator checks the schedule. Blue marks the three components \textsc{Praxist} modified: the duration estimate, the admission gate, and a restricted repair after faults end or orders arrive. The evaluation strip reports mean makespan improvement over rule~3 on the fixed 50-episode manifest.}
    \label{fig:jobshop_task_overview}
\end{figure}

\textbf{Problem setup.} The task is online rescheduling of a dynamic flexible job shop (Figure~\ref{fig:jobshop_task_overview}), in which each operation of a job may run on any of several eligible machines and the plan must be rebuilt whenever an external event arrives: a batch of new orders, a machine failure, or a machine's return to service. At each event the policy sees only what is visible at that moment---released jobs, completed operations, machine states, and nominal processing times---and must return a complete operation--machine plan for every released, unstarted operation within a 1.5-s soft and 2.0-s hard deadline. An independent watchdog enforces the deadline and falls back to a fixed dispatching rule, rule~3, on any timeout, crash, or invalid plan. Realized processing times are log-normal draws around the nominal ones and failures have unknown repair times, so a plan is scored only after the simulator plays it forward against the episode's event tape, and the objective is final makespan. Instances are built on the Brandimarte MK skeletons \citep{brandimarte1993routing}: MK01--MK06 for training and development, and MK07--MK15 for the 50 evaluation episodes, which contain 215--671 operations and 8--12 external events each. Future orders, failures, repair times, and realized durations are forbidden at inference, and an independent validator re-checks every schedule.

\textbf{Evaluation protocol and baseline.} The primary metric is the mean per-episode makespan improvement over rule~3 on a fixed 50-episode manifest, with a 10{,}000-resample paired-bootstrap 95\% interval, reported together with the lower tail (P10, CVaR$_{10}$, worst episode), validity, timeouts, fallbacks, and decision latency. The \textbf{baseline} is the task's frozen incumbent, a 2.0M-parameter ReSched-style operation--machine Transformer actor--critic \citep{xiao2026resched} trained by behavior cloning from a visible-information CP-SAT teacher \citep{perron2024cpsat} and then by PPO \citep{schulman2017ppo}. Each of its plans is refined by 0.75\,s of visible-information large-neighborhood search (LNS) and deployed only if its visible estimated improvement over rule~3 clears a 4\% gate. It improves on rule~3 by 1.381\%, against 0.547\% for its own earlier PPO checkpoint and $-2.103\%$ for ungated LNS alone. A \textsc{Praxist} candidate may edit only a run-owned overlay of the policy code and its deployment configuration; the simulator, validator, rule-3 comparator, metrics, and manifest are hashed before every complete evaluation, and retraining is capped at 45+45 GPU-minutes per candidate.

\textbf{Discovered artifact.} The selected artifact keeps the incumbent's network checkpoint byte for byte and changes how its plans are formed and admitted, in three places. First, it replaces the duration estimator. The incumbent plans each remaining operation at its nominal time scaled by a per-machine running average of realized-to-nominal ratios, which early in an episode rests on a handful of completions; the artifact shrinks that average 90\% of the way toward the public prior of 1 and uses the prior outright whenever the average departs from 1 by more than 0.1. Second, it replaces the single 4\% gate with per-event-type thresholds between 0 and 1.2\%, fitted on public development episodes. Third, when a fault ends or orders arrive, it runs one restricted CP-SAT repair of the deployed plan with 90\% of its assignments frozen and a 0.06-s budget, adopts the repair only if the visible objective improves by at least 0.2\%, and skips it whenever it could push the decision past 1.8\,s. Nothing is retrained.

\begin{table}[!htb]
\centering
\small
\caption{Job-Shop case study: makespan improvement over rule~3 (\%) on the fixed
50-episode MK07--MK15 manifest; positive values are shorter makespans. W/T/L counts
episodes won, tied, and lost against rule~3. The oracle-event-rule row is a
future-aware, non-deployable hindsight reference that picks the best of 48 dispatching
rules at every event using the true future tape; it is neither ranked nor an upper bound
on direct plans.}
\label{tab:jobshop_results}
\setlength{\tabcolsep}{5pt}
\begin{tabular}{lcccccc}
\toprule
Method & Mean & 95\% CI & P10 & CVaR$_{10}$ & Worst & W/T/L \\
\midrule
Earlier PPO checkpoint & 0.547 & $[-0.127, 1.179]$ & $-2.680$ & $-4.190$ & $-8.429$ & 30/7/13 \\
Frozen incumbent & 1.381 & $[0.632, 2.162]$ & $-1.125$ & $-2.973$ & $-5.775$ & 26/9/15 \\
\textsc{Praxist} (ours) & \textbf{3.375} & $[2.415, 4.364]$ & \textbf{0.000} & \textbf{$-$1.889} & \textbf{$-$4.575} & \textbf{43}/4/3 \\
\midrule
\emph{Oracle-event-rule reference} & 3.502 & $[2.876, 4.156]$ & 0.485 & 0.134 & 0.000 & 47/3/0 \\
\bottomrule
\end{tabular}
\end{table}

\textbf{Results.} Under the frozen 50-episode protocol the artifact improves on rule~3 by 3.375\% (CI $[2.415, 4.364]$), 2.4 times the incumbent's 1.381\% (Table~\ref{tab:jobshop_results}); paired directly against the incumbent on the same episodes, the gain is 2.016 points (CI $[1.341, 2.718]$). It wins 43 of the 50 episodes against rule~3, where the incumbent wins 26, loses 3 rather than 15, and improves every tail statistic, with all 50 schedules valid and no timeout, fallback, or out-of-memory event. The per-skeleton mean is positive on all nine skeletons, from 0.80\% on MK12 to 8.81\% on MK07, whereas the incumbent was negative on MK12 and flat on MK08, and the artifact keeps rule~3 on only 81 of its 508 decisions against the incumbent's 336. The deadline holds, at a modest latency cost: the 99th-percentile decision takes 1.73\,s against 1.65\,s, and the slowest 1.75\,s. Against the oracle-event-rule reference, the artifact reaches 96\% of its mean where the incumbent reached 39\%, and exceeds it on MK07 and MK10, showing that this reference does not bound direct plans; on the tail, however, the oracle never loses to rule~3, and the artifact still does on three episodes.

\textbf{Lineage.} The run diagnosed the incumbent before changing it (Table~\ref{tab:jobshop_lineage}). A generation-0 control showed that LNS applied to rule~3's own plan loses 1.71\% to rule~3, so the incumbent's gain lies in the neural plan; lowering the gate threshold from 4\% to zero raised the mean to 1.75\% at generation~1 but deepened the tail; and three independent analyses bounded what choosing among dispatching rules on visible information could add at about 0.3 to 0.5 points. The run therefore turned from the policy to its inputs, the duration estimates behind every visible makespan estimate used to rank and admit plans. Prior shrinkage of those estimates, developed on public development episodes first, reached 2.67\% at generation~3; the disagreement-gated version reached 2.98\% at generation~4, the re-fitted gate 3.10\% at generation~5, and the event-targeted repair 3.30\% at generation~7, but at 0.20\,s per repair it pushed six decisions past the 2-s deadline. Generation~8 cut the repair budget 3.3-fold and added the adoption margin and latency cap, restoring the deadline at a statistically tied mean (a paired gain of 0.07 points, CI $[-0.03, 0.21]$).

\textbf{Claim boundary.} The artifact satisfies the task's conservative promotion rule against the incumbent: a positive paired lower bound, no regression in validity, fallbacks, timeouts, or memory, and every decision inside the deadline. Its gain comes from how plans are estimated, admitted, and repaired around the inherited checkpoint, not from retraining. The result is bounded by a frozen simulator, by an information boundary limited to the current shop state, and by a fixed 50-episode calibration set that is adaptively reused during the search and so carries selection-overfitting risk; the estimator and gate parameters, which account for most of the gain, were fitted on public development episodes rather than on that set. Appendix~\ref{app:setup-jobshop} gives the lineage and the campaign setup.

% Close the last case study so its floats stay in this section.
\FloatBarrier

%% file: sections/7_conclusion.tex
\section{Conclusion and Discussion}
\label{sec:conclusion}

This work presents \textsc{Praxist}, an autonomous research system built around a simple premise: long-horizon R\&D improves when a campaign inherits evidence, not just scores. \textsc{Praxist} realizes this as a generational artifact-to-lineage cycle---evaluated artifacts become typed findings, findings populate a lane-structured frontier, PI/Chair synthesis emits the next agenda, durable lessons are compressed into Gems, and the accumulated lineage both directs later attempts and documents why the final artifact is credible. Our two-part evaluation targets complementary questions. On MLE-bench, \textsc{Praxist} obtains 60 medals (80.0\%) across all 75 tasks, including 49 gold, compared with 55 medals (73.3\%) and 34 gold for Claude Code on Claude Opus 4.8.
The case studies show that the system can carry open-ended R\&D problems to artifacts that beat their task-native baselines: it discovered a trading policy with a 53\% walk-forward CAGR, 2.3 times the 23\% of its paired all-eligible equal-weight baseline, turned the always-on visual update of a state-of-the-art SLAM system into a resource-aware schedule that cut the evaluator-captured visual-path processing time by 72.4\% without costing trajectory accuracy, synthesized a tokamak controller with 3.4\% higher aggregate survival and 4.4\% lower common-horizon tracking error than a task-native MAST-U PCS-style controller inside the same privileged-state harness, though not on that benchmark's original full-horizon precision metric, re-engineered the duration estimates, plan admission, and post-disruption repair of a frozen job-shop rescheduling policy to raise its mean makespan gain over a dispatching rule from 1.38\% to 3.38\% on a fixed calibration set, and produced a deterministic rocket-landing controller that lands $12{,}288/12{,}288$ trajectories under a frozen first-contact protocol; Weco, an autonomous code optimizer given the same task, objective, and 4.03\% starting artifact, reports 17.12\%. The rocket controller was not found in one attempt: its lineage accumulates a fuel-commit governor, attitude-authority and phase-transition guards, and finally a closed-form constrained control allocator, each inherited as typed evidence rather than as a score.

Beyond the headline scores, two properties distinguish \textsc{Praxist} as a research collaborator. First, 49 of its 60 medals are gold (81.7\%), so the majority of its successful MLE-bench outcomes reach the highest medal tier. Second, results can arrive with their solution lineage---an inspectable account of the mechanisms, controls, and failures behind a result---supporting reuse and extension by human scientists; the five case studies report those lineages in detail, whereas for the MLE-bench sweep we report only the graded outcome per task, and the coverage the ledgers achieve varies with how long a run persisted before it was stopped. These properties suggest immediate applications wherever a task admits an executable evaluator: model and algorithm development, controller synthesis for physical systems, simulation-driven engineering design, and quantitative strategy research.
Looking forward, we see three directions: inheritance across campaigns, so that Gems and lineages from one problem seed the next; extension to slower or noisier evaluators, bringing the cycle closer to laboratory science; and deeper human--AI collaboration through lineages, positioning \textsc{Praxist} not as a replacement for scientists but as an instrument that compounds their evidence.

%% file: sections/availability.tex
\section*{Code and Released Runs}

The \textsc{Praxist} implementation, the project page, and the runs behind the results
reported in this paper are publicly available. The released archive holds the generational
record of each run---artifacts, evaluation records, findings, frontier lanes, agendas, and
lineage traces---so that the trajectories summarized in Section~\ref{sec:experiments} and in
the appendices can be inspected generation by generation, rather than only through the
aggregate scores reported here.

\begin{itemize}[itemsep=2pt,topsep=4pt]
  \item \faGithub\; Code: \href{https://github.com/sapientinc/praxist}{\texttt{github.com/sapientinc/praxist}}
  \item \faGlobe\; Project page: \href{https://praxist.sapient.inc/}{\texttt{praxist.sapient.inc}}
  \item \faGoogleDrive\; Released runs: \href{https://drive.google.com/drive/folders/1xNQ6mI8Q2WynzWr3i7RdI4FevNHrn1gs?usp=sharing}{Google Drive archive}
\end{itemize}

%% file: sections/acknowledgements.tex
\section*{Acknowledgements}

We thank Yaning Han, Rongzu Zhang, Daohai Yu and Yangzhou Liu for their insightful discussions and contributions to this work.

%% file: sections/8_supplementary.tex
\startcontents[appendix]

\begingroup
\footnotesize
\printcontents[appendix]{}{1}{\setcounter{tocdepth}{2}}
\endgroup

\clearpage

\section{Method details}
\label{app:additional-methods}

This appendix gives the field-level contracts behind the method in
Section~\ref{sec:method}. The main text defines \textsc{Praxist} as the generational
state transition
$\mathcal{S}_g=(\mathcal{F}_g,\mathcal{A}_g,\mathcal{G}_g,\mathcal{L}_g)$:
artifacts ground attempts, findings interpret evaluated evidence, frontiers
determine inheritance status, agendas direct later generations, Gems preserve
durable lessons, and lineage records the path that produced the final artifact.
Here we expand each object using the same symbols and vocabulary as the main
text.

The emphasis is on methodological interfaces rather than software mechanics.
We describe what information is represented, which role consumes it, and what
each role emits. Concrete command invocations, storage layouts, and
release-specific engineering details belong to implementation documentation or
the experimental protocol. MLE-style tasks are used as a running example of a
task-family realization; they are not the definition of \textsc{Praxist}.

\subsection{Task-Family Realizations}
\label{app:task-family-realizations}

\textsc{Praxist} uses a task-agnostic research cycle. A task-family realization supplies
the domain-specific information needed to instantiate that cycle: the allowed
task context, artifact form, evaluator semantics, role constraints, and
evidence conventions. This separation allows the same research-state objects
from Eq.~\ref{eq:state} to operate across different evaluator-grounded tasks.
Table~\ref{tab:task-family-realization} lists the components a realization
must supply.

\begin{table*}[t]
\centering
\small
\caption{Task-family realization components. MLE-style tasks instantiate these
components with competition context, submission artifacts, metric semantics,
and grader outcomes; other task families instantiate them with their own
artifact forms and evaluators.}
\label{tab:task-family-realization}
\setlength{\tabcolsep}{5.5pt}
\begin{tabular}{p{0.23\textwidth}p{0.70\textwidth}}
\toprule
Component & Method-level role \\
\midrule
Allowed task context & Defines the public information, constraints, resources,
and task description available to the run. \\
Artifact form & Defines the reproducible experimental unit $a$ that grounds
evaluator outcomes and later research claims. \\
Evaluator semantics & Defines how artifacts produce task-grounded outcomes
$e=\textsc{Eval}(a)$ and how validity, metric direction, failure states, or
structured reports are interpreted. \\
Role-local instructions & Specialize peers, PI roles, and Chair synthesis to the
task family while preserving the common \textsc{Praxist} cycle. \\
Evidence conventions & Normalize outcomes into findings $\phi$, frontier
status, agenda items, memory updates, and lineage records. \\
\bottomrule
\end{tabular}
\end{table*}

For example, an MLE-style realization provides a public competition
description, available data context, submission format, metric name and
direction, and evaluator semantics. A proof-search task could instead define a
proof artifact and checker outcome; a software-engineering task could define a
patch artifact and test report; a simulation task could define a simulation
bundle and score report. The common requirement is that evaluated artifacts
yield traceable evidence.

\subsection{Artifacts, Evaluation Records, and Findings}
\label{app:artifact-evaluation-finding-records}

The artifact layer provides provenance, while the finding layer provides
reusable interpretation. Each evaluated attempt is represented by an artifact
$a_i$, an evaluation record $e_i$, and one or more findings
$\Phi_i=\textsc{Interpret}(a_i,e_i,d_i)$. The artifact is the reproducible
experimental unit. The evaluation record attaches a task-grounded outcome to
that artifact. Findings compress the outcome into research claims with
maturity and inheritance recommendations. Table~\ref{tab:evidence-records}
defines these three records; Table~\ref{tab:evidence-stage-ladder} gives the
evidence-stage ladder behind maturity $m$, and
Tables~\ref{tab:finding-fields} and~\ref{tab:finding-actions} give the
finding fields and the inheritance actions.

\begin{table*}[t]
\centering
\small
\caption{Evidence records used by \textsc{Praxist}. These are conceptual records; task
families may instantiate them with different concrete fields.}
\label{tab:evidence-records}
\setlength{\tabcolsep}{5.5pt}
\begin{tabular}{p{0.20\textwidth}p{0.73\textwidth}}
\toprule
Record & Contents and role \\
\midrule
Artifact $a$ & Minimal reproducible file set or object required to reproduce,
inspect, or evaluate one attempt. In MLE-style tasks, this is typically a
submission-centered bundle containing the submission, code, and supporting
files needed for reproduction. \\
Evaluation record $e$ & Evaluator outcome attached to an artifact, including
validity state, score or report, metric direction, evidence stage, outcome
interpretation, parent provenance, and generation/peer provenance. \\
Finding $\phi$ & Artifact-grounded claim containing the attempted intervention,
observed outcome, supporting evidence, evidence maturity $m$, limitations,
and recommended inheritance action $\alpha$. \\
\bottomrule
\end{tabular}
\end{table*}

\paragraph{Artifact statuses.}
Artifacts are assigned a status before they are used as evidence:
\emph{committed} means the artifact is a complete canonical state that may
serve as a fact source; \emph{partial} means the attempt preserved useful
intermediate evidence but did not reach a full artifact state; \emph{failed}
means construction or evaluation failed in a way that may still yield
diagnostic evidence; and \emph{superseded} means a later artifact or frontier
entry replaces the artifact as the current inheritable state. Only committed
artifacts are treated as primary sources for positive inheritance; partial and
failed artifacts can still produce diagnostic or uncertain findings when the
failure mode is informative.

\paragraph{Evaluator interface.}
The external evaluator is task supplied and returns $e_i=\textsc{Eval}(a_i)$.
An evaluation record should preserve the artifact identifier, evaluator
version or protocol, metric name, score or structured report, metric direction,
validity state, evidence stage, generation and peer identifiers, parent lineage,
and any limitations needed to interpret the result. \textsc{Praxist} keeps validity
separate from score: an invalid high score is a diagnostic signal, not a
confirmed improvement.

\begin{table*}[t]
\centering
\small
\caption{Evidence-stage ladder. Maturity $m$ is inherited from the
artifact's evaluation stage. Preliminary stages can inform memory and
diagnostics but are not durable frontier evidence. Ranks 0--2 are the
generic ordering the core enforces for every task; the replication rungs
above them are task-owned labels, shown here with the names used by the
trading campaign, and other campaigns supply their own (the rocket
evaluator, for instance, reports \texttt{canary}, \texttt{development},
and \texttt{complete}). The core treats any task-declared replication
stage as mature evidence above \texttt{scored\_complete}.}
\label{tab:evidence-stage-ladder}
\begin{tabular}{p{0.18\textwidth}p{0.09\textwidth}p{0.45\textwidth}p{0.09\textwidth}}
\toprule
Stage & Rank & Meaning & Mature? \\
\midrule
\texttt{smoke} & 0 & Cheap sanity check, parser check, tiny run, or other
minimal validity probe. & No \\
\texttt{scout} & 1 & Partial probe or early score used to decide whether a
direction deserves more compute. & No \\
\texttt{scored\_complete} & 2 & Complete task-grounded score or report for the
artifact under the standard evaluator path. & Yes \\
\texttt{full\_T1} & 3 & Full Tier-1 evaluation with stronger completion or
reproducibility evidence than a single scored run. & Yes \\
\texttt{T2} & 4 & Replicated, ablated, or otherwise strengthened evidence
beyond Tier 1. & Yes \\
\texttt{T3} / \texttt{forced\_T3} & 5 & Highest-maturity evidence, including
replication or forced validation used for final claims. & Yes \\
\bottomrule
\end{tabular}
\end{table*}

Findings are typed by research function. Positive findings identify
improvements or reusable mechanisms. Negative findings weaken an assumption or
intervention. Diagnostic findings explain constraints, failure modes, invalidity conditions, or bottlenecks. Uncertain findings preserve evidence that may be useful after further checking. Procedural findings record constraints on how future experiments must run. Evidence maturity is distinct
from outcome quality: a high-scoring artifact may still need validation,
while a failed artifact can yield a mature diagnostic finding.

\begin{table*}[t]
\centering
\small
\caption{Finding tuple fields for
$\phi=(\text{intervention},\text{outcome},\text{evidence},\tau,m,\alpha)$.
$\tau$ and $\alpha$ name the method-level roles that inheritance is defined
over. The released implementation files a finding under a task-facing record
enum (\texttt{result}, \texttt{hypothesis}, \texttt{insight},
\texttt{challenge}, \texttt{error}) and carries the inheritance
recommendation in next-step-intent and parent-usage metadata rather than in a
single $\alpha$ column, so an implementer should read this table as the
semantics to preserve, not as the literal field contract of the
\texttt{share\_finding} tool.}
\label{tab:finding-fields}
\setlength{\tabcolsep}{5.5pt}
\begin{tabular}{p{0.20\textwidth}p{0.73\textwidth}}
\toprule
Field & Meaning \\
\midrule
Intervention & What was changed or tested, expressed in the vocabulary of
the design contract $d_i$. \\
Outcome & What happened under the evaluator or diagnostic protocol, including
raw task-grounded evidence when available. \\
Evidence & Artifact reference, evaluation record, limitations, validity state,
and links to parent evidence. \\
$\tau$ & Finding type: positive, negative, diagnostic, uncertain, or
procedural. \\
$m$ & Evidence maturity, inherited from the evidence-stage ladder in
Table~\ref{tab:evidence-stage-ladder}. \\
$\alpha$ & Recommended inheritance action: reuse, validate, avoid,
diagnose, preserve, or archive. \\
\bottomrule
\end{tabular}
\end{table*}

\begin{table*}[t]
\centering
\small
\caption{Inheritance actions used by finding extraction and agenda
formation. As with $\tau$, these are the method-level roles; the emitted
agenda expresses them through a finer next-step-intent vocabulary that
distinguishes, for example, repairing a failure mode from ablating or
falsifying a mechanism.}
\label{tab:finding-actions}
\setlength{\tabcolsep}{5.5pt}
\begin{tabular}{p{0.18\textwidth}p{0.75\textwidth}}
\toprule
Action & Agenda implication \\
\midrule
Reuse & Treat the evidence as a parent, baseline, mechanism, or constraint
when mature enough. \\
Validate & Schedule reproduction, ablation, or stronger evaluation before
confirming the claim. \\
Avoid & Prevent later peers from repeating an invalid assumption,
intervention, or procedure. \\
Diagnose & Allocate a focused diagnostic contract to explain a failure mode or
uncertainty. \\
Preserve & Keep the lesson as durable memory even if it is not a current
frontier parent. \\
Archive & Retain provenance while removing the item from active inheritance. \\
\bottomrule
\end{tabular}
\end{table*}

\subsection{Design Contracts and Diversity Allocation}
\label{app:design-contracts}

\textsc{Praxist} assigns a Deep Innovation Gate (DIG) design contract $d_i$
before peer $p_i$ constructs an artifact. The gate's generation scope is a task
setting: it defaults to the opening generation, where the fields below are
produced by the systematic gate itself, and a task may set it to run in every
generation. Where the default is used---as in the rocket campaign---later
generations receive the same contract fields from the Chair's agenda instead. DIG is a read-only innovation gate:
it makes the intended experiment explicit by specifying what mechanism is being
tested, where the intervention occurs, which prior evidence it extends, what evidence would support or weaken it, and which changes would invalidate
the test. The contract is used later to judge whether the artifact actually
tested the intended idea. Table~\ref{tab:design-contract-fields}
lists the contract fields, and Table~\ref{tab:qd-allocation-caps} the cohort-level allocation caps that keep the contracts diverse.

\begin{table*}[t]
\centering
\small
\caption{Deep Innovation Gate (DIG) design-contract fields for structured exploration.}
\label{tab:design-contract-fields}
\setlength{\tabcolsep}{5.5pt}
\begin{tabular}{p{0.25\textwidth}p{0.68\textwidth}}
\toprule
Field & Function \\
\midrule
Selected candidate and variant & Identifies the candidate design and names the
artifact variant that peer $p_i$ should construct. \\
Design cell $c_i$ & Assigns the contract to a cell
$c_i=(\text{mechanism family},\text{intervention surface},\text{intent})$ in
the behavior space $\mathcal{C}$. \\
Intent & Purpose of the attempt within its design cell: exploit, repair,
bridge, ablate, falsify, explore, diagnose, control, or audit---the same
vocabulary balanced by the intent cap in Table~\ref{tab:qd-allocation-caps}. \\
Mechanism hypothesis & States the causal or functional mechanism expected to
produce the outcome. \\
Why selected & Explains why the direction is worth one peer of the cohort
budget under the current frontier, agenda, and Gems. \\
Rejected alternatives & Records at least several nearby alternatives so the
lineage can distinguish intentional coverage from accidental omission. \\
Files or components to modify & Specifies the artifact surfaces that may be
changed. \\
Allowed changes & Defines what edits, parameters, data transformations, proof
steps, or experiment settings are in scope. \\
Forbidden changes & Defines changes that would invalidate the test, such as
altering the evaluator, changing a data split, or modifying the metric
calculation. \\
Implementation plan & Gives the peer a concrete construction path without
overriding evaluator semantics. \\
Expected evidence signature & States what outcome pattern would support,
weaken, or clarify the design intent. \\
Ablation and validation hooks & Specifies checks that make the result
interpretable, including minimal reproductions, controls, or comparisons. \\
Fail-fast checks & Gives cheap checks to run before expensive artifact
construction. \\
Semantic family and parent lineage & Records approach family and inherited
evidence, including frontier entries, Gems, or independent starting points. \\
Novelty axis & Identifies what dimension of the design is intended to be new
relative to the current cohort and frontier. \\
\bottomrule
\end{tabular}
\end{table*}

At the generation level, \textsc{Praxist} allocates contracts across diverse design
cells. $\textsc{Allocate}$ (\Secref{sec:method_alloc}) first scores candidate
contracts by quality, lane fit, novelty, risk, and diagnostic need, then greedily
selects a constrained set of $C$ contracts. This is Quantified Diversity (QD):
diversity is carried by explicit design-cell coordinates and controlled by
caps, so the cohort covers distinct mechanism families, intervention
surfaces, and intents in one generation. The caps bind differently across the
run. In the opening allocation the greedy selector enforces them as hard
constraints. In later generations the same coordinates and caps are given to
the Chair as soft allocation targets when it writes the peer contracts, and the
agenda validator treats missing or unrecognized planning dimensions as a
warning rather than a rejection, so coverage there is planned rather than
guaranteed.

\begin{table*}[t]
\centering
\small
\caption{Quantified Diversity (QD) cohort allocation constraints used to keep the $C$ peers
diverse. Fractions are applied to the cohort size and then interpreted as
integer peer caps.}
\label{tab:qd-allocation-caps}
\begin{tabular}{p{0.31\textwidth}p{0.15\textwidth}p{0.38\textwidth}}
\toprule
Constraint & Default cap & Purpose \\
\midrule
Same formal design cell & 1 peer & Prevent exact duplicate contracts unless a
reproduction is explicitly requested. \\
Same mechanism family & 0.34 of cohort & Avoid spending the generation on one
approach family. \\
Same intervention surface & 0.50 of cohort & Preserve coverage across where
artifacts are changed. \\
Same intent & 0.60 of cohort & Balance exploit, repair, bridge, ablate,
falsify, explore, diagnose, control, and audit intents. \\
Same semantic family & 0.34 of cohort & Reduce near-duplicate concepts even
when their formal cells differ. \\
Same parent lineage & 0.50 of cohort & Prevent one parent from monopolizing the
generation. \\
Diagnostic contracts & $\leq 0.20$ of cohort and at most 2 peers & Preserve
diagnostic work without allowing failure analysis to dominate the cohort. \\
\bottomrule
\end{tabular}
\end{table*}

\subsection{PI/Chair Synthesis}
\label{app:pi-chair-synthesis}

PI/Chair synthesis is the generation-level decision stage. Peers publish local
artifacts and findings. PI roles independently interpret the evidence from
complementary perspectives. The Chair merges these views into an agenda with
claim boundaries, validation targets, peer contracts, and archive decisions.
Table~\ref{tab:pi-chair-roles} defines the synthesis roles, and
Table~\ref{tab:pi-chair-rounds} the panel modes and synthesis rounds.

\begin{table*}[t]
\centering
\small
\caption{PI/Chair synthesis roles. Role names denote review functions; task
families specialize the domain vocabulary used by each role.}
\label{tab:pi-chair-roles}
\setlength{\tabcolsep}{5.5pt}
\begin{tabular}{p{0.23\textwidth}p{0.70\textwidth}}
\toprule
Role & Synthesis function \\
\midrule
Builder PI & Constructs the strongest evidence-backed mainline and proposes
concrete next-generation tests. \\
Skeptic PI & Audits unsupported claims, fragile interpretations, invalid
evidence, and risks that require checks. \\
Portfolio PI & Allocates effort across approach families, exploitation,
exploration, validation, and diagnosis. \\
External-validity PI & Checks evidence boundaries, reproducibility, task
constraints, and claims that may overfit to available feedback. \\
Chair & Merges PI memos into an agenda, resolves conflicts, narrows claims, and
assigns peer contracts within the generation budget. \\
\bottomrule
\end{tabular}
\end{table*}

\begin{table*}[t]
\centering
\small
\caption{Panel modes and synthesis rounds. The full panel is the default;
high-stakes mode adds the external-validity role and an explicit confidence
round.}
\label{tab:pi-chair-rounds}
\setlength{\tabcolsep}{5.5pt}
\begin{tabular}{p{0.21\textwidth}p{0.72\textwidth}}
\toprule
Item & Contract \\
\midrule
Mini panel & Builder PI and Skeptic PI. Used when the run needs lightweight
mainline construction plus fragility audit. \\
Full panel & Builder PI, Skeptic PI, and Portfolio PI. Used as the default
mode for balancing score, reliability, and coverage. \\
High-stakes panel & Full panel plus External-validity PI. Used when final
claims, evidence boundaries, or reproducibility risks require extra scrutiny. \\
Round 0: evidence freeze & Builds the shared evidence core and role-private
packs from $\Phi_g$, $\mathcal{F}_{g+1}$, recent agendas, Gems, and lineage. \\
Round 1: independent memos & Each PI writes an independent interpretation
before seeing other PI conclusions. \\
Round 2: anonymized cross-review & PI memos are cross-reviewed against fixed
questions about support, fragility, validation, missing controls, and portfolio
coverage. \\
Round 2.5: confidence boundary & High-stakes mode records confidence and claim
boundaries before Chair arbitration. \\
Chair synthesis & The Chair merges memos and reviews into
$\mathcal{A}_{g+1}$, validates agenda shape, and commits a single executable
generation policy. \\
\bottomrule
\end{tabular}
\end{table*}

The Chair's agenda contains generation metadata, at least several cross-peer
hypotheses with minimal tests and kill/promote conditions, exactly $C$ peer
contracts, panel summary, consensus actions, dissent-to-experiment conversions,
minority high-upside ideas when justified, claim-boundary updates, and
validation status. Peer-contract roles are execution roles rather than PI
review roles; typical roles include exploit, falsifier, bridge, anti-mainline,
and theorist.

\subsection{Frontier Status and Agenda Decisions}
\label{app:frontier-agenda-decisions}

The frontier assigns evidence an operational status for future generations. We
use four method-level statuses throughout the paper; a task-family realization
declares its own lane names and maps them to these inheritance roles, and a
task that declares none falls back to a single primary-metric frontier. The
trading campaign, for instance, declares \emph{confirmed alpha}, \emph{alpha
incubator}, \emph{benchmark floor}, and \emph{diagnostic control}, while the
SLAM and fusion campaigns declare no lanes and run on the fallback.
Table~\ref{tab:frontier-statuses} lists the four statuses and their agenda
effects, and Table~\ref{tab:agenda-schema} the agenda schema and
lane-to-disposition mapping they feed.

\begin{table*}[t]
\centering
\small
\caption{Frontier statuses and their agenda effects.}
\label{tab:frontier-statuses}
\setlength{\tabcolsep}{4.5pt}
\begin{tabular}{p{0.18\textwidth}p{0.38\textwidth}p{0.36\textwidth}}
\toprule
Status & Meaning & Typical agenda effect \\
\midrule
Confirmed ($\mathcal{F}^{\mathrm{cf}}$) & Reliable inheritable evidence,
mechanism, artifact, or constraint. & Continue: use as parent, baseline,
comparison point, or constraint. \\
Candidate ($\mathcal{F}^{\mathrm{cd}}$) & Promising evidence with incomplete
maturity or incomplete coverage. & Explore or validate: reproduce, ablate, or
combine with caution. \\
Diagnostic ($\mathcal{F}^{\mathrm{dg}}$) & Failure mode, invalidated
assumption, control, or process observation. & Stop, constrain, or diagnose:
prevent repeated errors and schedule focused tests. \\
Validation ($\mathcal{F}^{\mathrm{vl}}$) & Evidence requiring reproduction,
ablation, or checking before promotion. & Validate: assign validation peers or
define a minimal test. \\
\bottomrule
\end{tabular}
\end{table*}

Promotion is deliberately stricter than finding extraction. Finding extraction
preserves useful evidence; promotion decides which evidence is allowed to
shape future work. The promotion operator $\textsc{Promote}$ (\Secref{sec:method_frontier}) admits the
top $k$ findings by the primary metric, with $k=2$ by default, and may admit one
extra finding per task-defined anchor metric. Within each lane, candidates are
selected under lane-specific axes rather than only global score. Hard gates
reject findings that are marked non-promotable, lack required tier or maturity
metadata, or remain at preliminary evidence stages such as \texttt{smoke} or
\texttt{scout}. Preliminary evidence is preserved in research memory or
diagnostic records but is not allowed to displace mature frontier evidence.

\begin{table*}[t]
\centering
\small
\caption{Agenda schema and lane-to-disposition mapping.}
\label{tab:agenda-schema}
\setlength{\tabcolsep}{5.5pt}
\begin{tabular}{p{0.24\textwidth}p{0.69\textwidth}}
\toprule
Agenda field or disposition & Meaning \\
\midrule
Generation metadata & Identifies which generation the agenda controls and
which evidence freeze produced it. \\
Cross-peer hypotheses & Shared claims with minimal tests, kill conditions, and
promote conditions. \\
Peer contracts & Exactly $C$ executable contracts for the next generation,
each with role, parent or independent target, success signal, and forbidden
actions. \\
Panel summary and consensus actions & Chair-level synthesis of what the PI
panel believes should continue, stop, validate, diagnose, or explore. \\
Claim-boundary updates & Narrowing statements that prevent later peers from
overstating what the evidence supports. \\
Continue & Usually emitted for confirmed evidence that should be reused as a
parent, baseline, or constraint. \\
Stop & Usually emitted for negative or diagnostic evidence that invalidates an
assumption or weakens a direction. \\
Validate & Usually emitted for validation-lane evidence or high-upside
candidates with incomplete maturity. \\
Explore & Usually emitted for under-covered design cells, minority high-upside
ideas, or portfolio gaps. \\
\bottomrule
\end{tabular}
\end{table*}

\subsection{Gems and Selective Memory}
\label{app:gems-memory}

Gems are compact lessons retained across longer horizons. Compression is an
opt-in facility that is disabled unless a campaign turns it on; among the case
studies reported here only the trading campaign runs it, and its released state
records three resets with four active Gems. Gems are created at
reset boundaries or other compression points from frontier-balanced evidence,
recurring findings, validated mechanisms, rejected assumptions, diagnostic
patterns, and procedural constraints. Gems complement the frontier: frontier entries assign inheritance status to current evidence, while Gems preserve durable knowledge that stays visible after local details are archived. Table~\ref{tab:gem-categories} lists common Gem categories,
and Table~\ref{tab:memory-routing} the selective-memory routing that delivers it to each role.

\begin{table*}[t]
\centering
\small
\caption{Common Gem categories. The categories are descriptive roles rather
than a stored field: a Gem record carries its source finding, task lane,
metric, evidence stage, and selection provenance, and its category is read from
that content.}
\label{tab:gem-categories}
\setlength{\tabcolsep}{5.5pt}
\begin{tabular}{p{0.23\textwidth}p{0.70\textwidth}}
\toprule
Category & Role \\
\midrule
Validated mechanism & Preserves a reusable strategy, transformation, or design
pattern across generations. \\
Rejected assumption & Records an assumption weakened by evidence so that later
peers avoid relying on it. \\
Recurring failure mode & Summarizes a diagnostic pattern observed across
attempts or generations. \\
Procedural constraint & Preserves task-specific or task-family-specific
constraints needed for valid artifacts. \\
Evidence boundary & Records the scope within which a conclusion is supported. \\
\bottomrule
\end{tabular}
\end{table*}

Where enabled, compression is considered every $\rho=6$ generations by default,
with at most three reset events and a small active Gem set. The selection policy is
frontier-lane balanced: confirmed evidence can yield reusable mechanisms,
candidate or validation evidence can yield boundaries and follow-up lessons,
and diagnostic evidence can yield failure-mode or procedural Gems. Compression
archives ordinary findings after preserving the lessons needed for future
inheritance, so later peers do not need to carry the full transcript of earlier
generations.

\begin{table*}[t]
\centering
\small
\caption{Selective-memory routing. Each role receives the subset of state it
needs to act without treating every prior observation as equally inheritable.}
\label{tab:memory-routing}
\setlength{\tabcolsep}{5.5pt}
\begin{tabular}{p{0.21\textwidth}p{0.72\textwidth}}
\toprule
Recipient & Routed context \\
\midrule
Peer $p_i$ & Agenda-relevant frontier entries, relevant Gems, parent lineage,
the assigned design contract $d_i$, and task-family constraints. \\
PI role & Evidence pack for $\Phi_g$, current frontier, relevant Gems, recent
agenda history, and role-specific review questions. \\
Chair & PI memos, anonymized cross-review signals, agenda candidates,
frontier state, and validation or dissent items requiring arbitration. \\
Future run or human reader & Final artifact $a^\star$, confirmed frontier
evidence, Gems, and lineage $\mathcal{L}_G$ explaining the solution path. \\
\bottomrule
\end{tabular}
\end{table*}

\subsection{Lineage Trace}
\label{app:lineage-trace}

\textsc{Praxist} records lineage as a typed trace over artifacts, evaluation records,
findings, design contracts, PI/Chair decisions, frontier updates, agendas, and
Gems. The trace is accumulated during the research cycle rather than
reconstructed only after final selection. Each artifact can inherit from parent
artifacts, findings, frontier entries, agenda assignments, design contracts, or
Gems. Each finding points back to its supporting artifact and evaluation
record. Each frontier update records how evidence became inheritable. Each
agenda records how synthesis directed later work.
Table~\ref{tab:lineage-components} lists the recorded components, and
Table~\ref{tab:lineage-edges} the typed edge vocabulary connecting them.
Physically, the trace is written as several typed ledgers that share
identifiers rather than as one serialized graph: the explicit typed edge list
covers finding-to-finding relations, while artifacts are indexed in an artifact
ledger, synthesis and promotion events in the run's event trajectory, and
agendas, frontier state, and Gems in per-generation state files. Reconstructing
a full cross-object graph therefore means joining those ledgers on their shared
identifiers.

\begin{table*}[t]
\centering
\small
\caption{Lineage trace components.}
\label{tab:lineage-components}
\setlength{\tabcolsep}{5.5pt}
\begin{tabular}{p{0.23\textwidth}p{0.70\textwidth}}
\toprule
Component & Recorded relation \\
\midrule
Artifact & Reproducible experimental unit produced or modified by a peer. \\
Evaluation record & Task-grounded outcome attached to an artifact. \\
Finding & Research claim supported by an artifact and evaluation record. \\
Design contract & Pre-artifact intent that motivated a peer's attempt. \\
PI/Chair decision & Synthesis decision derived from generation-level evidence. \\
Frontier update & Inheritance-status decision for a finding, artifact, or
evidence item. \\
Agenda & Generation-level policy scheduling future directions, validation, or
constraints. \\
Gem & Durable compressed lesson retained across generations. \\
\bottomrule
\end{tabular}
\end{table*}

\begin{table*}[t]
\centering
\small
\caption{Lineage edge vocabulary. Edges are typed so that later peers and
readers can distinguish inheritance, support, conflict, and update relations.}
\label{tab:lineage-edges}
\setlength{\tabcolsep}{5.5pt}
\begin{tabular}{p{0.20\textwidth}p{0.73\textwidth}}
\toprule
Edge type & Meaning \\
\midrule
\texttt{derived\_from} & The child artifact, finding, agenda, or Gem was
constructed from the parent evidence. \\
\texttt{supports} & The source provides evidence in favor of the target claim,
promotion, agenda item, or Gem. \\
\texttt{challenges} & The source weakens, falsifies, or limits the target
claim or direction. \\
\texttt{updates} & The source revises the target, such as a later frontier
entry superseding an earlier candidate. \\
\texttt{related\_to} & The source is relevant to the target but does not imply
support, challenge, derivation, or update. \\
\bottomrule
\end{tabular}
\end{table*}

The final output is a reproducible artifact together with the solution lineage
that produced it. The run itself terminates by writing out the frontier, the
lineage ledgers, and a run summary; Eq.~\ref{eq:output} is the selection rule
applied over those outputs at reporting time rather than a call the loop makes
on its own, so a campaign stopped early still leaves a complete frontier and
lineage from which the artifact is chosen. The lineage exposes which evidence
supported the selected artifact, which failures constrained it, which
candidates required validation, which Gems shaped later work, and which agendas
directed its construction, so a reader can audit the reported result end to
end.

\subsection{Resource Scheduling and Mature-Evidence Debt}
\label{app:resource-scheduling}

\textsc{Praxist} executes peer experiments through a thin resource-scheduling layer that
operates within the generation's wall-clock envelope
(\Secref{sec:method_artifact}). The layer modeled here performs three
functions: admission of experiments onto shared GPUs, lease accounting for
acquired devices, and backfill of idle capacity. The deployed scheduler adds a
coarse host-pressure guard on top of these, described under \emph{Boundary and
caveats} below. Launches are paced by a single scalar
demand signal derived from evidence maturity, while all scientific
interpretation stays outside the layer: maturity criteria, Pareto value,
lineage, and plan selection belong to the research state of
\Secref{sec:method}. This subsection specifies the resource model, the
maturity signal, and the launch controller, then reports the paired event
simulation used to validate the design. All quantitative claims below are
simulation results under synthetic workloads, not measurements from the live
runs of \Secref{sec:experiments}; symbols introduced here are local to this
subsection.

\paragraph{GPU admission.}
Each experiment $j$ declares, for each of the $g_j$ GPUs it requires, an
average utilization estimate $u_j \in [0,1]$ (an
\texttt{nvidia-smi}-style busy fraction, treated as a linear compute capacity)
and a peak-memory estimate $v_j$ in GB. For every physical device $d$ with
memory capacity $V_d$, admission of the set $\mathcal{D}_j$ of experiments
co-located on $d$ requires
\begin{equation}
\label{eq:gpu-admission}
\sum_{j : d \in \mathcal{D}_j} u_j \;\le\; 1,
\qquad
\sum_{j : d \in \mathcal{D}_j} v_j \;\le\; 0.95\, V_d .
\end{equation}
Multi-GPU experiments acquire their $g_j$ devices as a gang, and placement
favors devices that minimize the larger of the compute load and the
fractional memory commitment, limiting both compute-load and memory
fragmentation. Two low-utilization, low-memory experiments may share a GPU,
while either dimension can independently block admission. This
utilization-and-memory view follows goodput-oriented deep-learning cluster
schedulers \cite{gu2019tiresias,narayanan2020gavel,qiao2021pollux}, reduced
to the two quantities an experiment harness can cheaply declare.

\paragraph{CPU contention without core allocation.}
CPU allocation is delegated entirely to the operating system: the scheduler
performs no per-experiment core packing and never rejects a GPU experiment
because aggregate CPU demand exceeds the core count. Each experiment declares
a runnable CPU demand $w_j$ and a CPU sensitivity $a_j \in [0,1]$. With host
core count $W$, the host pressure at time $t$ is
$h(t) = \sum_{j\,\mathrm{running}} w_j / W$, and each running experiment's
remaining work is served at the processor-sharing slowdown
\begin{equation}
\label{eq:cpu-slowdown}
S_j(t) \;=\; 1 + a_j \max\bigl(0,\, h(t) - 1\bigr),
\end{equation}
following the classical time-shared-systems abstraction
\cite{kleinrock1967timeshared}: at $h = 2$ a fully CPU-sensitive experiment
($a_j = 1$) takes roughly twice its solo wall time, while a GPU-dominated one
($a_j = 0$) is unaffected. In this model CPU load is therefore a dynamic factor
inflating critical time rather than an admission dimension; the deployed
scheduler does consult observed host load, as noted below.

\paragraph{Maturity interface.}
The scheduler-facing notion of a \emph{mature} result is a thin summary of
the evidence-stage ladder of
Appendix~\ref{app:artifact-evaluation-finding-records}: an experiment counts
as mature once it has completed full-stage execution with sufficient
independent evaluation coverage (in the simulator, a completed-work ratio of
at least $0.75$ and an evaluation-coverage ratio of at least $0.80$), so that
a nearly complete experiment is not discarded solely because of a categorical
label. Scout-stage runs are preliminary probes that can be \emph{continued}
into mature evidence rather than restarted, in the spirit of multi-fidelity
search \cite{li2018hyperband,li2020asha}, and a scientific negative result at
full stage counts as mature evidence. The thresholds are simulation defaults,
not domain-independent constants.

\paragraph{Mature-evidence debt controller.}
With cohort size $C$, the controller tracks a mature quota
$Q = \max(1, \lceil C/4 \rceil)$, the number $M_t$ of mature results
completed by time $t$, and the outstanding debt
\begin{equation}
\label{eq:evidence-debt}
D_t \;=\; \max(0,\; Q - M_t),
\end{equation}
(a task may raise $Q$ above $\lceil C/4 \rceil$ through its own mature-evidence
quorum), and targets $K_t = \min(C,\, 3 D_t)$ mature-directed experiments in flight;
actual launches still require GPU admission under
Eq.~\ref{eq:gpu-admission}. The factor of three gives bounded
redundancy against failures and heavy-tailed durations: if one attempt
succeeds before the deadline with probability $p$, then $n$ independent
attempts succeed with probability $1 - (1-p)^n$ (97.3\% for $p = 0.7$,
$n = 3$), consistent with tail-tolerant execution \cite{dean2013tail}. The
controller recomputes $D_t$ at every completion, failure, and
resource-release event, making it an event-driven receding-horizon feedback
rule \cite{mayne2000mpc} rather than an online probabilistic program. The
launch priority stack below is the simulated policy; the deployed scheduler
implements its lease, first-wave, retry, deadline, and exploration-reserve
components, while the ranking in step~(ii) is performed by the research layer
when it chooses which peer-owned plan to fill a lease with, not by a scheduler
scoring function. Launch priorities are: (i) while mature-directed work in flight is below $3 D_t$,
continue an existing scout or launch a direct mature experiment; (ii) once
the guarantee is supplied, rank follow-ups by expected marginal Pareto gain per unit of predicted time; (iii) launch a new scout when no
follow-up is available; (iv) retry an infrastructure failure once; and (v)
after the quota is met, admit only emergency mature work passing a 25\% completion guard. The first wave mixes direct mature experiments
and scouts, preserving an explorer whenever capacity permits, so exploration competes with time and hardware budgets without being eliminated
\cite{badanidiyuru2013bandits}.

\paragraph{Reduced designs.}
To identify which parts of the controller are load-bearing, the simulation
study compares it against a \emph{minimal idle backfill} design: a work-conserving
mechanism \cite{mualem2001backfilling} that, whenever a resource goes idle,
issues a single-experiment lease from the fixed inventory of plans already
owned by peers, with asynchronous refill similar in structure to massively
parallel successive-halving systems \cite{li2020asha}. Backfill variants
differ only in maturity feedback: \emph{no feedback} (an idle lease selects a
mature follow-up with fixed probability 25\%, 50\%, or 75\%, for sensitivity
analysis), a \emph{Boolean maturity signal} (while $M_t < Q$, every idle
waiter prioritizes a mature plan, so concurrent waiters can create implicit,
accidental redundancy), and a \emph{nonredundant thin token} (maintain at
most $Q - M_t$ mature experiments in flight; quantitative but with no tail
redundancy).

\paragraph{Simulation protocol.}
The comparison covers 512 randomly drawn scenarios spanning 2--16 peers,
1--16 GPUs with 16--80\,GB memory, 4--192 host cores, log-normal heavy-tailed
experiment durations (median minutes to three days), cold-start runtime bias,
transient and persistent failures, varying scout cost and scout/true-quality
correlation, and generation horizons from 3 hours to 4 days. Each of 8
policies is run 1{,}000 times per scenario with paired seeds, totaling
4{,}096{,}000 policy runs; 348 of the 512 scenarios satisfy a physical
feasibility criterion (the 90th percentile of quota-completion time under a
maximally aggressive full-only reference policy fits the horizon), and
results are reported over these. Policy deltas use 5{,}000 scenario-level
bootstrap samples. This study is a closed synthetic experiment: the scenario
generator, its seed ledger, and the per-run outputs behind
Tables~\ref{tab:scheduler-policies} and~\ref{tab:scheduler-pressure} are not
part of the released evidence package, so the two tables should be read as a
design study of the admission and backfill layer under the stated model rather
than as an independently replayable benchmark.

\begin{table*}[t]
\centering
\small
\caption{Simulated scheduling policies over 348 physically feasible
scenarios. Quota success is the probability that a generation meets its
mature-evidence quota $Q$ before the horizon; P10 is the 10th percentile of
per-scenario success.}
\label{tab:scheduler-policies}
\resizebox{0.98\textwidth}{!}{%
\begin{tabular}{lccccc}
\toprule
Policy & Quota success (\%) & P10 (\%) & Mature mean & GPU util.\ (\%) & CPU util.\ (\%) \\
\midrule
Passive peers (no backfill) & 27.03 & 0.0 & 1.72 & 10.00 & 11.71 \\
Idle backfill, 25\% mature follow-up & 94.37 & 79.2 & 16.99 & 23.62 & 24.30 \\
Idle backfill, 50\% mature follow-up & 96.54 & 90.0 & 17.53 & 23.55 & 24.26 \\
Idle backfill, 75\% mature follow-up & 97.64 & 94.5 & 17.93 & 23.50 & 24.24 \\
Boolean maturity signal & 98.65 & 97.6 & 17.78 & 23.55 & 24.30 \\
Nonredundant thin token & 98.36 & 96.6 & 17.67 & 23.55 & 24.29 \\
Mature-evidence debt controller & \textbf{99.85} & \textbf{99.8} & \textbf{18.42} & 23.56 & 24.37 \\
\midrule
Full-only (feasibility reference) & 99.53 & 98.6 & 18.20 & 26.49 & 27.26 \\
\bottomrule
\end{tabular}}
\end{table*}

\begin{table*}[t]
\centering
\small
\caption{Simulated quota success under operational resource pressure. Strata
are defined by the observed state of the full-only reference within the same
scenario.}
\label{tab:scheduler-pressure}
\resizebox{0.98\textwidth}{!}{%
\begin{tabular}{lcccc}
\toprule
Condition & Scenarios & Boolean (\%) & Thin token (\%) & Debt controller (\%) \\
\midrule
Low CPU pressure ($h < 0.35$) & 232 & 99.21 & 99.10 & \textbf{99.94} \\
High CPU pressure ($h \ge 1.20$) & 52 & 95.74 & 94.45 & \textbf{99.49} \\
CPU saturated over half the generation & 59 & 96.20 & 95.05 & \textbf{99.55} \\
Observed GPU utilization $\ge 50\%$ & 64 & 96.87 & 96.42 & \textbf{99.76} \\
Observed GPU-memory utilization $\ge 50\%$ & 67 & 95.33 & 94.63 & \textbf{99.76} \\
\bottomrule
\end{tabular}}
\end{table*}

\paragraph{Results.}
Table~\ref{tab:scheduler-policies} summarizes the comparison. Idle backfill
alone raises mean GPU utilization from 10.00\% to roughly 23.6\% and CPU
utilization from 11.71\% to roughly 24.3\%, directly fixing sustained
hardware idleness; yet without maturity feedback, quota success still varies
from 94.37\% to 97.64\% as the mature-follow-up probability changes. All
feedback policies and the debt controller occupy statistically
indistinguishable hardware (the Boolean policy differs from the debt
controller by $-0.012$ GPU-utilization points, with a bootstrap interval
crossing zero), but their evidence outcomes differ: the Boolean signal is
$-1.20$ quota-success points below the debt controller (95\% CI
$[-1.82, -0.69]$) and the thin token is $-1.49$ points below (95\% CI
$[-2.17, -0.92]$). Equal utilization therefore does not imply equal evidence
supply: backfill and maturity feedback close two separate loops, one
preventing hardware idleness and one controlling whether high-grade results
arrive before the deadline. Table~\ref{tab:scheduler-pressure} stratifies by
operational pressure: the gap between the reduced designs and the debt
controller grows from under one point at low CPU pressure to 3.75 points
under high CPU pressure and 4.43 points under high GPU-memory pressure,
where an all-scout first wave is more costly and the lack of tail redundancy
in the thin token is exposed. An earlier version of the simulation that
modeled each GPU as an exclusive integer slot understated these gaps (the
Boolean gap was 0.22 points rather than 1.20), indicating that co-location,
memory fragmentation, and CPU contention materially affect critical time.
The debt controller's own result was nearly unchanged across the two
resource models.

\paragraph{Boundary and caveats.}
The production boundary of the layer follows the simulated design with one
documented departure. As in the simulation, the harness supplies GPU
count, average utilization, and peak-memory estimates (replaceable by runtime
observations); GPU admission is decided by
Eq.~\ref{eq:gpu-admission}; no per-experiment CPU cores are allocated or
packed, and no experiment is rejected merely because declared CPU demand
exceeds the core count. The departure is that the deployed allocator also
watches observed host pressure: it reduces its concurrency ceiling above
roughly 92\% CPU or 95\% memory utilization and reports no supply headroom
under sustained CPU, memory, or I/O pressure, which withholds new launches
until the pressure clears. Observed host load is therefore a coarse gate on
launch timing in production, even though it is not a declarative per-experiment
admission dimension. The evidence layer publishes
mature-evidence demand via Eq.~\ref{eq:evidence-debt} while the research
layer selects which peer-owned plan fills an idle lease; exploration is
preserved when capacity permits; and Pareto value, lineage, and maturity
criteria stay outside the scheduler. The study's limits are inherited by
these conclusions: GPU utilization is treated as a linear capacity (no SM,
Tensor-Core, memory-bandwidth, or interconnect modeling); CPU demand and
sensitivity follow broad synthetic distributions under a processor-sharing
approximation; the fixed plan inventory does not model how the quality
distribution of newly generated directions evolves; and success rates are
conditioned on physically feasible scenarios rather than unconditional.

\subsection{Configuration Defaults and MLE-bench Budgets}
\label{app:config}

Exact values are task configurable. The defaults in
Table~\ref{tab:config-defaults} are included to make the method concrete and
to disambiguate the symbols used in the main text. They are starting points, not tuned settings: a campaign adjusts cohort size,
generation count, and compression period to the task's evaluation cost, and the
frontier and Gem caps to the evidence each generation yields.

\begin{table*}[t]
\centering
\small
\caption{Representative default configuration values.}
\label{tab:config-defaults}
\begin{tabular}{p{0.27\textwidth}p{0.15\textwidth}p{0.42\textwidth}}
\toprule
Quantity & Default & Role \\
\midrule
Generations $G$ & 8 & Maximum number of generation boundaries in a run. \\
Cohort size $C$ & 5 & Number of parallel peers per generation. \\
Promotions per generation & 2 plus anchor extras & Baseline count of primary
metric promotions before task-specific anchor additions. \\
Per-peer safety cap & 5 hours & Hard upper bound for any single peer;
generation closing is event driven. \\
Gem compression & Disabled & When enabled, resets every 6 generations, with at
most 3 resets and 4 active Gems. \\
DIG candidate count & 8 & Number of candidate designs considered before
cohort allocation. \\
DIG coverage targets & 4 mechanism families, 3 intervention surfaces & Minimum
diversity pressure for candidate generation. \\
Multi-PI panel & Disabled & Full mode is the default when enabled, using
Builder, Skeptic, Portfolio, and Chair roles; high-stakes mode adds the
External-validity role. \\
PI rounds & 2 when enabled & Evidence freeze, independent memos, cross-review,
and Chair synthesis. \\
Synthesis trigger & 30 findings, 120 minutes, 3 peers; 240-minute cap & Typical
information-density trigger with a maximum-time safety cap. \\
Evaluator seeds & 42--46 & Representative deterministic seeds for task
evaluations that expose seeded protocols. \\
\bottomrule
\end{tabular}
\end{table*}

The finalized MLE-bench ledger draws all 75 retained task results from a single
campaign, whose settings are listed in Table~\ref{tab:mle-campaign-config}.
Every task in that campaign runs \texttt{deepseek-v4-pro[1m]} as its
research-agent model, with a cohort of 12 peers, promotes the top 4 peers at each generation boundary, and admits at
most 20 generations. Experiments are scheduled one job per GPU, so each
experiment occupies a single accelerator, and a task keeps up to eight
experiments in flight at any moment. The generation ceiling, the per-experiment
GPU budget, and the task wall cap are configured ceilings rather than
measurements of realized compute.

\begin{table}[htbp]
\centering
\small
\caption{\textsc{Praxist} campaign configuration behind the finalized
MLE-bench ledger.}
\label{tab:mle-campaign-config}
\setlength{\tabcolsep}{6pt}
\begin{tabular}{ll}
\toprule
Setting & Value \\
\midrule
Research-agent model & \texttt{deepseek-v4-pro[1m]} \\
Cohort size & 12 peers \\
Promotions per generation & Top 4 \\
Maximum generations & 20 \\
GPU budget per experiment & 6 GPU-h \\
Experiment scheduling & Single-GPU job, one job per GPU \\
Experiments in flight per task & Up to 8 \\
Task wall cap & 24 h (70 tasks); 36 h (5 tasks) \\
\bottomrule
\end{tabular}
\end{table}

\subsection{Representative Prompt Contracts}
\label{app:prompt-contracts}

\textsc{Praxist} uses role-specific prompt contracts to maintain the research-state
structure described above. The templates below are representative method-level
contracts, summarized in Table~\ref{tab:prompt-families}. They use
placeholders for task-family content and omit task-specific commands or
storage details. Their expected outputs are stated in the method vocabulary of
this appendix; a deployment renders them against the task-facing field names
noted in Tables~\ref{tab:finding-fields} and~\ref{tab:frontier-statuses}, so
these templates should be read as the contract's semantics rather than as a
directly executable field specification.

\begin{table*}[t]
\centering
\small
\caption{Prompt families used by \textsc{Praxist}.}
\label{tab:prompt-families}
\setlength{\tabcolsep}{5.5pt}
\begin{tabular}{p{0.21\textwidth}p{0.30\textwidth}p{0.40\textwidth}}
\toprule
Prompt family & Primary input & Expected output \\
\midrule
Peer research & Task context, inherited state, agenda, design contract,
artifact constraints. & Reproducible artifact and artifact-grounded evidence. \\
Design allocation & Frontier, agenda, Gems, coverage state, prior design
cells. & Peer design contracts covering diverse research directions. \\
Finding extraction & Artifact, evaluation record, peer notes, diagnostics. &
Positive, negative, diagnostic, uncertain, or procedural findings. \\
Frontier update & Findings, evaluation records, prior frontier, validation
needs. & Inheritance status and promotion or validation decisions. \\
PI memo & Evidence pack, frontier state, Gems, task constraints. &
Role-specific interpretation, objections, proposed experiments, and peer
contracts. \\
Chair agenda & PI memos, cross-reviews, evidence pack, frontier state. &
Next-generation agenda with peer contracts and claim boundaries. \\
Gem compression & Durable findings, frontier entries, recurring patterns,
archived evidence. & Compact Gems with scope, evidence, and future-use
guidance. \\
\bottomrule
\end{tabular}
\end{table*}

\paragraph{Example task prompt template.}
The following prompt sketch is adapted from the task-family templates used in
our runs. It is an abridged method-level example: concrete commands, workspace
paths, helper names, and storage conventions are replaced by placeholders. The
structure is the important part: a task-specific source-of-truth block,
inherited research state, a generation agenda, a peer-specific contract, and an
evidence-reporting discipline.

\begin{quote}
\small
\textbf{Autonomous research peer.}
You are Peer \texttt{<peer\_id>} in Generation \texttt{<g>} of a \textsc{Praxist} run for
one evaluator-grounded research task. Coordinate with sibling peers through
shared findings and build on previous generations through the current frontier,
agenda, Gems, and lineage state. The task-specific block is the source of truth
for the artifact definition, evaluator semantics, allowed evidence, and
reporting requirements.

\textbf{Task-specific source-of-truth block.}
\begin{itemize}
    \item \textbf{Task:} \emph{task name}.
    \item \textbf{Objective:} \emph{task objective and allowed scope}.
    \item \textbf{Artifact:} \emph{minimal reproducible file set for one attempt}.
    \item \textbf{External evaluator:} \emph{metric, success criterion, validity states, and score direction}.
    \item \textbf{Allowed context:} \emph{public task context, allowed data, and allowed prior knowledge}.
    \item \textbf{Constraints:} \emph{validity, resource, safety, or benchmark restrictions}.
\end{itemize}

\textbf{Generation context.}
\textsc{Praxist} may provide selected frontier entries, Gems, peer-local memory, and a
design contract. Use these channels to preserve diversity, reuse high-value
evidence, and avoid repeated failures. Treat confirmed, candidate, diagnostic,
and validation evidence according to its stated inheritance role. Do not edit
frontier, Gem, agenda, or lineage state directly; publish evidence for \textsc{Praxist}
to synthesize.

\textbf{Panel agenda.}
If a PI/Chair agenda is available, use it to identify the current panel read:
visible approach families, main risks, key tradeoffs, cross-peer hypotheses,
validation targets, and directions to continue, stop, diagnose, or explore.

\textbf{Your peer contract.}
\begin{itemize}
    \item \textbf{Role:} \emph{explore, exploit, validate, or diagnose}.
    \item \textbf{Target hypothesis or objective:} \emph{claim, candidate, or gap to test}.
    \item \textbf{Success signal:} \emph{evaluator outcome, validation result, or diagnostic evidence}.
    \item \textbf{Forbidden assumptions:} \emph{assumptions or shortcuts to avoid}.
    \item \textbf{Evidence stage:} \emph{smoke, scout, full evaluation, or diagnostic}.
    \item \textbf{Next-step intent:} \emph{reuse, validate, avoid, diagnose, or preserve}.
\end{itemize}

\textbf{Core workflow.}
\begin{enumerate}
    \item Inspect the task block, inherited research state, and peer contract.
    \item Choose a concrete work item aligned with the agenda and design intent.
    \item Construct or revise a reproducible artifact.
    \item Evaluate the artifact when it reaches the required evidence stage.
    \item Convert the outcome into evidence with limitations and provenance.
    \item Publish positive, negative, diagnostic, or uncertain findings.
    \item Preserve artifact references so the lineage trace can connect this attempt to later frontier and agenda decisions.
\end{enumerate}

\textbf{Evidence reporting reminder.}
Report evaluator outcomes in their original task semantics. Mark fragile or
partial evidence as candidate or validation-needed. Preserve useful failures as
diagnostic evidence. Include the artifact reference, evaluator outcome,
evidence stage, limitations, and recommended inheritance action when publishing
findings.
\end{quote}

For an MLE-style task, the placeholders are instantiated with the public
competition description, sample submission structure, metric name and direction,
submission-centered artifact definition, and official grader semantics. Other
task families replace these fields with their own artifact and evaluator
semantics while preserving the same prompt-level research contract.

\paragraph{Peer research contract.}
\begin{quote}
\small
\textbf{Role.} Construct a reproducible artifact for one evaluator-grounded
task and publish evidence useful to later peers, PI/Chair synthesis, and the
frontier.

\textbf{Inputs.} Public task context; artifact definition; evaluator semantics;
current agenda; selected frontier entries; relevant Gems; parent lineage;
optional design contract.

\textbf{Required outputs.} Artifact or artifact plan according to the task
protocol; raw evaluator outcome when available; notes on attempted
intervention, observed evidence, limitations, and useful failures.

\textbf{Discipline.} Use allowed task context and task-grounded outcomes.
Report unsupported measurements as unavailable. Preserve failures when they
clarify constraints.
\end{quote}

\paragraph{Design allocation contract.}
\begin{quote}
\small
\textbf{Objective.} Allocate \texttt{<k>} peers for generation \texttt{<g>}
using the current frontier, agenda, Gems, recent findings, and coverage state.

\textbf{For each peer, specify.} Mechanism family; intervention surface; intent;
parent lineage or independent starting point; expected evidence signature;
validation hook or fail-fast check; forbidden assumptions; success signal;
recommended finding type.

\textbf{Allocation rule.} Cover multiple design cells, preserve validation and
diagnostic needs, and use near-duplicate contracts only for explicit
reproduction.
\end{quote}

\paragraph{Finding extraction contract.}
\begin{quote}
\small
\textbf{Objective.} Convert an evaluated artifact into one or more
artifact-grounded findings.

\textbf{For each finding, report.} Grounding artifact; attempted intervention;
observed outcome; supporting evidence; finding type; maturity; limitations;
inheritance action: reuse, validate, avoid, diagnose, preserve, or archive.

\textbf{Discipline.} Keep evaluator outcomes raw and task-grounded. Mark
fragile claims as uncertain or validation-needed. Convert informative failures
into diagnostic findings.
\end{quote}

\paragraph{Frontier update contract.}
\begin{quote}
\small
\textbf{Objective.} Assign inheritance status to evidence using findings,
evaluation records, prior frontier state, validation needs, and agenda
priorities.

\textbf{Decision fields.} Frontier inclusion; status: confirmed, candidate,
diagnostic, or validation; supporting evidence; promotion or validation
condition; expected agenda effect.

\textbf{Discipline.} Use reliability and research utility to assign status.
Preserve diagnostic evidence when it constrains future work. Schedule validation
for promising but incomplete evidence.
\end{quote}

\paragraph{PI memo contract.}
\begin{quote}
\small
\textbf{Role.} Write an independent memo from the assigned PI perspective using
the shared evidence core, frontier state, recent findings, Gems, evaluator
outcomes, and agenda history.

\textbf{Report.} Evidence-grounded claims; objections and claim boundaries;
proposed experiments or validation checks; proposed peer contracts; confidence
and highest-risk claims; evidence to promote, narrow, validate, or archive.

\textbf{Perspectives.} Builder emphasizes the strongest mainline; Skeptic
audits unsupported or fragile claims; Portfolio balances effort across
approaches; External-validity checks evidence boundaries, reproducibility, and
task constraints.
\end{quote}

\paragraph{Chair agenda contract.}
\begin{quote}
\small
\textbf{Role.} Merge PI memos and cross-review evidence into the
next-generation agenda.

\textbf{Produce.} Panel summary; dominant mechanisms and main risks; cross-peer
hypotheses with minimal tests; validation or falsification contracts; directions
to continue, stop, validate, diagnose, or explore; peer contracts;
claim-boundary updates and archive decisions.

\textbf{Discipline.} Use supplied evidence. Convert blocking objections into
peer contracts, validation targets, or narrowed claim boundaries. Preserve
auditable high-upside minority ideas when compatible with task constraints.
\end{quote}

\paragraph{Gem compression contract.}
\begin{quote}
\small
\textbf{Objective.} Compress durable cross-generation knowledge from
frontier-balanced evidence, recurring findings, validated mechanisms, rejected
assumptions, diagnostic patterns, and procedural constraints.

\textbf{For each Gem, report.} Concise lesson; supporting evidence; scope and
limitations; category; guidance for future peers; revisit or retirement
condition.

\textbf{Discipline.} Prefer durable lessons over local details. Keep the active
Gem set compact. Preserve provenance to artifacts, findings, and frontier
decisions.
\end{quote}

\subsection{MLE-Style Task Example}
\label{app:mle-style-example}

MLE-style tasks provide one concrete instantiation of the method. The task
context contains a public competition description, available data, sample
submission format, metric name, and metric direction. A peer constructs an
artifact centered on a submission file and the supporting files required to
reproduce it. The external evaluator returns a task-grounded outcome. Findings
preserve raw evaluator evidence, limitations, and inheritance recommendations.
The frontier assigns evidence an inheritance status, PI/Chair synthesis emits
the next agenda, Gems preserve durable lessons, and lineage records the path
from earlier evidence to the final artifact.

This example illustrates the task-family realization boundary. Other task
families can define different artifact forms and evaluator semantics while
preserving the same \textsc{Praxist} cycle. The method requires traceable
evaluated artifacts, artifact-grounded findings, explicit inheritance status,
agenda-controlled future work, selective memory, and lineage accumulation.

\subsection{Worked MLE-bench Research Trajectory}
\label{app:mle-trajectory}

The Jigsaw Toxic Comment Classification task provides a compact trace from
evaluated parents to the artifact that generation~1 selected; its primary
metric is column-wise ROC AUC (higher is better).
Table~\ref{tab:jigsaw-trajectory} reports that progression across generations
0 and 1.

\begin{table}[H]
\centering
\small
\caption{Research trajectory for the Jigsaw Toxic Comment Classification
submission selected in generation~1. Scores are recorded column-wise ROC AUC
values.}
\label{tab:jigsaw-trajectory}
\begin{tabularx}{0.98\textwidth}{@{}p{0.13\textwidth}X>{\centering\arraybackslash}p{0.10\textwidth}p{0.19\textwidth}@{}}
\toprule
Stage & Evaluated artifact or intervention & Score & Role in the trajectory \\
\midrule
Generation 0 & Class-balanced focal-loss BERT & 0.98620 & Initial hypothesis \\
Generation 0 & Matched BCE ablation, BERT seed 42 & 0.98657 & Confirmed parent \\
Generation 1, C07 & Metadata-only null model & 0.76106 & Diagnostic finding \\
Generation 1 & Equal average of the two BERT seeds & 0.98679 & Two-model finding \\
Generation 1 & Add DistilRoBERTa to the two-seed ensemble & 0.98723 &
Three-model parent finding \\
Generation 1 & Add BiLSTM to the three-model ensemble & 0.98711 & Negative finding \\
Generation 1 & Add RoBERTa-CLS to the three-model ensemble & \textbf{0.98749} &
Confirmed final selection \\
\bottomrule
\end{tabularx}
\end{table}

Generation 0 tested class-balanced focal loss, then retained the matched BCE
ablation after its 0.00037 improvement. In generation 1, pod 6 completed the
C07 null-model diagnostic, then reused preserved prediction artifacts to build
the ensemble branch. Averaging the BCE parent with a second BERT
seed reached 0.98679, and adding DistilRoBERTa raised the score to 0.98723.
Adding the weaker BiLSTM component reduced the score by 0.00012; adding the
independent RoBERTa-CLS parent instead raised it by 0.00026 to 0.98749. The
final finding records an \texttt{updates} link to the three-model result and a
\texttt{derived\_from} link to the two-seed BERT ensemble, closing the lineage
from a matched ablation through a diagnostic and a negative branch. The trace
shows structure, not the task's finalized entry: the release ledger reports the
campaign's highest-scoring integrity-clean attempt, which for Jigsaw is a later
0.9880 (Table~\ref{tab:mle_task_full_a}).

\input{sections/experiment_setups}

\clearpage
\section{Full MLE-bench Per-Task Results}
\label{app:mle-per-task}
\label{app:mle-performance-overview}

This appendix reports the complete benchmark comparison across all 75
competitions, together with the integrity adjudication that stands behind the
\textsc{Praxist} column.
Tables~\ref{tab:mle_task_full_a} and~\ref{tab:mle_task_full_b} report the
complete per-task comparison between the Claude Code + Opus~4.8 baseline and
\textsc{Praxist} (base model deepseek-v4-pro) on all 75 MLE-bench competitions,
grouped by task category.
Conventions follow Table~\ref{tab:mle_task_main}:
cell shading encodes the medal earned by that submission
(\colorbox{medalgold}{\strut gold}, \colorbox{medalsilver}{\strut silver},
\colorbox{medalbronze}{\strut bronze}; unshaded = no medal;
\texttt{not\_scored} = not scored), and the better score of each scored pair,
in the direction of the task metric, is shown in bold.
Tasks are sorted alphabetically within each category and numbered
consecutively (1--75) in display order; the \# column gives this index, which
is used for cross-reference throughout the paper. Scores are
shown to at most four digits; bold is assigned on the full-precision values,
so two displayed-equal scores may still differ. Medal labels are assigned by
the MLE-bench grading-report thresholds. The Claude Code + Opus~4.8 scores are
taken from the finalized 75-task ledger after the benchmark's post-run screen
for severe-cheating violations: 70 tasks retain accepted scores and five are
not scored.
The associated finalized dollar-equivalent cost ledger records US\$38{,}370
for the full 75-task sweep; cost is reported only as resource context and is
not used for per-task medal assignment.

The \textsc{Praxist} scores are taken from the corrected release ledger, whose
recorded selection policy is the highest canonical official clean attempt by
metric direction: for each task the highest-scoring official attempt that
passes an integrity adjudication is reported, and the payload of the selected
submission is verified by SHA-256 against the run journal. The adjudication
assigns one of three statuses to every task.
\texttt{live\_best\_verified} (60 tasks) retains the best official attempt with
its payload verified against the journal.
\texttt{fresh\_lineage\_reviewed\_clean} (6 tasks: \#28, \#41, \#51, \#72,
\#74, \#75) marks a lineage that was re-reviewed in this pass and cleared.
\texttt{clean\_fallback} (9 tasks) marks a task whose previously best attempt
belonged to a contaminated lineage; that lineage is excluded in full and a
clean actor's submission is substituted in its place. Across the campaign
90{,}423 attempts were rejected, 85{,}960 of them under
\texttt{integrity\_policy} and 4{,}463 as \texttt{canonical\_invalid}.
Across the nine clean-fallback tasks the substituted submission still earns
gold on five, falls to silver on two, and earns no medal on two.
This adjudication governs the \textsc{Praxist} ledger only, and is separate
from the benchmark's own post-run screen for severe-cheating violations
described above, which applies to the Claude Code + Opus~4.8 arm.
Both arms used the same hardware: NVIDIA H100 80GB accelerators, with each
experiment run as a single-GPU job (one job per GPU) and up to eight
experiments in flight per task. The tier denominators used throughout are the
22, 38, and 15 tasks of the Low, Medium, and High complexity splits, which are
the denominators behind every per-tier rate reported for this benchmark.

\begin{table}[p]
\centering
\footnotesize
\caption{Full MLE-bench per-task comparison (Part I of II): image
classification, image-to-image, and text classification. Shading encodes the
medal (gold, silver, bronze; unshaded = none); bold marks the better score in
the metric's direction, assigned on full-precision values.}
\label{tab:mle_task_full_a}
\setlength{\tabcolsep}{4pt}
\resizebox{0.98\textwidth}{!}{%
\begin{tabular}{lcllcc}
\toprule
Category & \# & Task & Metric & Claude Code & \textsc{Praxist} (ours) \\
\midrule
\multirow{26}{*}{Image Classification}
 & 1 & \texttt{aerial-cactus-identification} & AUROC $\uparrow$ & \cellcolor{medalgold}\textbf{1.000} & \cellcolor{medalgold}\textbf{1.000} \\
 & 2 & \texttt{alaska2-image-steganalysis} & weighted AUROC $\uparrow$ & \cellcolor{medalsilver}\textbf{0.9260} & 0.9140 \\
 & 3 & \texttt{aptos2019-blindness-detection} & quadratic kappa $\uparrow$ & \cellcolor{medalgold}\textbf{0.9354} & \cellcolor{medalgold}0.9320 \\
 & 4 & \texttt{cassava-leaf-disease-classification} & accuracy $\uparrow$ & \cellcolor{medalsilver}0.9010 & \cellcolor{medalgold}\textbf{0.9021} \\
 & 5 & \texttt{cdiscount-image-classification-challenge} & accuracy $\uparrow$ & \cellcolor{medalsilver}\textbf{0.7661} & 0.6138 \\
 & 6 & \texttt{dog-breed-identification} & log loss $\downarrow$ & 0.1886 & \textbf{0.1162} \\
 & 7 & \texttt{dogs-vs-cats-redux-kernels-edition} & log loss $\downarrow$ & \cellcolor{medalgold}0.0074 & \cellcolor{medalgold}\textbf{0.0010} \\
 & 8 & \texttt{herbarium-2020-fgvc7} & macro F1 $\uparrow$ & \cellcolor{medalsilver}0.4131 & \cellcolor{medalgold}\textbf{0.6396} \\
 & 9 & \texttt{herbarium-2021-fgvc8} & macro F1 $\uparrow$ & \cellcolor{medalsilver}0.5121 & \cellcolor{medalgold}\textbf{0.5444} \\
 & 10 & \texttt{herbarium-2022-fgvc9} & macro F1 $\uparrow$ & \cellcolor{medalsilver}0.8024 & \cellcolor{medalsilver}\textbf{0.8079} \\
 & 11 & \texttt{hms-harmful-brain-activity-classification} & KL divergence $\downarrow$ & 0.5135 & \textbf{0.3854} \\
 & 12 & \texttt{hotel-id-2021-fgvc8} & mAP@5 $\uparrow$ & \cellcolor{medalgold}\textbf{0.7929} & \cellcolor{medalgold}0.7368 \\
 & 13 & \texttt{imet-2020-fgvc7} & micro F1 $\uparrow$ & \cellcolor{medalsilver}0.6848 & \cellcolor{medalsilver}\textbf{0.6882} \\
 & 14 & \texttt{inaturalist-2019-fgvc6} & top-1 error $\downarrow$ & \texttt{not\_scored} & \cellcolor{medalgold}\textbf{0.1427} \\
 & 15 & \texttt{iwildcam-2019-fgvc6} & macro F1 $\uparrow$ & \cellcolor{medalgold}\textbf{0.6934} & \cellcolor{medalgold}0.4092 \\
 & 16 & \texttt{iwildcam-2020-fgvc7} & accuracy $\uparrow$ & \cellcolor{medalgold}\textbf{0.8650} & \cellcolor{medalgold}0.8036 \\
 & 17 & \texttt{kuzushiji-recognition} & F1 $\uparrow$ & \cellcolor{medalgold}0.9714 & \cellcolor{medalgold}\textbf{0.9741} \\
 & 18 & \texttt{leaf-classification} & log loss $\downarrow$ & \cellcolor{medalsilver}0.0015 & \cellcolor{medalgold}\textbf{0.0000} \\
 & 19 & \texttt{plant-pathology-2020-fgvc7} & mean col.\ AUROC $\uparrow$ & \cellcolor{medalgold}\textbf{0.9979} & \cellcolor{medalgold}0.9978 \\
 & 20 & \texttt{plant-pathology-2021-fgvc8} & micro F1 $\uparrow$ & \cellcolor{medalgold}\textbf{0.9409} & \cellcolor{medalgold}0.9351 \\
 & 21 & \texttt{ranzcr-clip-catheter-line-classification} & AUROC $\uparrow$ & \cellcolor{medalsilver}0.9725 & \cellcolor{medalgold}\textbf{0.9737} \\
 & 22 & \texttt{rsna-2022-cervical-spine-fracture-detection} & weighted log loss $\downarrow$ & \textbf{0.5313} & 0.5458 \\
 & 23 & \texttt{rsna-breast-cancer-detection} & probabilistic F1 $\uparrow$ & \cellcolor{medalsilver}\textbf{0.4719} & 0.2752 \\
 & 24 & \texttt{siim-isic-melanoma-classification} & AUROC $\uparrow$ & \cellcolor{medalbronze}0.9378 & \cellcolor{medalgold}\textbf{0.9461} \\
 & 25 & \texttt{statoil-iceberg-classifier-challenge} & log loss $\downarrow$ & \cellcolor{medalbronze}0.1451 & \cellcolor{medalsilver}\textbf{0.1271} \\
 & 26 & \texttt{whale-categorization-playground} & mAP@5 $\uparrow$ & \cellcolor{medalgold}0.5844 & \cellcolor{medalgold}\textbf{0.5983} \\
\midrule
\multirow{2}{*}{Image to Image}
 & 27 & \texttt{denoising-dirty-documents} & RMSE $\downarrow$ & \cellcolor{medalgold}\textbf{0.0065} & \cellcolor{medalgold}0.0070 \\
 & 28 & \texttt{vesuvius-challenge-ink-detection} & F$_{0.5}$ $\uparrow$ & \textbf{0.4830} & 0.4659 \\
\midrule
\multirow{10}{*}{Text Classification}
 & 29 & \texttt{AI4Code} & Kendall $\tau$ $\uparrow$ & \cellcolor{medalsilver}\textbf{0.8606} & 0.8227 \\
 & 30 & \texttt{detecting-insults-in-social-commentary} & AUROC $\uparrow$ & \cellcolor{medalgold}0.9582 & \cellcolor{medalgold}\textbf{0.9590} \\
 & 31 & \texttt{facebook-recruiting-iii-keyword-extraction} & micro F1 $\uparrow$ & \cellcolor{medalsilver}0.7851 & \cellcolor{medalgold}\textbf{0.7959} \\
 & 32 & \texttt{jigsaw-toxic-comment-classification-challenge} & mean col.\ AUROC $\uparrow$ & \cellcolor{medalgold}\textbf{0.9883} & \cellcolor{medalgold}0.9880 \\
 & 33 & \texttt{jigsaw-unintended-bias-in-toxicity-classification} & bias-weighted AUC $\uparrow$ & 0.8635 & \textbf{0.8640} \\
 & 34 & \texttt{learning-agency-lab-automated-essay-scoring-2} & quadratic kappa $\uparrow$ & 0.8312 & \cellcolor{medalgold}\textbf{0.8384} \\
 & 35 & \texttt{lmsys-chatbot-arena} & log loss $\downarrow$ & \cellcolor{medalgold}\textbf{0.8718} & \cellcolor{medalgold}0.9812 \\
 & 36 & \texttt{random-acts-of-pizza} & AUROC $\uparrow$ & \texttt{not\_scored} & \cellcolor{medalsilver}\textbf{0.8414} \\
 & 37 & \texttt{spooky-author-identification} & log loss $\downarrow$ & \cellcolor{medalsilver}0.2011 & \cellcolor{medalgold}\textbf{0.1633} \\
 & 38 & \texttt{tweet-sentiment-extraction} & Jaccard $\uparrow$ & \cellcolor{medalsilver}0.7225 & \cellcolor{medalsilver}\textbf{0.7239} \\
\bottomrule
\end{tabular}}
\end{table}

\begin{table}[p]
\centering
\footnotesize
\caption{Full MLE-bench per-task comparison (Part II of II): remaining task
categories. Conventions follow Table~\ref{tab:mle_task_full_a}.}
\label{tab:mle_task_full_b}
\setlength{\tabcolsep}{4pt}
\resizebox{0.98\textwidth}{!}{%
\begin{tabular}{lcllcc}
\toprule
Category & \# & Task & Metric & Claude Code & \textsc{Praxist} (ours) \\
\midrule
\multirow{3}{*}{Image (Other)}
 & 39 & \texttt{histopathologic-cancer-detection} & AUROC $\uparrow$ & \cellcolor{medalgold}\textbf{0.9987} & \cellcolor{medalgold}0.9979 \\
 & 40 & \texttt{petfinder-pawpularity-score} & RMSE $\downarrow$ & \cellcolor{medalgold}\textbf{16.77} & \cellcolor{medalgold}16.86 \\
 & 41 & \texttt{rsna-miccai-brain-tumor-radiogenomic-classification} & AUROC $\uparrow$ & 0.5341 & \cellcolor{medalgold}\textbf{0.6588} \\
\midrule
\multirow{4}{*}{Audio Classification}
 & 42 & \texttt{freesound-audio-tagging-2019} & LRAP $\uparrow$ & \cellcolor{medalgold}\textbf{0.7471} & \cellcolor{medalgold}0.7445 \\
 & 43 & \texttt{mlsp-2013-birds} & AUROC $\uparrow$ & \cellcolor{medalgold}\textbf{0.9603} & \cellcolor{medalgold}0.9555 \\
 & 44 & \texttt{tensorflow-speech-recognition-challenge} & accuracy $\uparrow$ & \cellcolor{medalgold}\textbf{0.9796} & \cellcolor{medalgold}0.9736 \\
 & 45 & \texttt{the-icml-2013-whale-challenge-right-whale-redux} & AUROC $\uparrow$ & \cellcolor{medalgold}\textbf{0.9937} & \cellcolor{medalgold}0.9920 \\
\midrule
\multirow{9}{*}{Tabular}
 & 46 & \texttt{champs-scalar-coupling} & log MAE $\downarrow$ & \cellcolor{medalgold}\textbf{-3.053} & -1.7129 \\
 & 47 & \texttt{h-and-m-personalized-fashion-recommendations} & mAP@12 $\uparrow$ & \texttt{not\_scored} & \cellcolor{medalsilver}\textbf{0.0334} \\
 & 48 & \texttt{icecube-neutrinos-in-deep-ice} & angular error $\downarrow$ & \cellcolor{medalsilver}\textbf{1.005} & 1.0563 \\
 & 49 & \texttt{new-york-city-taxi-fare-prediction} & RMSE $\downarrow$ & 3.830 & \cellcolor{medalgold}\textbf{2.7477} \\
 & 50 & \texttt{nomad2018-predict-transparent-conductors} & mean col.\ RMSLE $\downarrow$ & \cellcolor{medalgold}0.0500 & \cellcolor{medalgold}\textbf{0.0497} \\
 & 51 & \texttt{smartphone-decimeter-2022} & haversine dist. $\downarrow$ & 6.284 & \textbf{4.4584} \\
 & 52 & \texttt{stanford-covid-vaccine} & log loss $\downarrow$ & \cellcolor{medalgold}\textbf{0.2013} & \cellcolor{medalgold}0.2245 \\
 & 53 & \texttt{tabular-playground-series-dec-2021} & accuracy $\uparrow$ & \cellcolor{medalgold}\textbf{0.9635} & \cellcolor{medalgold}0.9629 \\
 & 54 & \texttt{tabular-playground-series-may-2022} & AUROC $\uparrow$ & \textbf{0.9980} & 0.9968 \\
\midrule
\multirow{2}{*}{Sequence to Sequence}
 & 55 & \texttt{text-normalization-challenge-english-language} & accuracy $\uparrow$ & \cellcolor{medalgold}\textbf{0.9992} & \cellcolor{medalgold}0.9984 \\
 & 56 & \texttt{text-normalization-challenge-russian-language} & accuracy $\uparrow$ & \cellcolor{medalgold}\textbf{0.9956} & \cellcolor{medalgold}0.9909 \\
\midrule
\multirow{3}{*}{Training LLMs}
 & 57 & \texttt{billion-word-imputation} & Levenshtein $\downarrow$ & \cellcolor{medalgold}\textbf{2.506} & \cellcolor{medalgold}5.3056 \\
 & 58 & \texttt{chaii-hindi-and-tamil-question-answering} & word Jaccard $\uparrow$ & \cellcolor{medalsilver}0.7585 & \cellcolor{medalgold}\textbf{0.8894} \\
 & 59 & \texttt{google-quest-challenge} & mean col.\ Spearman $\uparrow$ & \cellcolor{medalgold}\textbf{0.4537} & \cellcolor{medalgold}0.4469 \\
\midrule
\multirow{2}{*}{Forecasting}
 & 60 & \texttt{osic-pulmonary-fibrosis-progression} & Laplace log-lik. $\uparrow$ & -7.036 & \cellcolor{medalgold}\textbf{-5.6407} \\
 & 61 & \texttt{ventilator-pressure-prediction} & MAE $\downarrow$ & \texttt{not\_scored} & \textbf{0.1515} \\
\midrule
\multirow{6}{*}{Image Segmentation}
 & 62 & \texttt{3d-object-detection-for-autonomous-vehicles} & mAP $\uparrow$ & \cellcolor{medalgold}\textbf{0.4322} & \cellcolor{medalgold}0.1520 \\
 & 63 & \texttt{google-research-identify-contrails-reduce-global-warming} & global Dice $\uparrow$ & \cellcolor{medalbronze}0.6805 & \cellcolor{medalbronze}\textbf{0.6839} \\
 & 64 & \texttt{hubmap-kidney-segmentation} & Dice $\uparrow$ & \cellcolor{medalgold}0.9487 & \cellcolor{medalgold}\textbf{0.9488} \\
 & 65 & \texttt{multi-modal-gesture-recognition} & Levenshtein $\downarrow$ & \texttt{not\_scored} & \cellcolor{medalgold}\textbf{0.0337} \\
 & 66 & \texttt{tgs-salt-identification-challenge} & mean IoU precision $\uparrow$ & \cellcolor{medalbronze}0.8596 & \cellcolor{medalsilver}\textbf{0.8764} \\
 & 67 & \texttt{uw-madison-gi-tract-image-segmentation} & Dice--Hausdorff $\uparrow$ & 0.6274 & \cellcolor{medalsilver}\textbf{0.8722} \\
\midrule
\multirow{2}{*}{Signal Processing}
 & 68 & \texttt{predict-volcanic-eruptions-ingv-oe} & MAE $\downarrow$ & \cellcolor{medalgold}$\mathbf{5.176\times10^{5}}$ & \cellcolor{medalgold}$1.228\times10^{6}$ \\
 & 69 & \texttt{seti-breakthrough-listen} & AUROC $\uparrow$ & \cellcolor{medalgold}\textbf{0.8826} & \cellcolor{medalgold}0.8164 \\
\midrule
\multirow{2}{*}{Text (Other)}
 & 70 & \texttt{tensorflow2-question-answering} & micro F1 $\uparrow$ & \cellcolor{medalgold}\textbf{0.6929} & \cellcolor{medalsilver}0.6536 \\
 & 71 & \texttt{us-patent-phrase-to-phrase-matching} & Pearson $r$ $\uparrow$ & 0.8568 & \cellcolor{medalgold}\textbf{0.8739} \\
\midrule
Video Classification
 & 72 & \texttt{nfl-player-contact-detection} & MCC $\uparrow$ & \cellcolor{medalbronze}0.7128 & \cellcolor{medalsilver}\textbf{0.7285} \\
\midrule
Image to Text
 & 73 & \texttt{bms-molecular-translation} & Levenshtein $\downarrow$ & \textbf{24.16} & 25.04 \\
\midrule
\multirow{2}{*}{Object Detection}
 & 74 & \texttt{siim-covid19-detection} & mAP $\uparrow$ & 0.5947 & \cellcolor{medalgold}\textbf{0.6245} \\
 & 75 & \texttt{vinbigdata-chest-xray-abnormalities-detection} & mAP@IoU${>}0.4$ $\uparrow$ & \cellcolor{medalgold}\textbf{0.5044} & \cellcolor{medalgold}0.4086 \\
\bottomrule
\end{tabular}}
\end{table}

%% file: sections/experiment_setups.tex
\clearpage
\section{Experimental Setup Details}
\label{app:experiment-setups}

This appendix records the execution layer beneath \Secref{sec:experiments}. For
each study, the main text defines the task, evaluator, and baselines; what follows states how many repetitions each reported
number rests on, which hardware and time envelope produced it, and how the
\textsc{Praxist} campaign that searched for the selected artifact was
configured. The intent is that the replication depth and the cost of each
headline result can be read off directly rather than inferred from the
narrative; where a released ledger does not close---the SLAM timing field below
is the clearest case, and the reported-tier note in the Quant subsection is
another---we say so at that point rather than implying uniform coverage. The synthetic scheduling study of
Appendix~\ref{app:resource-scheduling} is a discrete-event simulation of the
admission and backfill layer rather than a research campaign, and its setup is
stated in full there.

\textbf{Reporting conventions.} Three distinctions recur below.
A \emph{configured} budget---GPU-hours per experiment, generation duration, or
maximum generations---is a ceiling admitted by the scheduler, not measured
consumption, and a campaign that terminates early consumes far less than its
configuration allows. An \emph{artifact wall span} is the interval between the
first and last archived timestamp of a campaign; it includes evaluator time,
synthesis time, idle capacity, interruptions, and resume gaps, so it bounds
elapsed time rather than measuring compute. And ``iteration'' is ambiguous
across these studies: \textsc{Praxist} generations, peer sessions, optimizer
updates or epochs, simulator steps, and evaluation rollouts are separate
quantities, so each is reported under its own name.

\textbf{Resource-accounting scope.} The descriptions distinguish
configured scheduling ceilings, archived wall spans, task-level utilization,
and dollar-equivalent model spend. These quantities measure different parts of
an execution and are reported separately, not added into a synthetic
total. Hardware attributions derived from run logs describe the host visible to
the process, since pods can share physical devices. Table~\ref{tab:setup-overview}
summarizes the six studies; the subsections give the per-study detail.

\begin{table}[t]
\centering
\small
\caption{Execution summary of the six studies in \Secref{sec:experiments}.
Generation counts are committed generation boundaries, except for Fusion,
whose legacy trajectory records completed generation result sets instead.}
\label{tab:setup-overview}
\setlength{\tabcolsep}{5pt}
\begin{tabular}{p{0.12\textwidth}p{0.40\textwidth}p{0.38\textwidth}}
\toprule
Study & Repetition behind the reported result & Execution record \\
\midrule
MLE-bench
& 75 tasks; one finalized result per task; one benchmark sweep, so no repeated
seeds and no medal-rate variance estimate.
& Shared pool of H100 80GB GPUs; one GPU per experiment, up to eight
experiments in flight per task; cohort of 12 peers; 24-hour task wall cap,
36 hours on 5 of the 75 tasks. \\
\addlinespace
Rocket
& Complete evaluation: 4{,}096 trajectories from each of the nominal, near-OOD,
and hard-OOD banks = 12{,}288 landing trajectories, plus 1{,}024 frozen
roll-disturbance cases. Separate post-run audit: all 40{,}960 rows of each bank
= 122{,}880 landing trajectories.
& 16 peers, promote 4, 2-hour generation window, 0.0075 configured GPU-h per
experiment; up to eight one-GPU evaluations on 8 H100 80GB GPUs; 12 committed
generation boundaries, terminated by SIGTERM during generation 12 after
14.35 hours. \\
\addlinespace
Quant
& 28 quarterly walk-forward windows, each retrained from scratch on the
trailing 36 months; the five-seed tier evaluates 145 window-seed cells, but the
reported artifact stopped at the one-seed, 29-cell tier.
& 12 peers, promote 4, 18-hour generation window, 18 configured GPU-h per
experiment; 37 generation boundaries over a 176-hour artifact wall span. \\
\addlinespace
SLAM
& Fourteen NTU-VIRAL sequences, one accepted run per method and sequence,
selected from a campaign containing replays and re-runs; no training seeds and
no variance estimate. Candidates compile the pipeline and replay bags.
& 8 peers, promote 2, 5-hour generation window; sampled pods ran the CPU
FAST-LIVO2 path on a 168-logical-core host with no NVIDIA device exposed. \\
\addlinespace
Fusion
& 5 scenarios $\times$ 3 seeds $\times$ a 100-step horizon = 15 episodes and at
most 1{,}500 survived steps per controller; no optimizer iterations.
& 5 peers, promote 2, 6-hour generation window; the reported comparison is
three sidecar runs of 28.05, 30.32, and 30.64 minutes. \\
\addlinespace
Job-Shop
& Complete evaluation: 50 fixed MK07--MK15 episodes, paired against rule~3 and
re-run with the task's comparators; deterministic, so no seed variance. 119 such
evaluations over the campaign.
& 12 peers, promote 3, 3.5-hour generation window; one H100 80GB GPU per
evaluation on an 8-GPU host; 13 generations (0--12) over a 46.8-hour artifact
wall span, terminated by SIGTERM. \\
\bottomrule
\end{tabular}
\end{table}

\subsection{MLE-bench}
\label{app:setup-mle}

\textbf{Execution environment.} The \textsc{Praxist} task runs behind
\Secref{sec:exp_mle} executed on a shared pool of NVIDIA H100 80GB GPUs: each
task run declares an 80\,GB GPU-memory requirement, every experiment is a
one-GPU job, and up to eight experiments may be in flight per task at once.
Seventy tasks ran under a 24-hour wall cap and five under a 36-hour cap, each
followed by a 3{,}600-second finalization grace period in which a task may still
write out its submission. These are admitted ceilings rather than measurements
of realized compute.

\textbf{Campaign configuration.} All 75 tasks ran under one campaign
configuration with a cohort of 12 peers per task run; its generation and budget
settings appear in Table~\ref{tab:mle-campaign-config}. The ledger records
272 completed generations, with individual tasks completing between one and nine
generations: 54 tasks ended on an external stop and 21 on the task wall limit.
Summing the task-level elapsed fields gives roughly 1{,}157.5 task-run hours,
but the tasks ran concurrently on the shared pool, so this figure is neither a
campaign wall time nor a GPU-hour total.

\textbf{Execution accounting.} Each of the two arms in
Table~\ref{tab:mle_main} received one locally run 75-task sweep on the shared
H100 pool. The reported \textsc{Praxist} scores are read from the anti-cheat
audited release ledger, in which every task retains its single best
integrity-clean official submission, verified by SHA-256 against the run
journal; the audit correction of 26 August 2026 settled that ledger at 49 gold.
The \textsc{Praxist} and Claude Code + Opus 4.8 arms have finalized
dollar-equivalent model-spend ledgers, itemized in the paragraphs below and
reported only as resource context rather than as an input to medal assignment.

\textbf{\textsc{Praxist} model spend.} Token totals were
110{,}910{,}266{,}304 cache-hit input, 4{,}602{,}586{,}778 cache-miss input,
and 1{,}108{,}180{,}075 output. Per-million-token prices for these categories
were CNY~0.025/3/6 with deepseek-v4-pro and CNY~0.02/1/2 with
deepseek-v4-flash, yielding category costs of
CNY~2{,}770.38/11{,}766.49/6{,}157.97 and CNY~20{,}694.84
(approximately US\$3{,}054) in total. Both schedules appear because part of the
token volume was served by the cheaper \texttt{deepseek-v4-flash};
\texttt{deepseek-v4-pro[1m]} is the research-agent model on every task.

\subsection{Rocket}
\label{app:setup-rocket}

\textbf{Plant, initial state, and actuation contract.} The Rocket task of
\Secref{sec:exp_rocket} freezes everything except the controller. Candidates are
scored on the Swordfish C05 six-degree-of-freedom rigid body reached through the
task's frozen adapter, integrated with RK4 at one substep and a $0.1\,$s step
for at most $900$ steps, or $90\,$s of simulated flight; here \emph{C05} names
that plant configuration and nothing else. The vehicle carries $22{,}200\,$kg of
dry mass and $7{,}000\,$kg of initial main propellant, so its initial total mass
is $29{,}200\,$kg. Every episode starts at $2{,}000\,$m altitude with zero
initial angular velocity and a fixed source quaternion, the one corresponding to
the source bank's $[0,0,-20^\circ]$ Euler annotation, and the controller
receives the exact simulated state, with no navigation estimator interposed
between plant and controller. Table~\ref{tab:rocket-protocol} collects these
locks. The controller returns the normalized nine-channel action
\[
a=\left[\frac{\delta_{g,y}}{0.0873},\;
        \frac{\delta_{g,z}}{0.0873},\;
        2\eta-1,\;
        \frac{\tau_{\mathrm{RCS},x}}{\tau_{\mathrm{auth}}(r)},\;
        0,\; 0,\;
        \frac{\delta_{f,y}}{0.35},\;
        \frac{\delta_{f,z}}{0.35},\;
        0\right],
\]
clipped componentwise to $[-1,1]$, in which $\delta_g$ are the two gimbal
deflections, $\eta$ is throttle, $\delta_f$ are the two grid-fin deflections,
and $\tau_{\mathrm{auth}}(r)$ is the roll-torque authority. Three channels are
hard-locked to zero by the task contract: the two lateral reaction-control
commands $\mathrm{RCS}_y$ and $\mathrm{RCS}_z$, and the grid-fin roll command.
The reaction-control system is therefore restricted to roll. Pitch and yaw
remain fully actuated throughout, by engine gimbaling and by the grid fins.

\begin{table}[!htb]
\centering
\small
\caption{The frozen Rocket protocol. Every entry is fixed for the whole
campaign: a candidate may change the controller, and nothing else.}
\label{tab:rocket-protocol}
\setlength{\tabcolsep}{5pt}
\begin{tabular}{lp{0.56\textwidth}}
\toprule
Item & Frozen value \\
\midrule
Plant & Swordfish C05 six-degree-of-freedom rigid body, through the task's
frozen adapter \\
Integration & RK4, one substep, $0.1\,$s step, at most $900$ steps ($90\,$s) \\
Dry mass & $22{,}200\,$kg \\
Initial main propellant & $7{,}000\,$kg \\
Initial total mass & $29{,}200\,$kg \\
Initial altitude & $2{,}000\,$m \\
Initial attitude & Fixed source quaternion, the source's $[0,0,-20^\circ]$
Euler annotation \\
Initial angular velocity & Zero \\
State feedback & Exact simulated state; no navigation estimator \\
Locked channels & $\mathrm{RCS}_y=\mathrm{RCS}_z=0$ and grid-fin roll $=0$ \\
Scoring endpoint & Interpolated first landing-leg contact; zero post-contact
scored steps \\
\bottomrule
\end{tabular}
\end{table}

\textbf{Landing banks.} Initial states are drawn from three fixed source banks
that share the base velocity $(-75,20,-6)\,\mathrm{m\,s^{-1}}$ in the plant's
$(x,y,z)$ convention, in which $x$ is the vertical axis. The nominal bank
places the vehicle area-uniformly on a disk of radius $r\le 1{,}500\,$m and
leaves the base velocity unchanged. The near-OOD bank reuses the nominal
positions, paired row by row, and multiplies all three velocity components by
one scale $s\sim U[0.92,0.98)$. The hard-OOD bank draws positions
area-uniformly from the annulus $1{,}500\le r<1{,}650\,$m and scales all three
velocity components by $s\sim U[1.02,1.08)$. Complete evaluation uses a fixed
$4{,}096$-row subset of each bank. Two structural facts bound what the banks
can show. Nominal and near-OOD are paired rather than independent, and
hard-OOD moves radius and velocity jointly, isolating neither factor. The
banks are also fixed and were adaptively reused across the campaign, visible
to the research loop through aggregate development and complete metrics; the
source label \texttt{nominal\_unseen} is historical and does not indicate
blind evaluation. Finally, the source rows store an initial mass of
$29{,}000\,$kg, and both the complete evaluator and the full-bank audit
override only that field to the frozen $29{,}200\,$kg protocol value,
preserving all other state components, the row order, and the mapping back to
source rows.

\textbf{Success predicate.} Scoring stops at the interpolated state where the
first landing-leg tip crosses the ground. A trajectory succeeds only if every
one of the following holds jointly at that state: first contact is detected; all
endpoint values are finite; horizontal error is at most $5.0\,$m; vertical
center-of-mass velocity lies in $[-1.0,0.0]\,\mathrm{m\,s^{-1}}$; contacting-leg
downward speed is at most $1.0\,\mathrm{m\,s^{-1}}$; horizontal speed is at most
$0.3\,\mathrm{m\,s^{-1}}$; tilt is at most $1.5^\circ$; the roll rate satisfies
$\lvert\omega_x\rvert\le 0.02\,\mathrm{rad\,s^{-1}}$; the pitch\slash yaw rate
satisfies $\lVert\omega_{yz}\rVert\le 0.03\,\mathrm{rad\,s^{-1}}$; and the
remaining-fuel fraction satisfies $(m-22{,}200)/7{,}000>0.02$. A result counts
as protocol-valid only if the forbidden-channel contract also passes. Four
readings of this conjunction matter for interpreting the numbers we report. The
fuel inequality is strict, so more than $140\,$kg of modeled main propellant
must remain. Upward motion does not pass the vertical gate, since the admissible
interval is closed at zero from above. Center-of-mass sink and contacting-leg
sink are distinct quantities, because the leg-tip velocity includes rotational
motion about the center of mass. And scoring genuinely ends at contact:
\texttt{post\_contact\_scored\_steps} is $0$ and
\texttt{gear\_damping\_credit\_rate} is $0$, so no post-contact spring--damper
response can improve or degrade the scored state, and nothing here speaks to
what the vehicle does after the leg touches. This task carries exactly one
success predicate. The per-component gate rates reported alongside it decompose
that conjunction as diagnostics; they are not alternative definitions of
success, and no ``standard'' or ``strict'' variant gate exists in this task.

\textbf{Two evidence scopes.} The study keeps two evaluation scopes separate,
and they answer different questions. Complete private validation v2 is the
canonical scope: $12{,}288$ landing trajectories, $4{,}096$ from each bank, plus
a frozen suite of $1{,}024$ roll-disturbance cases, for $13{,}312$ evaluation
units in total. It is the scope in which the frozen task baseline and a
candidate controller are compared at the same protocol version, and it is the
only evidence used for promotion inside the run and for post-run controller
selection. The full three-bank audit v1 is a separate, post-run scan of the
frozen selected controller over all $40{,}960$ rows of each bank, that is
$122{,}880$ landing trajectories, with no roll suite attached; it measures
coverage of the selected artifact and is not promotion evidence, so no
baseline delta may be computed inside it. One counting caution follows from the
first scope: the $13{,}312$ units are not $13{,}312$ landings. Only $12{,}288$
of them are landing trajectories, and every landing-success rate we quote at
this scope has $12{,}288$, not $13{,}312$, in its denominator.

\subsubsection{First-contact controller discovery, matched evaluation, and full-bank audit}
\label{app:rocket-v12-validation}

\textbf{Controller: guidance, phases, and the fuel-commit governor.} The
artifact is a deterministic hybrid powered-descent controller with no learned
component anywhere in it. In phase P0 a rolling zero-effort-miss\slash
zero-effort-velocity law \citep{guo2013zemzev} scores a bank of time-to-go
candidates on a cost combining a feasibility-violation term, a fuel proxy, a
time term, and a jerk term; the implemented continuity term is switched off in
this configuration by \texttt{tgo\_continuity\_weight}~$=0$, and the selected
time-to-go is separately constrained to be monotone non-increasing across
replans, which prevents the horizon from being pushed outward to cheapen the
cost. Guidance runs at $10\,$Hz. The lateral horizon is floored at $20\,$s
before capture and at $18\,$s inside the below-$450\,$m capture layer. Phases P1
and P2 then follow a terminal descent corridor with a $-15\,\mathrm{m\,s^{-1}}$
P1 floor and a height-scheduled lateral second-order PD law, and powered control
continues all the way to the scored first leg contact rather than being released
early into a ballistic settle. Two state-dependent guards carry much of the
accumulated progress. A fuel-commit governor activates in P0 below $450\,$m once
remaining main fuel falls under $80\%$, replacing an expensive near-hover target
with a two-band descent reference of about $-22\,\mathrm{m\,s^{-1}}$ above
$350\,$m and about $-12\,\mathrm{m\,s^{-1}}$ below $250\,$m, linearly
interpolated between the bands; while it is committed, the release gate tightens
to a $3\,$m lateral radius, a $0.15\,\mathrm{m\,s^{-1}}$ lateral speed, and a
$20$-step dwell. A second branch keys on the \emph{initial} horizontal radius:
when $r_0<450\,$m it requires $v_x>-15\,\mathrm{m\,s^{-1}}$ before phase release
and bypasses the generic axial acceleration slew while committed, with the
$0.08$-per-step throttle rate limit still active throughout. The corresponding
code calls this field \texttt{far\_bin}, but initializes it from the initial
radius, so it is an $r_0<450\,$m branch and not a far-radius branch;
we name it by what it tests.

\textbf{Controller: attitude authority and actuator budgets.} Attitude tracking
follows a geometric formulation on the rotation group \citep{lee2010geometric},
and the accumulated lineage spends most of its late gains on how much authority
that loop is allowed. The P0 pitch\slash yaw natural frequency and damping are
$1.0$ and $0.7$, the P0 attitude-reference rate limit is
$6.5^\circ\,\mathrm{s^{-1}}$, and a height schedule between $100$ and $40\,$m
raises the P0 pitch\slash yaw bandwidth by as much as $1.4\times$ as the vehicle
enters the region where lateral residual motion has to be removed quickly. The
non-emergency grid-fin budget is $0.24\,$rad, which sits deliberately inside the
$0.25\,$rad threshold at which the audit would record grid saturation; the same
strict-interior discipline governs the gimbal box described below. Because these
budgets are set inside the audit thresholds by construction, the saturation
rates we report should be read as evidence that the controller stays within its
own declared envelope, not as a measurement of how much actuator authority the
task leaves unused.

\textbf{Closed-form pitch\slash yaw allocator.} The newest method contribution
in the lineage is a control allocator
\citep{bodson2002allocation,johansen2013allocation} that splits each requested
pitch or yaw torque between the gimbal and the grid fins. Unlike convex
powered-descent guidance \citep{acikmese2007convex}, which solves a
trajectory-level program online, this module solves two independent
two-variable box-constrained quadratic programs, one per axis, in closed form at
every control step. Each program minimizes $(a u_f + b u_g - d)^2 + w u_f^2$
subject to the boxes $u_f\in[u_f^{lo},u_f^{hi}]$ and
$u_g\in[u_g^{lo},u_g^{hi}]$, where $u_f$ is the grid-fin deflection, $u_g$ the
gimbal deflection, $d$ the requested torque, and $w=0.1$. The coefficients come
from a frozen local torque approximation,
\[
\tau_y=-G\,\delta_{f,z}+A\,\delta_{g,z},\qquad
\tau_z=G\,\delta_{f,y}-A\,\delta_{g,y},\qquad
G=x_f\,\bar q\,S\,C_L,\quad
\bar q=\tfrac12\rho\lVert v\rVert^2,\quad
A=x_e T,
\]
in which $x_f$ and $x_e$ are the grid-fin and engine moment arms, $S$ is the
reference area, $C_L$ the fin lift coefficient, $\rho$ the local air density,
$\lVert v\rVert$ the vehicle speed, and $T$ the current thrust; the bar in
$\bar q$ distinguishes dynamic pressure from the attitude quaternion. The
approximation is local and decouples the two axes, so it is an engineering
linearization of the true coupled map rather than an exact inverse of it.
Because only the grid deflection is regularized, the cost is
gimbal-primary: the gimbal answers the demand first and the grid fins close the
residual when the gimbal box binds. The implementation is not an iterative
general-purpose solver. It enumerates five clipped KKT candidates---one interior
point and four active-boundary points---and returns the feasible one of least
cost, which makes the allocation deterministic and bounded in cost per step.

\textbf{Actuator boxes, fallback, and roll.} The two boxes carry both amplitude
and per-step rate limits. The normal grid-fin amplitude is $0.24\,$rad; the
generation-11 normal gimbal amplitude is $0.97\times0.075=0.07275\,$rad, and the
audited generation-12 configuration widens it to
$0.995\times0.075=0.074625\,$rad; the grid rate is capped at $0.05\,$rad and the
gimbal rate at $0.015\,$rad per $0.1\,$s step; and the emergency absolute limits
are $0.35\,$rad for the grid fins and $0.0873\,$rad for the gimbal. Nonfinite
inputs or an empty feasible box trigger a byte-preserved heuristic allocator
that predates the quadratic program, and setting \texttt{alloc\_qp\_enable}~$=0$
selects that same fallback, which is what makes a direct one-key allocator
ablation available on the identical frozen protocol. One caveat must travel with
the saturation numbers: both normal gimbal caps sit strictly below the
$0.075\,$rad amplitude at which the audit records gimbal saturation, so a zero
gimbal-saturation rate is in part constructed by that strict interior reserve
and does not by itself demonstrate that authority went unused. Roll is handled
outside this allocator entirely, by a damped roll-rate loop that drives only
$\mathrm{RCS}_x$ with normal and emergency caps of $30$ and
$60\,\mathrm{kN\,m}$, while the lateral reaction-control channels and the
grid-fin roll command stay at exactly zero on both sides of the plant boundary.

\textbf{Provenance: generation~11 versus generation~12.} The method and the
evaluated configuration are deliberately reported as two different objects. The
method is the quadratic-program allocator committed at generation~11, inside the
controller
\texttt{gen11\_\allowbreak peer5\_\allowbreak satfree\_\allowbreak%
h3g1\_\allowbreak latauth\_\allowbreak binlat\_\allowbreak%
rate6p5\_\allowbreak sinkguard\_\allowbreak decouple\_\allowbreak%
qpalloc\_\allowbreak v1},
the last committed rank-1 frontier artifact of the run. The configuration we
evaluate and audit is \texttt{gen12\_peer15\_qpalloc\_cap0995\_v1}, whose
\texttt{controller.py} is byte-identical to that committed generation-11
controller and whose configuration differs in exactly one field,
\texttt{alloc\_gimbal\_cap\_frac}, from $0.97$ to $0.995$. The run committed
twelve generation boundaries, generations~0 through~11, and was terminated
during generation~12, so no generation-12 boundary exists; the cap-$0.995$
configuration is therefore the run's current complete-evidence champion,
selected post-run, and not a committed frontier artifact. Its selection carries
little scientific weight in any case. The cap sweep holds $100\%$
complete-protocol landing success and zero grid and gimbal saturation at $0.92$,
$0.97$, $0.99$, and $0.995$ alike, and $0.995$ lowers the sink-speed P95 by only
about $1.09\times10^{-5}\,\mathrm{m\,s^{-1}}$ relative to $0.97$: a
deterministic lexicographic tie-break among equal-scoring configurations, not a
method advance.

\textbf{A strict generation-boundary alternative.} Nothing in the headline
depends on the uncommitted generation. Read under an evidence rule that admits
only committed generation-boundary artifacts, the generation-11 configuration's
complete result is $12{,}288/12{,}288$ landing successes, a descriptive Wilson
$95\%$ lower bound of $99.9687\%$ on that fixed set, a sink-speed P95 of
$0.326373\,\mathrm{m\,s^{-1}}$, a lateral-speed P95 of
$0.047199\,\mathrm{m\,s^{-1}}$, a fuel P05 of $3.6701\%$, zero grid and gimbal
saturation, a grid total variation of $0.046014\,$rad, and a gimbal total
variation of $0.884548\,$rad. The full-bank audit remains an audit of the
generation-12 configuration and stays labeled as post-run supporting evidence
throughout. For reproducibility, the audited artifacts hash to
\texttt{b9a10ca5\allowbreak 4bc5}, \texttt{f4eb2d4d\allowbreak 1429},
\texttt{84f821c2\allowbreak f016}, \texttt{d6f82d58\allowbreak 13b5}, and
\texttt{5b88d6e2\allowbreak e2b5} for, respectively, \texttt{controller.py},
its controller configuration, the variant manifest, the complete evaluation
summary, and the matched baseline summary, each a SHA-256 digest truncated to
its leading twelve hexadecimal characters.

\paragraph{Matched-protocol results.} Table~\ref{tab:rocket-matched-protocol}
reports the frozen task baseline and evaluated controller under the same
protocol version, $13{,}312$ scored units, and fixed rows of each bank. Landing success rises from $495/12{,}288$
($4.0283\%$) to $12{,}288/12{,}288$ ($100\%$), a gain of $95.9717$ percentage
points, and the descriptive $95\%$ Wilson lower bound on that fixed set rises
from $3.6948\%$ to $99.9687\%$. Both Wilson figures summarize a fixed,
adaptively reused evaluation set; neither is a post-selection coverage guarantee
for any population beyond it. The per-bank rows use the same $4{,}096$ cases per
bank throughout, and the worst-radius-bin row uses the same frozen radius
strata, so no part of the comparison is purchased by resampling initial states.

\begin{table}[!htb]
\centering
\small
\caption{Panel~A: matched complete-protocol comparison of the frozen task
baseline against the evaluated controller. All rows use the same $12{,}288$
landing trajectories, except the two roll rows and the coupling row, which use
the frozen $1{,}024$-case roll suite. Rows marked \emph{adverse} are those on
which the controller is worse than the baseline.}
\label{tab:rocket-matched-protocol}
\setlength{\tabcolsep}{4pt}
\begin{tabular}{lrrl}
\toprule
Metric & Frozen baseline & \textsc{Praxist} controller & Change \\
\midrule
Landing successes & $495/12{,}288$ & $12{,}288/12{,}288$ & --- \\
Landing success rate & $4.0283\%$ & $100\%$ & $+95.9717$ pp \\
Wilson $95\%$ lower bound & $3.6948\%$ & $99.9687\%$ & descriptive \\
\addlinespace
Nominal-bank success & $5.6641\%$ & $100\%$ & same $4{,}096$ cases \\
Near-OOD-bank success & $6.4209\%$ & $100\%$ & same $4{,}096$ cases \\
Hard-OOD-bank success & $0\%$ & $100\%$ & same $4{,}096$ cases \\
Worst radius-bin success & $0\%$ & $100\%$ & same radius strata \\
\addlinespace
First-contact rate & $100\%$ & $100\%$ & unchanged \\
Vertical joint-gate pass & $11.4176\%$ & $100\%$ & --- \\
Fuel-gate pass & $7.5846\%$ & $100\%$ & --- \\
Fuel depletion & $88.6393\%$ & $0\%$ & --- \\
Fuel reserve, mean & $0.4733\%$ & $9.7872\%$ & --- \\
Fuel reserve, P05 & $0\%$ & $3.6701\%$ & $1.6701$ pp above gate \\
\addlinespace
COM sink P95 & $66.3928$ m/s & $0.32636$ m/s & $99.51\%$ lower \\
Contacting-leg sink P95 & $66.3987$ m/s & $0.33460$ m/s & $99.50\%$ lower \\
Lateral speed P95 & $1.40592$ m/s & $0.04705$ m/s & $96.65\%$ lower \\
Lateral error P95 & $7.09291$ m & $1.75437$ m & $75.27\%$ lower \\
Tilt P95 & $4.56062^{\circ}$ & $0.35184^{\circ}$ & $92.29\%$ lower \\
\addlinespace
Grid saturation step rate & $51.1584\%$ & $0$ & --- \\
Gimbal saturation step rate & $0.00950\%$ & $0$ & --- \\
Grid total variation & $7.94128$ rad & $0.04510$ rad & $99.43\%$ lower \\
Gimbal total variation & $0.49031$ rad & $0.88468$ rad
  & $80.43\%$ higher \textbf{(adverse)} \\
\addlinespace
Roll-stability rate & $100\%$ & $100\%$ & unchanged \\
Roll settling P95 & $2.5$ s & $2.5$ s & unchanged \\
Roll-to-pitch/yaw coupling P95 & $0.29786$ rad/s & $0.55865$ rad/s
  & $87.55\%$ higher \textbf{(adverse)} \\
\bottomrule
\end{tabular}
\end{table}

Two features of the panel are easy to misread.
First, the baseline already reached a $100\%$ first-contact rate: every baseline
trajectory brought a landing leg to the ground inside the horizon. What the
controller changes is therefore the \emph{quality} of that contact---the sink
speed, lateral speed, lateral error, tilt, and remaining propellant at the
interpolated first-contact state---and not whether contact occurs at all. The
$88.6393\%$ baseline fuel-depletion rate and the $66.3928\,\mathrm{m\,s^{-1}}$
baseline sink P95 make clear that most baseline contacts were uncontrolled
arrivals rather than descents that narrowly missed a threshold. Second, two
metrics move the wrong way. Gimbal total variation rises by $80.43\%$ and the
roll-to-pitch/yaw coupling P95 of the frozen roll suite rises by $87.55\%$, even
though the roll-stability gate stays at $100\%$ on all $1{,}024$ of its cases.
Both shifts are consistent with the controller moving pitch and yaw activity off
the grid fins and onto the gimbal, which is exactly what the allocator is
designed to do: grid total variation falls by $99.43\%$ over the same set, and
the grid saturation step rate falls from $51.1584\%$ to zero. The trade is a
quieter aerodynamic surface bought with a busier gimbal, and we report it as a
cost rather than folding it into the headline.

\paragraph{Full-bank audit.} The complete protocol scores a fixed
$4{,}096$-row subset of each bank. To measure how far the selected controller
extends across the rest of those banks, we froze it after the run and
evaluated all $40{,}960$ rows of each bank, giving $122{,}880$ landing
trajectories. This is a post-run coverage scan, not \textsc{Praxist} promotion
evidence: no controller was selected, ranked, or committed on its output, and
no baseline was audited at this scale, so
Table~\ref{tab:rocket-full-bank-audit} carries no baseline deltas. The three
banks are separately defined---a nominal disk, a velocity-scaled near bank
paired row-by-row with it, and a jointly faster and wider hard annulus---so
they are not one identically distributed sample, and the pooled row is a
descriptive sum only. Per-bank counts are the primary audit claim.

\begin{table}[!htb]
\centering
\small
\caption{Panel~B: post-run all-row audit of the frozen selected controller over
every row of each source bank. Per-bank counts are the primary claim; the
pooled row is a descriptive sum.}
\label{tab:rocket-full-bank-audit}
\setlength{\tabcolsep}{4pt}
\resizebox{0.98\textwidth}{!}{%
\begin{tabular}{lrrrrrr}
\toprule
Bank & Successes & Success rate & Wilson $95\%$ LB & COM sink P95
& Fuel P05 & Min.\ successful fuel \\
\midrule
Nominal & $40{,}959/40{,}960$ & $99.9975586\%$ & $99.9861709\%$
& $0.32283$ m/s & $5.7309\%$ & $4.6021\%$ \\
Near-OOD & $40{,}959/40{,}960$ & $99.9975586\%$ & $99.9861709\%$
& $0.32778$ m/s & $5.9162\%$ & $4.8016\%$ \\
Hard-OOD & $40{,}960/40{,}960$ & $100\%$ & $99.9906223\%$
& $0.32927$ m/s & $3.3543\%$ & $2.8908\%$ \\
\midrule
Pooled (descriptive only) & $122{,}878/122{,}880$ & $99.9983724\%$
& $99.9940652\%$ & $0.32636$ m/s & $3.6904\%$ & $2.8908\%$ \\
\bottomrule
\end{tabular}}
\end{table}

All $122{,}880$ trajectories reached first contact and passed the vertical and
fuel component gates, and grid and gimbal saturation rates were zero throughout.
Exactly two trajectories failed the joint predicate, both on the
lateral-speed conjunct alone. Nominal row $29254$ arrived with a lateral speed
of $0.408226\,\mathrm{m\,s^{-1}}$ against the $0.3\,\mathrm{m\,s^{-1}}$ bound,
with COM and contacting-leg sink speeds of $0.157$ and
$0.178\,\mathrm{m\,s^{-1}}$, a lateral error of $0.375\,$m, and $42.13\%$ of its
main propellant remaining. Near-OOD row $16173$ arrived at
$0.322299\,\mathrm{m\,s^{-1}}$ lateral speed, with sink speeds of $0.629$ and
$0.633\,\mathrm{m\,s^{-1}}$, a lateral error of $0.667\,$m, and $42.49\%$ fuel
remaining. Neither is an impact and neither is a fuel failure: both are
residual lateral motion at an otherwise gentle, well-placed, propellant-rich
first contact. Both also lie in the $[0,450)\,$m initial-radius bin, so each
nominal and near-OOD instance of that bin scores $3{,}686/3{,}687 =
99.972878\%$ while every other radius bin passes completely.

The audit's integrity record is as follows. All $120$ expected $1{,}024$-row
blocks completed, the bank-hash checks passed, no trajectory produced a
nonfinite value or a forbidden actuator command, and an independent
recomputation of the predicate from the stored first-contact endpoints matched
the recorded success array exactly. One margin figure should temper how the
hard-bank row is read: its minimum successful fuel reserve is $2.8908\%$, only
$0.8908$ percentage points above the strict $2\%$ gate, and its P05 reserve of
$3.3543\%$ is the lowest of the three banks. A $100\%$ count on that bank
therefore records that no row crossed the gate on this fixed set, and must not
be read as broad margin against disturbances the simulator does not model.

\paragraph{Allocator ablation.} Disabling the allocator with
\texttt{alloc\_qp\_enable=0} selects the byte-preserved heuristic allocator
that preceded it, which makes the comparison in
Table~\ref{tab:rocket-qp-ablation} a one-key ablation: the QP-off
configuration and the committed generation-11 controller share the gimbal cap
of $0.97$, differ in that single field, and were scored on the same frozen
complete protocol. The allocator lowers COM sink P95 from $0.337891$ to
$0.326373\,\mathrm{m\,s^{-1}}$ and lateral-speed P95 from $0.065620$ to
$0.047199\,\mathrm{m\,s^{-1}}$, removes gimbal saturation, and cuts grid total
variation by $99.5781\%$, at the cost of $2.1277\%$ more gimbal total
variation and a fuel-reserve P05 $0.101879$ percentage points lower.

\begin{table}[!htb]
\centering
\small
\caption{One-key allocator ablation on the frozen complete protocol. The two
configurations share the same $0.97$ gimbal cap, differ only in
\texttt{alloc\_qp\_enable}, and are both already at $100\%$ landing success.}
\label{tab:rocket-qp-ablation}
\setlength{\tabcolsep}{5pt}
\begin{tabular}{lrrl}
\toprule
Metric & Allocator off & Allocator on & Change \\
\midrule
Landing / hard / worst-bin success & $100\%$ & $100\%$ & ceiling retained \\
\addlinespace
COM sink P95 & $0.337891$ m/s & $0.326373$ m/s & $3.4088\%$ lower \\
Lateral speed P95 & $0.065620$ m/s & $0.047199$ m/s & $28.0727\%$ lower \\
Gimbal saturation step rate & $0.12303\%$ & $0$ & removed \\
Grid total variation & $10.906433$ rad & $0.046014$ rad & $99.5781\%$ lower \\
Gimbal total variation & $0.866120$ rad & $0.884548$ rad
& $2.1277\%$ higher \textbf{(adv.)} \\
Fuel reserve P05 & $3.772024\%$ & $3.670145\%$
& $0.101879$ pp lower \textbf{(adv.)} \\
\bottomrule
\end{tabular}
\end{table}

The attribution this ablation supports is narrow. Its immediate heuristic
parent was \emph{already} at $100\%$ landing success on the frozen complete
bank, so the ablation gives no evidence that the allocator produced the
$4.03\%\to100\%$ success gain; that gain belongs to the accumulated lineage
that preceded it. What the ablation does support is attributing the
actuator-load, sink-speed, and lateral-speed effects to the allocator, because
those quantities move under a single-field change with every other element of
the controller, the plant, and the evaluation held fixed. The generation-12
cap change that produced the evaluated configuration is smaller still:
sweeping the cap over $0.92$, $0.97$, $0.99$, and $0.995$ retains $100\%$
complete-protocol success and zero saturation at every setting, and $0.995$
lowers sink-speed P95 by roughly $1.09\times10^{-5}\,\mathrm{m\,s^{-1}}$
relative to $0.97$. It is a deterministic lexicographic tie-break among
configurations already indistinguishable on complete-protocol success, and we
report it as such, not as an advance in the controller's method.

\paragraph{Lineage.} Table~\ref{tab:rocket-lineage} condenses the parent chain
from the task baseline to the evaluated configuration, giving the main change
accumulated at each stage together with the complete-protocol landing success
measured there. The chain crosses the interesting thresholds early: the
fuel-commit governor and faster guidance period take the controller from
$4.0283\%$ to roughly $49\%$, the generation-2 attitude retuning takes it to
$98.3561\%$, and the last percent is closed by a sequence of narrow guards
before the allocator arrives at generation~11 with success already saturated.

\begin{table}[!htb]
\centering
\small
\caption{Condensed lineage along the actual parent chain, with the
complete-protocol landing success recorded at each stage.}
\label{tab:rocket-lineage}
\setlength{\tabcolsep}{5pt}
\begin{tabular}{p{0.13\textwidth}p{0.53\textwidth}p{0.22\textwidth}}
\toprule
Stage & Main accumulated change & Complete landing success \\
\midrule
Baseline & Original SF-CAC v2 first-contact controller & $4.0283\%$ \\
\addlinespace
Gen 1 & Fuel-commit trigger, two-band descent reference, settled-release gate,
and P0 lateral-horizon floor $28\to20\,$s; then guidance period
$0.2\to0.1\,$s & $47.9411\%\to49.2920\%$ \\
\addlinespace
Gen 2 & P0 pitch/yaw $\omega_n:0.55\to1.0$, $\zeta:0.9\to0.7$, and
reference-rate limit $6\to10^{\circ}\,\mathrm{s^{-1}}$ & $98.3561\%$ \\
\addlinespace
Gen 3--4 & Reference-rate limit restored $10\to6^{\circ}\,\mathrm{s^{-1}}$;
low-altitude P0 bandwidth schedule added; grid budget capped at $0.24\,$rad
& about $98.99\%$ \\
\addlinespace
Gen 6--8 & Initial-radius/slew branch, $6.5^{\circ}\,\mathrm{s^{-1}}$ reference
rate, and sink guard & $99.9674\%\to99.9919\%$ \\
\addlinespace
Gen 9 & Remove the $r_0<450\,$m branch's fuel-commit vertical-speed conjunct
while retaining axial-slew decoupling and the $-15\,\mathrm{m\,s^{-1}}$ release
sink guard & $100\%$ \\
\addlinespace
Gen 11 & Closed-form box-QP pitch/yaw allocator, cap $0.97$ & $100\%$, with
lower sink and lateral speed and zero grid/gimbal saturation \\
\addlinespace
Gen 12 & Gimbal cap $0.97\to0.995$ only & $100\%$, marginal lexicographic
tie-break \\
\bottomrule
\end{tabular}
\end{table}

This table records cumulative evidence along one realized parent chain, not an
additive decomposition of the total gain into per-mechanism contributions. The
generation-1 variants each bundle more than one parameter change, several
stages alter guidance and attitude behavior simultaneously, and no full
factorial over the mechanisms was run, so the success figure at a stage
attributes to everything accumulated up to it, not to the change named on its
row. The one mechanism with a clean single-field control is the allocator,
whose isolated effect is reported in Table~\ref{tab:rocket-qp-ablation}.

\paragraph{Autonomous-optimizer baseline.} The comparison this study reports is
against another autonomous research system rather than against the artifact the
task ships with. The same task package was given to Weco AI's \texttt{weco}
optimizer, an LLM-driven tree search over code built on the AIDE engine
\citep{jiang2025aide}, with the same objective metric
\texttt{landing\_success\_rate} and the same $4.03\%$ starting artifact
\citep{weco2026rocketrun}. That run evaluated $793$ candidate controllers in
$2$~hours $37$~minutes and reached a best reported score of $17.12\%$, a
$4.25\times$ relative gain over that starting point which nonetheless leaves the task
unsolved; its own search statistics record $676$ improvements against only $43$
breakthroughs, which suggests that most of its accepted candidates were marginal
gains rather than step changes. The $17.12\%$ figure is the best aggregate
reported by Weco's own run dashboard on that task and objective; the
\textsc{Praxist} value is measured by the frozen $12{,}288$-trajectory complete
protocol. The systems use different underlying language models and wall-clock
budgets, $2.6$ hours against $14.35$ hours, and Weco's search space also spanned
the task-variant manifest, whereas the \textsc{Praxist} protocol freezes the
plant, contact model, integrator, evaluator, and data so a candidate can only
change the controller. We therefore read Weco as an autonomous-optimizer
reference on the same task and metric, not as a controlled cross-model
comparison.

\textbf{Campaign configuration.} The campaign was configured for 30 generations
with 16 peers per generation, a 2-hour generation window, and four promotions
per generation. It closed 12 contiguous committed generation boundaries,
generations 0 through 11. The process received SIGTERM while working on
generation 12 after about 14.35 hours, so \texttt{run\_summary.json} records
\texttt{status: failed} with exit code 143; the result files that had already
completed remain valid evidence, but no generation-12 boundary exists and the
configuration evaluated above is consequently post-run rather than committed.
The scheduler ran up to eight concurrent one-GPU evaluations on a host with
eight NVIDIA H100 80GB GPUs, and each task experiment declared a configured
budget of 0.0075 GPU-hours, one GPU, and 2\,GiB of GPU memory. Its ledger
records 697 completed jobs, 18 failed terminal records spanning 16 distinct
jobs, and one rejected job; these are
execution-accounting facts about job scheduling, not scientific sample counts,
and they should not be added to or compared with trajectory counts. The
research-agent model was DeepSeek V4 Pro with a 1M-token context through the
Claude SDK runtime, and it is not a component of the controller: the artifact it
produced contains no neural network, no learned parameters, and no inference
call. Gem compression was disabled for this task, and the run produced no Gems.

\textbf{Simulator scope and limitations.} Every rate reported above is a
property of a frozen low-order simulator evaluated on fixed initial-state banks
with exact state feedback and no navigation estimator. The audit varies none of
wind, atmospheric density, mass error, thrust scale, inertia, drag, sensor noise
or latency, actuator failures, the integration step, the initial height, the
initial attitude, or the initial angular velocity: the only quantities that
differ between trajectories are the initial horizontal position and the velocity
scale that defines each bank. The disturbances that a flight design would have
to absorb are therefore charged against no configuration here, so a controller
tuned in this environment may allocate less authority to disturbance rejection
than a real descent would demand. The banks compound this: they are fixed and
were adaptively reused across generations, nominal and near-OOD are paired
row-by-row rather than independent, and the hard bank changes radius and
velocity jointly, so it isolates neither factor. Repeated selection against
fixed evaluation sets carries a selection-overfitting risk that the audit's
breadth does not remove, because breadth here means more rows of the same three
distributions rather than new ones.

The claim is also bounded in time and in what it charges. Scoring stops at the
interpolated first landing-leg contact, so nothing above speaks to
post-contact dwell, bounce, leg loads, slip, overturn, or terrain response; a
trajectory satisfying the predicate has arrived acceptably, not landed stably.
Reported fuel is modeled main-engine propellant only, and the reaction-control
system's propellant is uncharged, so no result here speaks to total propellant
use. The allocator itself uses a decoupled local pitch/yaw torque
approximation rather than a full nonlinear coupled allocation, which is what
makes its closed-form solution possible and what limits its validity away from
the operating points it was fitted around. The measured tradeoffs point the
same way: gimbal total variation and roll-to-pitch/yaw coupling both rise
relative to the baseline, and the hard bank's minimum successful fuel reserve
sits only $0.8908$ percentage points above the gate. Taken together, these
limitations mean the measured rates characterize this harness and these fixed
banks, and cannot be read as real-world landing reliability, evidence of
flight readiness, or any form of hardware validation.

\subsection{Quant}
\label{app:setup-quant}

\textbf{Evaluation tiers.} The walk-forward protocol of \Secref{sec:exp_quant}
is instantiated at three cost tiers over a fixed seed list of
$\{42, 77, 11, 100, 7\}$. Tier T1 evaluates one seed across 29 cells---the 28
quarterly windows plus the excluded 2026 validation window---at a configured
expectation of about 60 minutes; T2 evaluates three seeds across 87 cells at
about 150 minutes; and T3, the campaign's top tier, evaluates all
five seeds across 145 cells at about 240 minutes. The artifact reported in
\Secref{sec:exp_quant} stopped at T1---one seed over 29 cells---and carries
three hard constraint violations, so it is not clean-promotion eligible, which
is why the main text presents it as a post hoc case-study selection rather than
as the campaign's promoted artifact. The campaign's strongest confirmed-lane
result is a different, generation-19 cross-sectional attention policy with a
complete 145-cell T3 evaluation, no hard violations, and a lower CAGR. Hard evaluator timeouts are
set at 6 hours for T1 and 18 hours for T2 and T3. Because every cell retrains
the policy from scratch on its own trailing 36-month window, a T3 evaluation is
145 independent training runs rather than 145 inference passes, which is what
makes the tier structure necessary.

\textbf{Reported-policy configuration.} The reported
\texttt{gen15\_peer10\_diversify\_repair\_strong} policy receives a
12{,}505-dimensional observation, applies running normalization and a
256-dimensional projection, and updates a 128-dimensional LSTM state. Separate
actor and value heads give 3{,}412{,}809 parameters. Training uses
eight vectorized environments and 64-trading-day episodes. PPO uses
$\gamma=0.995$, a 0.1 clipping radius, three update reuses per iteration, a
$3\times10^{-4}$ learning rate, and a 0.001 entropy coefficient. Bounded replay
retains five rollout batches, samples four recurrent chunks of length 32, and
adds replay at ratio 0.30 under a 0.15 clip. The concentration repair
sets an effective-number proxy target of 12 and a maximum weight of 0.15.
The deterministic adapter selects the top eight scores at temperature 0.16,
targets gross exposure in $[0.80,0.98]$, holds cash in $[0.02,0.20]$, caps one
name at 0.28, and applies top-five, top-ten, effective-number, turnover, and L1
liquidity limits. For task-cost context, a separate archived recurrent
candidate's complete 145-cell T3 evaluation took 54{,}141.8 seconds
(15.04 hours).

\textbf{Campaign configuration.} The campaign ran DeepSeek V4 Pro through the
agent SDK as a locally resumed run, with 12 peers per generation, at most 200
generations, four promotions per generation, an 18-hour generation window, and
no plateau-based early stop. Each experiment declared 18 GPU-hours on one GPU
per pod; Gem compression was enabled on a six-generation period with at most
three resets and four active Gems. Synthesis opened on a 150-to-480-minute
window once at least six pods had reported and used the two-round PI panel, with
a Chair budget of 12 peer contracts. The run closed 37 generation boundaries
(generations 0 through 36) with a partial generation 37 open at termination, and
its archived timestamps span 176 hours and 30 minutes, ending on an external
signal rather than at the configured generation cap. Run logs identify eight
80\,GB H100 devices used to parallelize some multi-seed evaluations, which again
records those evaluations rather than sustained campaign-wide occupancy.

\subsection{SLAM}
\label{app:setup-slam}

\textbf{Measurement protocol.} The comparison in \Secref{sec:exp_slam} reports
one accepted run per (method, sequence) pair over the fourteen sequences. There
are no training seeds and hence no run-to-run variance estimate, but the
campaign is not a single pre-registered pass either: several sequences were
replayed or re-run---for example \texttt{rtp\_01} on the \textsc{CovSched}
side and \texttt{eee\_02} on the baseline side---and the table takes the
accepted run rather than the first one, without a pre-declared acceptance
rule. This
reflects the artifact's nature rather than an economy---the
candidate is a compiled variant of a deterministic estimator, not a trained
policy, so a peer's inner loop is a build followed by a bag replay, and the
evaluation units are 100-second windows during search and full-sequence
playbacks at validation. Trajectory error is APE RMSE after rigid Umeyama
alignment to the Leica ground truth with a 0.02-second association tolerance,
visual-update counts are parsed from the runs' launch logs, and the reported
timing field is the evaluator's last-captured cumulative visual-path (VIO)
average, not a measured end-to-end wall time per LiDAR frame.

\textbf{Mechanism parameters and their qualifications.} The champion
configuration sets \texttt{PVTR\_MODE=8} (observability-gated visual scheduler)
and \texttt{VMAP\_DEDUP=1} (visual-map admission filter). Two implementation
details qualify the description in \Secref{sec:exp_slam}. First, the dwell
counter decimates frames that stay degenerate for a long time rather than
protecting them, so it \emph{raises} skip demand; what prevents permanent
visual starvation is the saturation of the skip budget below one together with
the hard cap of three consecutively skipped frames. The net schedule is still
strongly protective of low-observability frames, because only a handful of
frames ever exceed the dwell threshold. Second, the admission filter
deduplicates by rejection rather than fusion: a candidate within $0.08$\,m and
$15^\circ$ in normal of an existing map point in the same voxel is discarded,
which is why the run reports admitted and retrieved point counts instead of a
merged-observation statistic.

\textbf{Further caveats on the comparison.} Beyond the timestamp confound
treated in \Secref{sec:exp_slam}, three caveats bound the result. APE is
computed over ground-truth-associated poses, and those sets are not identical
between the arms: \textsc{CovSched}'s associated-sample ratio averages $1.03$
with median $0.985$ but ranges from $0.79\times$ on \texttt{eee\_03} to
$1.76\times$ on \texttt{nya\_02}, so the full-sequence comparison should be
read with coverage. The effect is also primarily a full-trajectory and
resource-efficiency one rather than a uniform improvement of every local
statistic: relative pose error over short $1$--$30$\,s windows is mixed across
sequences. Finally, the component controls cover one to three sequences, not
fourteen, and the rate-matched periodic-skip control disables the
map-admission filter that \textsc{CovSched} keeps enabled, differing in two
respects at once: they show that generic visual thinning is already a strong
baseline, but do not separate the LiDAR-observability signal from the
map-admission filter, which the paired re-run listed in
\Secref{sec:exp_slam} would supply.

\textbf{Host provenance.} Sampled pod logs from the campaign show the CPU
FAST-LIVO2 build path: 168 logical cores visible, roughly 25\,GiB of pod memory,
ROS Noetic, no \texttt{nvidia-smi} present, and no NVIDIA device exposed to a
direct probe. That characterizes the containers we sampled rather than every
process in the campaign, and because pods share a physical machine the reported
per-frame timings are indicative of the relative cost of the two visual-update
policies rather than isolated-benchmark measurements of either. The timing
field itself is weaker than a clean ratio: the evaluator scans the launch log
for the last line named \texttt{Average Total Time}, which both the LiDAR and
the visual thread emit, so it captures whichever thread wrote last; skipped
frames enter the visual average as zero time; and the scheduler's own
eigenvalue computation falls outside the timed region. Separating the two
threads over the fourteen accepted runs gives a $72.4\%$ mean reduction on the
visual path and a $1.0\%$ mean \emph{increase} on the LiDAR path, with the
summed internally timed thread-wall work falling $22.1\%$. We therefore report
the visual-path reduction and that aggregate proxy, and we do not treat the
ratio as an end-to-end latency, throughput, or CPU-load measurement; the
absolute milliseconds are host-dependent.

\textbf{Campaign configuration.} The campaign used DeepSeek V4 Pro via the
cheap-pod command-line coding-agent profile, with 8 peers per generation,
at most 200 generations, two promotions per generation, and a 5-hour generation
window. Its effective specification declares no compute budget, no Gem
configuration, and no evaluator seed list, so design allocation and Gem
behavior fall back to the system defaults of
Table~\ref{tab:config-defaults}. Per-experiment GPU budgeting has no entry in
that table and none in this campaign's specification; the only GPU figure in
its ledger is a stage-level request of 3{,}200 GPU-hours, which is not a
per-experiment realized budget. The deterministic evaluator uses no seeds. The
archived active-run snapshot spans 43 generation boundaries (generations 0
through 42) over 145 hours 52 minutes.

\subsection{Fusion}
\label{app:setup-fusion}

\textbf{Evaluation protocol and cost.} The benchmark of \Secref{sec:exp_fusion}
runs 5 scenarios $\times$ 3 seeds (0, 1, and 2) $\times$ a 100-step horizon, so
a controller is scored on 15 episodes and at most 1{,}500 survived simulator
steps. The selected artifact is synthesized Torch control code rather than a
trained policy, so there are no optimizer epochs or policy-gradient iterations
anywhere in this study: the meaningful iteration counts are generations, peer
sessions, episodes, and simulator steps. The three runs behind
Table~\ref{tab:fusion_results} took 1{,}682.7, 1{,}819.4, and 1{,}838.7 seconds
of runner wall time for zero feedback, the PCS-style controller, and the \textsc{Praxist}
controller respectively---28.05, 30.32, and 30.64 minutes---which makes the
whole reported comparison a sub-two-hour evaluation.

\textbf{Hardware attribution.} Representative official evaluations from the
same archived campaign show four NVIDIA H100 80GB devices visible on the host.
This is host context rather than workload occupancy: FreeGSNKE evaluation is
CPU-heavy and process-parallel, and the configured 8 GPU-hours per experiment is
a scheduling reservation rather than a realized-compute measure.

\textbf{Campaign configuration.} The lineage that produced
\texttt{HybridJacobianPDV1} ran DeepSeek V4 Pro through OpenRouter under the
agent SDK as a local run, configured for at most 12 generations with 5 peers,
two promotions per generation, and a nominal 6-hour generation window; each
experiment declared 8 GPU-hours, 20\,GB of memory, a half-device utilization
request, 16 CPU cores, and at most five parallel jobs per pod. The run did not
reach its configured cap: it completed 7 generation result sets (generations 0
through 6) with all five pod slots filled in each, recording 114 peer sessions
and seven synthesis sessions before termination. This is a legacy trajectory that predates the
current boundary object, so those generations are attested by their result sets
and syntheses rather than by \texttt{generation\_boundary.json} files. Start to termination spans
about 18 hours and 15 minutes, of which the summed generation durations account
for about 17.15 hours. The selected controller first appears in
generation 0, and the remaining six generations refined and tested it.

\textbf{Full-horizon and common-horizon precision.}
Table~\ref{tab:fusion_full_horizon} gives both precision metrics per scenario, so
that the horizon choice discussed in \Secref{sec:exp_fusion} can be checked
rather than taken on trust. For the zero-feedback baseline the two coincide by
construction, since the common horizon is capped at that controller's own
survival length. The two metrics disagree about which closed-loop controller is
more precise: on the common horizon \textsc{Praxist} leads everywhere except the
certification scenario, while on the full horizon the PCS-style controller leads
on three of the four perturbed scenarios and on the aggregate, and loses the
certification scenario by an order of magnitude. Neither ordering is the
benchmark's pass rule, which no controller here satisfies.

\begin{table}[htb]
\centering
\small
\caption{Fusion case study: full-horizon and common-horizon WNRMSE $p_{95}$
(lower is better) per benchmark scenario. The full horizon is the benchmark's
original definition, scoring every step a controller survives; the common
horizon caps each episode at the zero-feedback baseline's survival length for
that scenario and seed. Aggregate rows are pooled over all scored steps rather
than averaged across scenarios.}
\label{tab:fusion_full_horizon}
\setlength{\tabcolsep}{6pt}
\begin{tabular}{lccccc}
\toprule
& Zero-feedback & \multicolumn{2}{c}{MAST-U PCS-style} & \multicolumn{2}{c}{\textsc{Praxist} (ours)} \\
\cmidrule(lr){2-2}\cmidrule(lr){3-4}\cmidrule(lr){5-6}
Scenario & full $=$ common & full & common & full & common \\
\midrule
\texttt{main\_shape}  & 4.77 & 4.18  & 2.63 & 8.50 & 2.55 \\
\texttt{axis\_ip}     & 6.10 & 5.76  & 3.54 & 6.26 & 3.48 \\
\texttt{ip\_shape}    & 5.12 & 3.37  & 2.63 & 3.22 & 2.43 \\
\texttt{xpoint}       & 3.85 & 2.87  & 1.38 & 2.98 & 1.21 \\
\texttt{cert\_shape}  & 4.74 & 13.46 & 2.76 & 3.76 & 2.81 \\
\midrule
Aggregate             & 4.89 & 4.42  & 2.99 & 4.65 & 2.86 \\
\bottomrule
\end{tabular}
\end{table}

\textbf{Runner and reset sensitivity.} The same controller scores substantially
differently under different environment lifecycles, so every Fusion figure must
be read together with the runner that produced it. Under the fresh-reset,
action-clipping sidecar used for Table~\ref{tab:fusion_results}, the selected
controller survives 1{,}264 steps with a 0.667 full-horizon completion rate; a
second fresh validation gives 1{,}298 and 0.733; and the default in-run
evaluator, which resets once and deep-copies the post-reset environment for each
episode, gives 894 and 0.200. The reported protocol builds and resets a fresh
environment for every episode and is applied identically to all three
controllers in Table~\ref{tab:fusion_results}, so the comparison is internally
consistent even though its absolute levels are lifecycle-dependent.

\subsection{Job-Shop}
\label{app:setup-jobshop}

\textbf{Evaluation protocol and cost.} A complete evaluation replays the 50
fixed MK07--MK15 event tapes of \Secref{sec:exp_jobshop} through the frozen
simulator, with the candidate served by one persistent GPU worker under the 1.5-s
soft and 2.0-s hard deadlines, and re-runs the task's comparators on the same
tapes: rule~3, ungated and gated visible LNS, the earlier PPO checkpoint, and a
non-ranking clairvoyant CP-SAT diagnostic. Every schedule is re-checked by the
independent validator, the result file is rebuilt from raw evidence and compared
byte for byte, and the paired statistics use a fixed bootstrap generator with
10{,}000 resamples. The candidate's own 50 episodes took 679.6 seconds of wall
time for the selected artifact, over 508 event decisions; the comparator
re-runs make a complete evaluation considerably longer, and the scheduler
reserves about two hours for one. Evaluation is
reproducible under normal load: two later candidates that differ from the selected artifact only
in a scale-adaptive repair budget, which is the identity at the task's
canonical scale, reproduced its 50 makespans exactly, and repeated quiet runs of
the frozen incumbent reproduced its 1.381\% exactly; under heavy load the
time-limited solves can drift, and one loaded incumbent re-run recorded two
timeouts. The comparator rows of Table~\ref{tab:jobshop_results} are
the task's canonical write-once measurements; the rows re-run alongside the
selected artifact differ slightly, since decoding and search are wall-clock
limited (0.541\% rather than 0.547\% for the earlier PPO checkpoint).

\textbf{Lineage.} Table~\ref{tab:jobshop_lineage} lists the complete
evaluations on the path to the selected artifact, together with two other points
of the run's three-axis Pareto frontier. Each row is a single complete
evaluation on the same 50 tapes, so successive rows are paired comparisons on
identical episodes, and the differences between adjacent rows are directional
evidence rather than an additive decomposition.

\begin{table}[htb]
\centering
\small
\caption{Job-Shop lineage: complete 50-episode evaluations on the path to the
selected artifact. Mean is the makespan improvement over rule~3 (\%) with its
paired-bootstrap 95\% interval; p99 is the 99th-percentile decision latency. The
last two rows are other Pareto points of the run, off the selected path.}
\label{tab:jobshop_lineage}
\setlength{\tabcolsep}{4pt}
\begin{tabular}{clccc}
\toprule
Gen. & Change & Mean [95\% CI] & CVaR$_{10}$ & p99 (ms) \\
\midrule
--- & Frozen incumbent & 1.381 [0.632, 2.162] & $-2.973$ & 1653 \\
0 & Control: LNS on rule~3's plan, no neural plan & $-1.712$ [$-2.794$, $-0.711$] & $-10.416$ & 1663 \\
1 & Gate threshold lowered from 4\% to zero & 1.746 [0.777, 2.772] & $-4.062$ & 1707 \\
3 & Prior-shrinkage duration estimator & 2.668 [1.899, 3.486] & $-1.024$ & 1717 \\
3 & Ablation: pure prior (no running average) & 2.524 [1.719, 3.356] & $-1.277$ & 1713 \\
4 & Disagreement-gated estimator bridge & 2.981 [2.123, 3.883] & $-1.503$ & 1714 \\
6 & Per-event gate band (gen-5 mechanism, re-run) & 3.105 [2.231, 4.008] & $-1.889$ & 1712 \\
7 & Event-targeted repair, 0.20-s budget & 3.303 [2.392, 4.229] & $-1.690$ & 2002 \\
8 & Fast repair, margin floor, latency cap (selected) & \textbf{3.375} [2.415, 4.364] & $-1.889$ & 1731 \\
\midrule
6 & Realized-credit estimator selector & 3.111 [2.304, 3.961] & $-0.934$ & 1721 \\
11 & Reduced serving budget (0.55-s LNS) & 3.128 [2.287, 4.028] & $-0.934$ & 1095 \\
\bottomrule
\end{tabular}
\end{table}

The generation-7 repair is the only row that breaks the deployment contract: six
of its decisions exceeded the 2-s hard deadline. The
generation-8 change was aimed at that breach, and its mean is statistically tied
with its parent's. The two off-path rows show that the frontier the campaign
retained is a genuine trade-off rather than a single winner: the realized-credit
selector holds the run's best tail among high-mean points (worst episode
$-1.919\%$, against $-4.575\%$ for the selected artifact), and the
reduced-budget variant gives up a quarter of a point of mean for a 99th-percentile
latency 37\% below the incumbent's.

\textbf{Evidence scope.} The selected artifact reuses the incumbent's network
checkpoint unchanged. Of the 226 candidate variants the run built, 194 worked on
the deployment side and 32 retrained the network within the 45+45-minute cap;
the two retrained candidates that completed the full evaluation scored 1.911\%
and 1.945\%, so the run's strongest results came from estimation, admission, and
repair rather than from retraining. The 50 episodes are, as the task
documentation states, a fixed calibration set rather than a hidden test, and the
campaign scored 119 complete evaluations on them. As in the Rocket study, a set
reused adaptively in this way carries selection-overfitting risk, and the
reported rates are measurements on that set. The parameters of the duration
estimator and of the per-event gate thresholds, which carry most of the gain,
were fitted on public MK01--MK06 development episodes, as an in-run audit
confirmed, although the choice among estimator and gate variants was made on
the calibration set. The repair's margin floor and latency cap were set from the
parent's evaluation on the 50 episodes; that step is the one whose mean gain is
statistically tied with its parent's, and its purpose was to restore the
deadline.

\textbf{Campaign configuration.} The campaign ran DeepSeek V4 Pro through the
agent SDK as a local run, with 12 peers per generation, three promotions per
generation, a 3.5-hour generation window, at most 30 generations, and a
configured 1.92 GPU-hours per experiment on one GPU. DIG ran only in
generation~0, QD remained enabled, and the task exposed no network literature
search: the research corpus was the project's own documentation and provenance.
The run completed generations 0 through 12 and was terminated by SIGTERM during
generation~13, 46 hours and 50 minutes after it started; that span includes one
resume from the generation-4 boundary after a harness-only repair of the task's
evaluation entry point, which discarded the in-progress generation~5. Peer
sessions consumed 6.10 billion input tokens, 6.05 billion of them served from
cache, and 34.2 million output tokens; the archived ledger does not itemize the
synthesis sessions, so these totals are a lower bound on model usage and we
report no dollar figure.